\documentclass[preprint,floatfix] {revtex4} 
\usepackage{graphicx}
\usepackage{subfigure}

\begin{document}

\title{Information entropy as a measure of tunneling and quantum confinement 
in a symmetric double-well potential}

\author{Neetik Mukherjee}
\altaffiliation{Corresponding author. Email: neetik.mukherjee@iiserkol.ac.in.}
\affiliation{Department of Chemical Science \\
Indian Institute of Science Education and Research (IISER) Kolkata, 
Mohanpur-741246, Nadia, West Bengal, India}

\author{Arunesh Roy}
\altaffiliation{Present address: Zernike Institute for Advanced Materials, 
University of Groningen, 9747 AG Groningen, The Netherlands.}
\affiliation{Department of Physical Science\\
Indian Institute of Science Education and Research (IISER) Kolkata, 
Mohanpur-741246, Nadia, West Bengal, India}

\author{Amlan K.~Roy}
\altaffiliation{Corresponding author. Email: akroy@iiserkol.ac.in.}
\affiliation{Department of Chemical Sciences\\
Indian Institute of Science Education and Research (IISER) Kolkata, 
Mohanpur-741246, Nadia, West Bengal, India}

\begin{abstract}
Information entropic measures such as Fisher information, Shannon entropy, Onicescu energy and
Onicescu Shannon entropy of a symmetric double-well potential are calculated in both position 
and momentum space. Eigenvalues and eigenvectors of this system are obtained through a 
variation-induced exact diagonalization procedure. The information entropy-based uncertainty 
relation is shown to be a better measure than conventional uncertainty product in interpreting 
purely quantum mechanical phenomena, such as, tunneling and quantum confinement in this case. 
Additionally, the phase-space description provides a semi-classical explanation for this feature. 
Total information entropy and phase-space area show similar behavior with increasing barrier height. 

\vspace{5mm}

\vspace{5mm}

\end{abstract}
\maketitle

\section{Introduction}
Quantum  confinement is a measure of particle localization \cite{razavy2003quantum}. In a system, 
where tunneling is present, amplitude of tunneling depends on strength of the barrier of potential 
energy landscape, and so is quantum confinement. Tunneling is essentially related to the probability 
of finding a particle even in classically forbidden regions. It also indicates delocalization of a 
particle through a barrier \cite{schumm2005matter}. In other words, tunneling increases the 
uncertainty of finding a particle under confinement \cite{niquet2000quantum}. The well-known 
uncertainty relation \cite{griffiths1995introduction}, 

\begin{equation}\label{uncer}
\triangle x \triangle p \geq \frac{1}{2}, 
\end{equation}
($\hbar = 1$) implies that the momentum of a quantum particle becomes increasingly more uncertain 
with an increase in certainty of finding the particle in spatial confinement and vice-versa. 
Equation~(1) has been successfully used to study uncertainty-like relationship in case of 
spherically symmetric potentials including 
radial position-momentum uncertainties in Klein-Gordon H-like
atoms \citep{tsapline1970expectation,majernik1997entropic,grypeos2004hvt, kuo2005uncertainties,qiang2006radial,
qiang2008radial}. Uncertainty in position 
space is known to provide a good measure of spatial delocalization \cite{vidal2000interaction} of 
a quantum particle. However, although it provides a lower bound on the information content, it does 
not quantify the full information content of a particle in a given quantum state. 

Arguably, the double-well (DW) represents one of the most important potentials in 
quantum mechanics as it offers much valuable insight towards the understanding of various 
quantum phenomena. It has been a subject of considerable research activity from the dawn of quantum mechanics 
till today (see, for example \cite{griffiths1995introduction,merzbacher98,jelic2012}). 
A particle in a symmetric DW potential is a prototype of a system for which 
an understanding of full quantum mechanical description of confinement needs a more complete 
theory than that of the traditional uncertainty product relation, Eq.~(\ref{uncer}). It has 
relevance in several physical, chemical phenomena in which the potential can be modeled as a 
two-state system \cite{verguilla1993tunneling}, wherein one may identify two wells 
as two different states of a quantum particle. These applications range from hydrogen bonding 
\cite{somorjai1962double}, umbrella flipping of ammonia molecule \cite{campoy1989inversion,
Tserkis2014497},
anomalous optical lattice vibrations in HgTe \cite{kozyrev2010}, 
proton transfer in DNA \cite{RevModPhys.35.724} to model brain micro-tubules 
\cite{faber2006information}, internal rotation \cite{hammons1979electron}, etc., to name a few. 
Physics of solid-state devices \cite{holmes1980addendum}, solar cells and electron tunneling 
microscopes \cite{holmes1980addendum} constitute some other important applications. These constitute
some of the major motivations behind choosing this potential.

In case of a DW, there exists an interesting interplay between localization and delocalization 
effects. These competing effects lead to a number of quasi-degenerate pair states. An 
increase in the positive term (strength) reduces spacing between classical turning points but at 
the same time, reduces barrier area as well as barrier height (refer to Eq.~(2) later). Conversely, 
the negative term increases spacing between classical turning points but also increases barrier 
area and barrier height. This invites a careful analysis of these two responses from a more complete 
viewpoint. In this context, the primary motivation is to study DW potentials using information
entropy (IE)-based measures, leading to a fuller description of such contrasting effects. It is
worth mentioning that, throughout the article, unless otherwise specified, IE implies any one of the 
four quantities defined later in Section III. 

At first, Eq.~(1) is used to interpret quantum confinement in a symmetric 1D DW potential (main focus of 
this work), which is written in the following form,  

\begin{equation}\label{dw}
V(x) = \alpha x^{2n}-\beta x^{2m} +h,
\end{equation}
where $n, m$ are two positive integers such that $n>m$, and $h=\frac{\beta^2}{4\alpha}$ signifies barrier height. In 
what follows, we consider the specific case of $n=2,m=1$, unless mentioned otherwise.
It is because, this choice corresponds to the simplest possible DW case within the polynomial potential family in 
Eq.~(2). One requires a quartic and quadratic term; former should be positive while latter 
negative. \emph{Exact} analytical solution of this model potential has not been reported as yet; many approximate 
methods were developed over the years. Energy 
spectra of this and central DW potentials along with many interesting facets have been reported in a number works.
Literature is vast; some representative ones are given here \cite{child2000quantum,zappala2001,turbiner2005,
zhao2005,pedram2010,amadou2015}. It is appropriate to mention here that the 
characteristic feature of quasi-degeneracy 
\cite{pathak1995eigenstates} present in a DW potential can not be examined from commonly used 
uncertainty relation, Eq.~(\ref{uncer}). 

Now a plot of traditional uncertainties (in position and 
momentum space and their products) of above potential with respect to $\beta$ is given in left, 
right panels of Fig.~(1) for ground, first excited state respectively. It reveals that, 
for both states $\triangle{x}$, $\triangle{x}\triangle{p}$ increase monotonically with $\beta$. 
However, $\triangle{p}$, with increase in $\beta$, at first tends to decrease slightly from its 
starting value, attains a minimum at a particular $\beta$ and finally goes on increasing (although 
at a much lower rate compared with the other two). But these trends cannot be used 
to interpret either confinement (trapping of particle within one of the wells) or tunneling. 

\begin{figure}                         
\begin{minipage}[c]{0.40\textwidth}
\centering
\includegraphics[scale=0.45]{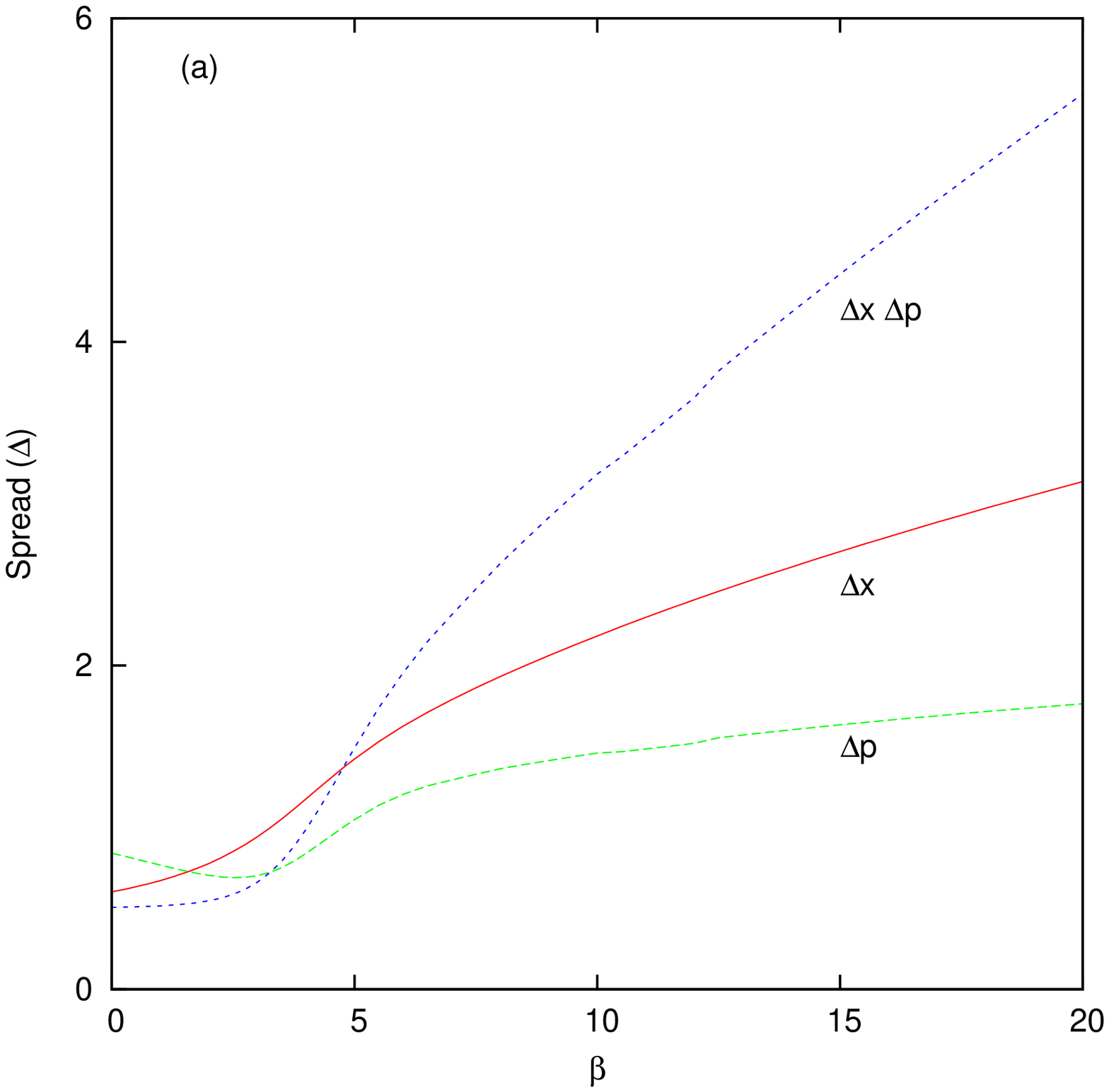}
\end{minipage}%
\hspace{0.9in}
\begin{minipage}[c]{0.40\textwidth}
\centering
\includegraphics[scale=0.45]{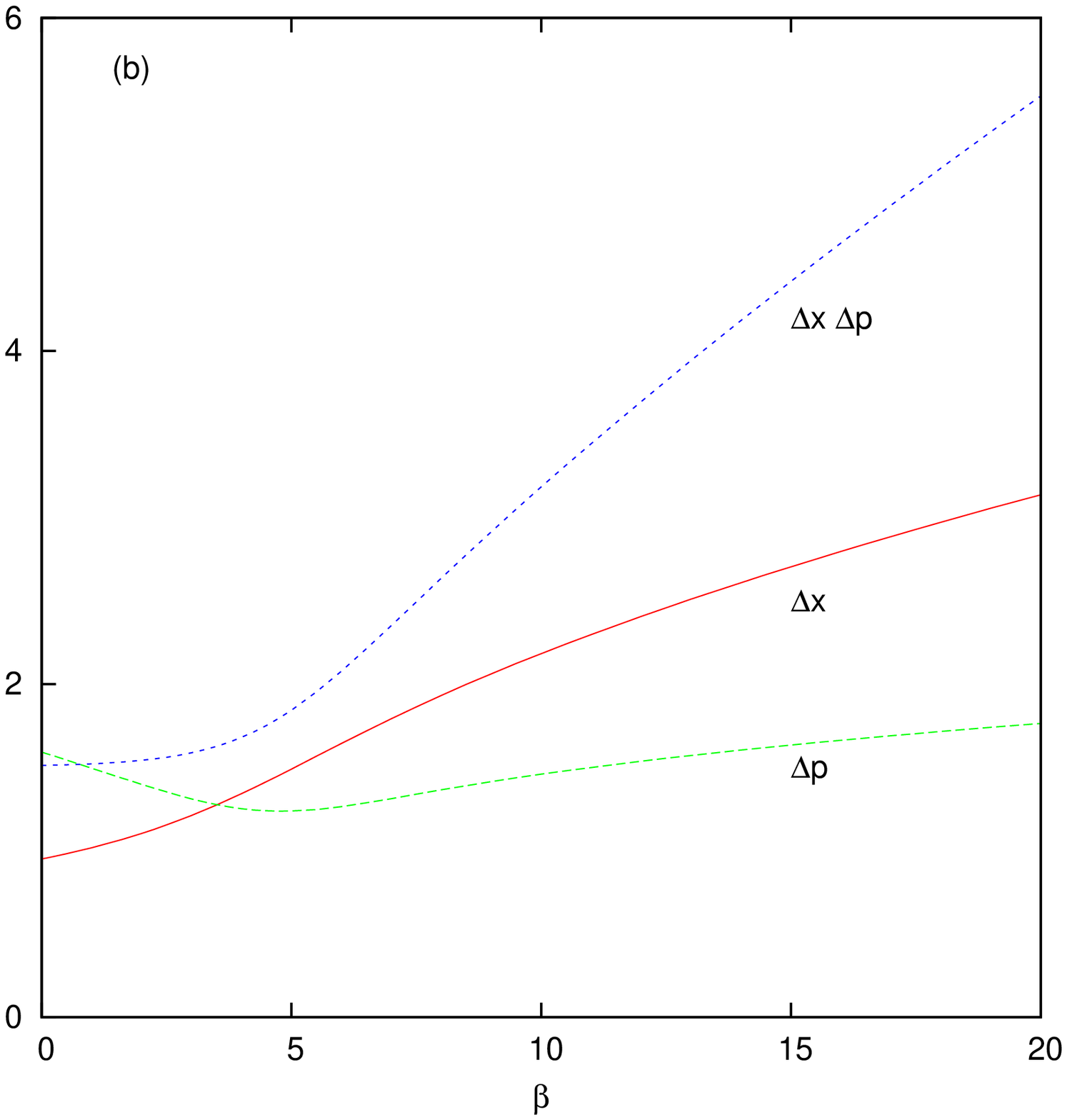}
\end{minipage}%
\caption{$\triangle{x}$, $\triangle{p}$ and $\triangle{x}\triangle{p}$ plotted against barrier 
strength $\beta$ for (a) ground and (b) first excited states of the DW potential in Eq.~(2). 
$\alpha=1, m=1, h=\frac{\beta^2}{4\alpha}$.}
\end{figure}

During the past three decades information measures have evolved in to a subject of considerable 
topical interest 
\citep{cover2012elements}. Ever since the inception, these have provided major impetus not only 
in the domain of statistical mechanics but also in many diverse fields. IE has been used to 
handle divergent perturbation series \citep{bender1987maximum}, image reconstruction and spectral 
interpretation \citep{skilling1991maximum}, polymer science \citep{la2003constrained}, 
thermodynamics \citep{poland2000maximum, singer2004maximum, antoniazzi2007maximum}, etc. Later,  
numerical \citep{agmon1979algorithm, mead1984maximum} and practical implementation of various 
related quantities has attracted significant attention from many researchers. Also, in a 
purely quantum-mechanical context, these have been discussed covering a broad range of topics 
\citep{bialynicki1975uncertainty, atre2004quantum, plastino1995quantum}. Primarily, the curiosity 
behind these entropic measurements grew out of the notions of IE in position and momentum spaces 
that led to entropic uncertainty relations \citep{bialynicki1975uncertainty, aydiner2008quantum}. 
Lately, IE has been extensively studied for confined quantum systems ranging from confined 
harmonic oscillator \citep{laguna2014quantum} to confined atom \citep{patil2007characteristic, 
romera2005fisher, toranzo2014pauli}. The interest in the subject continues to grow day by day. 

Initially, maximization of IE with known values of first few moments was carried out exhaustively 
for a number of problems to find ground quantum stationary states \citep{canosa1989ground, 
plastino1995quantum}. Later it was found that, in place of supplying values of individual moments 
it is more profitable to employ principle with moment recursion relations as constraints 
\citep{bandyopadhyay2001maximum, bandyopadhyay2002exploiting}, which are
obtainable by analyzing a given problem. Such a route yields nice results for both stationary 
quantum states and classical chaotic states \citep{bandyopadhyay2001maximum}. A neat WKB 
analysis was imitated to find variation of IE with a quantum number for stationary states 
in $x^{2M}$ potentials \citep{dehesa2002quantum, bandyopadhyay2002exploiting}. This study reveals 
an increase in IE with state index $n$, except for confined systems. Later, spectroscopic relevance 
of IE in detection of avoided crossings has been highlighted \citep{gonzalez2003shannon}, taking 
explicit example of a H-atom in presence of an external perturbation. 

It is well known that, IE offers a stronger bound \citep{patil2007characteristic} than that given 
by the traditional uncertainty relation, Eq.~(1). One can study the measures in conjugate position 
and momentum space. 
In past few years there has been an upsurge of interest in Shannon entropy for a variety
of potentials, such as P\"oschl-Teller-like \cite{sun2013quantum}, Rosen-Morse \cite{sun2013quantum1}, 
squared tangent well \cite{dong2014quantum}, position-dependent mass Schr\"odinger equation \cite{yanez2014quantum}, 
hyperbolic \cite{valencia2015quantum}, infinite circular well \cite{song2015shannon}, hyperbolic DW 
\cite{sun2015shannon} potential, etc.
In this endeavor, our objective is threefold. First, to study various measures 
such as, Fisher Information (I) \citep{frieden2004science, romera2005fisher}, Shannon Entropy (S) 
\citep{shannon1951prediction}, Onicescu Energy (E) \citep{agop2014implications} and 
Onicescu-Shannon Information (OS) \citep{LpezRuiz1995321, chatzisavvas2005information, 
panos2007comparison}, in a symmetrical DW potential, in both position and momentum space. Besides, 
we also calculate net information measures of all the above said entropy-based quantities to 
analyze the behavior of a quasi-degenerate pair in DW. Subsequently, these IEs are employed to 
interpret the confinement and trapping \cite{karski2009quantum} of particle within one of the 
potential wells. Secondly, we have made an attempt to explain probability ($T$) of finding the 
particle within the barrier using IE measures. For this, $T$ is calculated following a 
semi-classical approximation \cite{maslov2001semi}, 

\begin{equation}
\left[V(x)-E_{n}\right]=0, \ \ \ \ \ \ \ \ \ \ \ 
T =\int^{x_{m}}_{-x_{m}} |\psi(x)|^2 \ dx.
\end{equation}
Here, $x_{m}$ denotes classical turning point of the particle within the barrier.
Finally, the rate of change of these IEs with respect to $\beta$ is examined to get further 
insight into the possibility of transfer of information from position to momentum space.
There are three possibilities,

\begin{equation}\label{di}
\left|\frac{d(IE)_{x}}{d\beta}\right|=\left|\frac{d(IE)_{p}}{d\beta}\right|, \ \ \ \ \ \ \ 
\left|\frac{d(IE)_{x}}{d\beta}\right|>\left|\frac{d(IE)_{p}}{d\beta}\right|, \ \ \ \ \ \ \ 
\left|\frac{d(IE)_{x}}{d\beta}\right|<\left|\frac{d(IE)_{p}}{d\beta}\right|.
\end{equation}
Equation~(\ref{di}) indicates that the rate of change of $(IE)_{x}$ w.r.t. $\beta$ is 
\emph{exactly equal} to rate of change of $(IE)_{p}$ w.r.t. $\beta$; thus total IE is 
conserved and one gets the harmonic trend. The remaining two inequalities imply that 
the rate of change of $(IE)_{x}$ w.r.t. $\beta$ is \emph{not} equal to rate of change of 
$(IE)_{p}$ w.r.t. $\beta$; then we can expect an extremum in $\frac{d(IE)}{d\beta}$ at certain 
values of $\beta$. Additionally, we make an attempt to correlate total IE with phase-space 
area ($A_{p}$),
\begin{equation}
A_{n}=\int \sqrt{(V(x)-E_{n})} \ dx.
\end{equation}
Phase-space description is a semi-classical concept \cite{ceausu2014phase}. Thus, while IE offers 
a purely quantum mechanical view, $A_{n}$ can provide a semi-classical angle in the study of DW 
\cite{manfredi2000entropy}. Recently, phase-space IE calculation for a DW has already been 
done using Wigner probability distribution function \citep{ceausu2014phase}. Thus in this 
work, our objective is to relate the IE results with phase-space calculations. Additionally, 
efforts are made to examine variation of phase-space area with respect to (a) the onset of 
tunneling (b) changes in $\beta$ and (c) confinement.  

\section{Methodology}
For our purposes, the DW potential is conveniently written as, 
\begin{equation}
V(x) = \alpha x^{4}-\beta x^{2} + \frac{\beta^{2}}{4 \alpha}.
\end{equation}
An increase in the quartic parameter, $\alpha$ reduces separation between classical turning points 
and decreases barrier strength. Thus, on one hand, it increases localization and on the other hand, 
decreases the confinement of particle within a well. An increase in $\beta$ also causes competing 
effects on the particle. It leads to an increase in the separation of classical turning points 
implying added delocalization of particle, whereas at the same time, barrier area and barrier 
height increase, promoting localization into one of the wells.

It may be noted that, the lowest states where effect of $\beta$-variation is observed most 
prominently, are ground and first excited state; hence our study mostly focuses on these two 
states. However, 
we also analyze the effect of $\beta$ on second and third excited state, to verify if the 
qualitative trends remain unaltered from first two states. It is also worthwhile to study 
how the successive appearance of nodes in wave function (as state index increases) impacts the 
behavior of IE. Note that the last term in Eq.~(2) shifts the minimum at zero to make energy 
value positive. The potential is symmetric around $x = 0$ and location of two minima are at 
$x_{0} = \pm \sqrt{\frac{\beta}{ 2 \alpha}}$, while the maximum value of potential becomes 
$h=\frac{\beta^2}{4\alpha}$ \cite{jelic2012}. 
 
The Hamiltonian in position space is:
\begin{equation}
\hat{H_{x}}=-\frac{d^{2}}{dx^{2}}+\alpha x^{4}-\beta x^{2} + \frac{\beta^{2}}{4 \alpha},
\end{equation}
whereas in momentum space, this reads, 
\begin{equation}
\hat{H_{p}}=p^{2}+\alpha\frac{d^{4}}{dp^{4}}+\beta\frac{d^{2}}{dp^{2}}+ \frac{\beta^{2}}{4 \alpha}.
\end{equation}
For sake of convenience, we choose $m=\frac{1}{2}$ and $\hbar=1$.

\subsection{Variation-induced exact diagonalization}
One needs eigenvalues, eigenfunctions to calculate IE of respective states of DW potential. This 
was initially attempted using variation method for a particle-in-a box basis set. Although we were 
successful in obtaining position-space IE in this way, this choice of basis was found to 
be unsuitable for the same in momentum space, as eigenstates of Hamiltonian corresponds to a shifted 
Dirac-Delta function. This is because, momentum eigenstates are obtained by performing a Fourier 
transform of position eigenstates, which for particle-in-a box potential are simply plain waves. 
Therefore, a different basis was desirable; present calculation was done using an Exact 
diagonalization (ED) method with a Hermite basis. This approach \cite{griffiths1995introduction} 
can be applied for both eigenvalues, eigenfunctions of a physical potential characterized by a 
polynomial form. \emph{In principle,} exact energies can be obtained 
provided the basis set is \emph{complete}. Our Hamiltonian is represented in terms of raising and 
lowering operators of a conventional quantum harmonic oscillator (QHO). Thus we choose QHO 
number-operator basis with a single non-linear parameter $\gamma$, 
which is related to the force constant, $\kappa$ of the QHO, $\gamma^2=\frac{1}{8}\kappa$ (see Eq.~(16) later).
Dimension of the Hamiltonian 
matrix is set at $N=100$ to guarantee convergence of eigenvalues. Table~I gives a sample
calculation for $\alpha=0.01, \beta=1.0$, 
where ground, first and fifth excited-state energies 
are given. As number of basis function is increased, eigenvalues steadily improve. With
$N=100$, our results excellently match with the accurate estimates of \cite{pedram2010}. They
employed trigonometric basis functions satisfying periodic boundary conditions within a 
variational framework. This clearly demonstrates that very accurate and reliable solutions could be achieved
for both ground and excited states. Also, \emph{very small energy difference} between two successive energy levels 
is nicely reproduced; the table illustrates that between the lowest two states while for other pairs
this is verified and not produced here to save space. 
In such a \emph{near-complete} basis, matrix elements of our Hamiltonian are evaluated as, 
\begin{equation}
h_{mn} = \langle m|\hat{\mbox{H}}|n\rangle.
\end{equation}  
Diagonalization of the matrix {\bf h} leads to energy eigenvalues and corresponding eigenvectors, 
which is accomplished by the MATHEMATICA program package.

\begingroup
\squeezetable
\begin{table}                    
\caption{Convergence of eigenvalues for ground and first excited state of potential in Eq.~(6) 
taking $\alpha=0.01$ and $\beta=1.0$. Accurate reference values are taken from \citep{pedram2010}.}
\centering
\begin{ruledtabular}
\begin{tabular}{clll}
$N$ &  E$_0^{\S}$  &  E$_1^{\dag}$   &  E$_5^{\ddag}$          \\
\hline
25  &  $-$23.595951394689         &  $-$23.595951394587         &   $-$18.035962277239317         \\
50  &  $-$23.5959513947022885     &  $-$23.5959513947022765     &   $-$18.12991112369145          \\   
75  &  $-$23.5959513947022929177  &  $-$23.5959513947022929105  &   $-$18.129911166285982         \\  
100 &  $-$23.595951394702293117574292  &  $-$23.59595139470229311739743  & $-$18.129911166285953197575             \\
\end{tabular}
\end{ruledtabular}
\begin{tabbing}             
$^\S$Reference value is: $-$23.595951394702293117574292. \hspace{25pt} \=
$^\dag$Reference value is: $-$23.59595139470229311739743.  \hspace{25pt} \\
$^\ddag$Reference value is: $-$18.129911166285953197575. 
\end{tabbing}
\end{table}
\endgroup

A Manifold-Energy minimization method \cite{hendekovic1983reply} has been invoked to minimize the 
energy functional. In this approach \cite{pathak1994nonlinear}, instead of minimizing a particular 
energy state, trace of the matrix with respect to non-linear parameter, $\gamma$ is minimized, which
remains invariant under diagonalization of the matrix; thus one diagonalizes the matrix at 
that particular value of $\gamma$ for which trace becomes minimum. 
As a result, one obtains all desired eigenstates in a single diagonalization step. Table~I has
demonstrated the accuracy obtained by this simple approach. Thus, while several analytical and numerical 
schemes are available for such potentials, our present prescription is sufficiently good enough for the 
purpose at hand. 
Same minimization principle holds true in momentum space as well, because energy of the 
particle is same in both spaces necessitating trace to be same as well. This is given by, 

\begin{widetext}
\begin{equation}
Tr[h]=\sum_{m}h_{mm} = \sum_{m}\left[\frac{3\alpha}{16\gamma^{2}}(2m^{2}+2m+1)-
\frac{(\beta + 4\gamma^2)(2m+1)}{4\gamma}+2\gamma(2m+1)\right].
\end{equation}  
\end{widetext}
Now, minimization of the trace leads to a cubic equation in $\gamma$,
\begin{equation}\label{gen_eq}
8{\gamma^3}+2\gamma\beta-\alpha \ \frac{(2{N^2}+4N+3)}{(N+1)}=0.
\end{equation}
Use of parity, transforms Eq.~(11), for even $m$ as,
\begin{equation}\label{min_even}
8{\gamma^3}+2\gamma\beta-\alpha(2N+1)=0,
\end{equation}
while for odd $m$ this gives,
\begin{equation}\label{min_odd}
8{\gamma^3}+2\gamma\beta-\alpha(2N+3)=0. 
\end{equation}
It can be easily shown that $\gamma$ will have a single real root in Eqs.~(\ref{gen_eq}),
(\ref{min_even}) and (\ref{min_odd}), for which the trace will be minimum for respective cases. 
Now, one can minimize the matrix using the value of non-linear parameter $\gamma$.

The eigenvectors in position space then turn out to be, 
\begin{equation}
\psi_n(x)=\sum_{m=0}b_{m}^n\left(\frac{2\gamma}{\pi}\right)^
{\frac{1}{4}}\frac{1}{\sqrt{2^{m}m!}} \ H_{m}(\sqrt{2\gamma}x) \ e^{-{\gamma}x^{2}}
\end{equation}
while in momentum space, they are given as,
\begin{equation}
\psi_n(p)=\sum_{m=0}c_{m}^n\left(\frac{1}{2\gamma\pi}\right)^{\frac{1}{4}}\frac{1}{\sqrt{2^{m}m!}}
\ H_{m} \left( \frac{p}{\sqrt{2\gamma}} \right) \ e^{-\frac{p^{2}}{4\gamma}}.
\end{equation}

\section{Uncertainty-like information measures: some analytic expressions for a QHO}
Generally speaking, IEs are some appropriately weighted measures of quantum probability distribution 
function, $\rho(x) = |\psi(x)|^2$. Unlike uncertainty in position space, which just contains the 
second moment of $\rho(x)$, IE in position space contains contribution from all the moments that 
are relevant and present in quantum probability distribution \cite{diosi2011short}. Thus, 
intuitively IE gives a better and more complete description of all the competitive moments 
present in wave function. $(IE)_{x}$ helps in predicting localization of a particle in position 
space accurately. This description is also applicable for momentum-space. The quantity $IE$ then 
corresponds to total information available for a system, which remains conserved under uniform 
scaling of particle coordinates \cite{gadre1985some}.
In this section, we first present some analytical results of IE. After that, we calculate such IEs 
for first four energy states of a 1D QHO and discuss their changes with respect to force constant, 
$\gamma$ \citep{aguiar2015shannon, majernik1996entropic, yanez1994position}. 

The pertinent Schr\"{o}dinger equation for a QHO is written in following form, 
\begin{equation}
 \left[ -\frac{\hbar^2}{2m}\frac{d^2}{dx^2}+4\gamma^{2}x^2  \right] \psi_n (x)
= 2\gamma\left(2n+1\right) \ \psi_n(x).
\end{equation}   

\subsection{Fisher Information}
The Fisher information \citep{cover2012elements}, in position ($I_{x}$) and momentum ($I_p$) space, 
are given as, 
\begin{equation}
I_{x} =  \int \left[\frac{|\nabla\rho(x)|^2}{\rho(x)}\right] dx, \ \ \ \ \ \ \ \ 
I_{p} =  \int \left[\frac{|\nabla\rho(p)|^2}{\rho(p)}\right] dp. 
\end{equation}
The net Fisher Information ($I$) is then expressed as, 
\begin{equation}
I=I_{x}I_{p}.
\end{equation}
In case of a QHO, $I_{x}$s for first four energy states are obtained as, 
\begin{eqnarray}
I_{x}^{0}= 4 \gamma,  \hspace{10mm}
I_{x}^{1}=12 \gamma,  \hspace{10mm}
I_{x}^{2}=20 \gamma,  \hspace{10mm} 
I_{x}^{3}=28 \gamma
\end{eqnarray}
$I_{x}^{n}$ increases with increase of $\gamma$ (linearly) and state index ($n$). The difference 
of $I_{x}$ between two adjacent energy levels remains constant, with the value being $8\gamma$.

Likewise, $I_{p}$s for first four energy states of a QHO are as follows:
\begin{eqnarray}
I_{p}^{0}=\frac{1}{\gamma}, \hspace{5mm}
I_{p}^{1}=\frac{3}{\gamma}, \hspace{5mm}
I_{p}^{2}=\frac{5}{\gamma}, \hspace{5mm}
I_{p}^{3}=\frac{7}{\gamma}.
\end{eqnarray}
$I_{p}^{n}$ is inversely proportional to $\gamma$. The proportionality constant increases with 
$n$. In this case also, the change in $I_{p}$ between two successive states is constant, with 
the value being $\frac{2}{\gamma}$. 

Finally, the net Fisher Information (I) of first four states of a QHO are,
\begin{eqnarray}
I^{0}=4\gamma \frac{1}{\gamma}=4, \hspace{5mm}
I^{1}=12\gamma \frac{3}{\gamma}=36, \hspace{5mm}
I^{2}=20\gamma \frac{5}{\gamma}=100, \hspace{5mm}
I^{3}=28\gamma \frac{7}{\gamma}=196.
\end{eqnarray}
$I^{n}$ is independent of $\gamma$ and increases with an increase in $n$. However, unlike $I_x$ 
and $I_p$, $\left[I^{n}-I^{(n-1)}\right]$ is not constant. Instead, 
$\left[\sqrt{I^{n}}-\sqrt{I^{(n-1)}}\right]$ is stationary with a value of 4.  

\subsection{Shannon entropy}
Next, the Shannon entropy \citep{shannon1951prediction}, in position ($S_{x}$) and momentum ($S_p$) 
space is defined as, 
\begin{equation}
S_{x} =  -\int \rho(x) \mbox{ln}\left[\rho(x)\right] \ dx, \ \ \ \ \ \ \ \ 
S_{p} =  -\int \rho(p) \mbox{ln}\left[\rho(p)\right] \ dp. 
\end{equation}
The total Shannon Entropy ($S$) is then given by, 
\begin{equation}
S=S_{x}+S_{p}.
\end{equation}
For the lowest four states of a QHO, $S_{x}$s are given by following expressions, 
\begin{eqnarray}
S_{x}^{0} =\frac{1}{2}\left[1+\mbox{ln}\left(\frac{\pi}{2\gamma}\right)\right], \hspace{10mm} 
S_{x}^{1} =\left[0.77036+\frac{1}{2}\mbox{ln}\left(\frac{\pi}{2\gamma}\right)\right], \nonumber \\
S_{x}^{2} =\left[0.92624+\frac{1}{2}\mbox{ln}\left(\frac{\pi}{2\gamma}\right)\right], \hspace{10mm}
S_{x}^{3} =\left[1.03735+\frac{1}{2}\mbox{ln}\left(\frac{\pi}{2\gamma}\right)\right].
\end{eqnarray} 
As can be seen, $S_{x}^{n}$ decreases with increase in $\gamma$ but increases with increase in 
$n$. However, $\left[{S_{x}^{n}}-{S_{x}^{(n-1)}}\right]$ decreases as $n$ increases.
 
The corresponding quantities in momentum space, $S_{p}$, for a QHO, are obtained as,
\begin{eqnarray}
S_{p}^{0} =\frac{1}{2}\left[1+\mbox{ln} \left({2\gamma\pi}\right)\right], \hspace{10mm}
S_{p}^{1} =\left[0.77036+\frac{1}{2} \ \mbox{ln}\left({2\gamma\pi}\right)\right], \\ \nonumber 
S_{p}^{2} =\left[0.92624+\frac{1}{2} \ \mbox{ln}\left({2\gamma\pi}\right)\right], \hspace{10mm}
S_{p}^{3} =\left[1.03735+\frac{1}{2} \ \mbox{ln}\left({2\gamma\pi}\right)\right]. 
\end{eqnarray} 
Thus, $S_{p}^{n}$ increases with $\gamma$ and $n$. Here again,  
$\left[{S_{p}^{n}}-{S_{p}^{(n-1)}}\right]$ decreases with increase in $n$. 

Finally, the total $S$ for these four states can be written as, 
\begin{eqnarray}
S_{0}=1+\mbox{ln}\pi, \hspace{5mm}
S_{1}=1.54072+\mbox{ln}\pi,\\  \nonumber 
S_{2}=1.85248+\mbox{ln}\pi, \hspace{5mm}
S_{3}=2.07470+\mbox{ln}\pi
\end{eqnarray}
$S^{n}$ is seen to be independent of $\gamma$ but increases as $n$ increases.

At this stage, it is worth mentioning that, Shannon entropy measures have been extensively used
in the context of confined harmonic oscillator (CHO) as well, where many fascinating features 
have been reported. It is well established that, the behavior of 
$S_{x}$, $S_{p}$ and $S$ for a 1D CHO is substantially different from its corresponding \emph{free} 
counterpart \cite{laguna2014quantum}. For example, $S_{x}$ increases with increase of box length 
$x_{c}$ and then finally converges to a constant value. Similarly, $S_{p}$ initially decreases as
$x_{c}$ increases and then attains a constant value. For, higher states such as with $n=4,5$, 
$S_{p}$ attains a minimum before convergence. $S^{0}$ decreases with increase of $x_{c}$ and 
again finally converges to a constant. Whereas $S$ for other states ($n=1-5$) rise with the 
relaxation of confinement and finally reaches a constant value, which is fixed for a particular state. 

\begingroup
\squeezetable
\begin{table}     
\caption{Ground and first three excited-state energies of a 1D CHO at four different $x_c$ values. Numbers in the
parentheses denote reference values, quoted from \citep{campoy2002}.}
\centering
\begin{ruledtabular}
\begin{tabular}{ccccc}
$x_{c}$ &  $E_0$  &  $E_1$  &  $E_2$   &  $E_3$    \\
\hline
0.5  & 4.951 129 323 254        & 19.774 534 179 208        & 44.452 073 829 740        & 78.996 921 150 747        \\
     & (4.951 129 323 254)      & (19.774 534 179 208)      & (44.452 073 829 740)      & (78.996 921 150 747)  \\
1.0  & 1.298 459 832 032        & 5.075 582 015 227         & 11.258 825 781 482        & 19.899 696 650 183        \\
     & (1.298 459 832 032)  & (5.075 582 015 226)   & (11.258 825 781 482)  & (19.899 696 650 183)  \\
2.0  & 0.537 461 209 282        & 1.764 816 438 780         & 3.399 788 241 107         &  5.584 639 079 031        \\
     & (0.537 461 209 281)  & (1.764 816 438 780)   & (3.399 788 241 107)   &  (5.584 639 079 031)  \\
5.0  & 0.500 000 000 078        &  1.500 000 003 672        &  2.500 000 084 018        &  3.500 001 221 456        \\
     & (0.500 000 000 076)  &  (1.500 000 003 671)  &  (2.500 000 084 018)  &  (3.500 001 221 456)  \\
\end{tabular}
\end{ruledtabular}
\end{table}
\endgroup

In this work, while we have not attempted to reproduce all these quantities, however, as an 
additional objective, only one of them ($S_x$) has been calculated for a CHO inside an impenetrable
box. These results very well corroborate the reference results
of \cite{laguna2014quantum}. For this, the eigenvalues and eigenfunctions were obtained using 
a variation principle with particle in a box basis set. For sake of completeness and to demonstrate 
the quality of our wave function employed, Table~II compares energies of lowest four states, at 
four selected $x_c$ values. As clearly seen, for a broad region of confinement, present results 
offer up to 11-12 decimal place agreement with reference values (given in parentheses) quite 
easily. All energies practically coincide with the accurate literature results. 

\subsection{Onicescu energy} 
The Onicescu energy \citep{agop2014implications, alipour2012onicescu}, in position and momentum 
space, is given by, 
\begin{equation}
E_{x} =  \int \left[|\rho(x)|^2\right] dx, \ \ \ \ \ \
E_{p} =  \int \left[|\rho(p)|^2\right] dp, 
\end{equation}
whereas the total Onicescu energy ($E$) is written as, 
\begin{equation}
E=E_{x}E_{p}.
\end{equation}
In case of a QHO, $E_{x}$ for the lowest four energy states are obtained as, 
\begin{eqnarray}
E_{x}^{0}= \sqrt{\frac{\gamma}{\pi}}, \hspace{15mm}
E_{x}^{1}=\frac{3}{4}\sqrt{\frac{\gamma}{\pi}},\\    \nonumber 
E_{x}^{2}=\frac{41}{64}\sqrt{\frac{\gamma}{\pi}}, \hspace{15mm}
E_{x}^{3}=\frac{147}{256}\sqrt{\frac{\gamma}{\pi}.}
\end{eqnarray}
$E_{x}^{n}$ increases with increase in $\gamma$ and \emph{decreases} as $n$ increases.

$E_{p}$ for the four states of a QHO are given accordingly, 
\begin{eqnarray}
E_{p}^{0}= \frac{1}{2\sqrt{\gamma\pi}}, \hspace{15mm}
E_{p}^{1}=\frac{3}{8\sqrt{\gamma\pi}},\\  \nonumber 
E_{p}^{2}=\frac{41}{128\sqrt{\gamma\pi}}, \hspace{15mm}
E_{p}^{3}=\frac{147}{512\sqrt{\gamma\pi}}.
\end{eqnarray}
Thus, $E_{p}^{n}$ decreases with increase in both $\gamma$ and $n$.

The corresponding total $E^{n}$s of a QHO, are expressed as, 
\begin{eqnarray}
E^{0}=\left[\frac{1}{2\pi}\right], \hspace{15mm}
E^{1}=\frac{9}{16}\left[\frac{1}{2\pi}\right],\\  \nonumber 
E^{2}=\frac{1681}{4096}\left[\frac{1}{2\pi}\right], \hspace{15mm}
E^{3}= \frac{21609}{65536}\left[\frac{1}{2\pi}\right].
\end{eqnarray}
$E^{n}$ is evidently independent of $\gamma$ and decreases with increase of $n$.

\subsection{Onicescu-Shannon information measure}
The Onicescu-Shannon (OS) entropy in position, momentum space 
\citep{LpezRuiz1995321, chatzisavvas2005information, panos2007comparison}, are given by, 
\begin{equation}
OS_{x} = \exp\left[\frac{2}{3}{S_{x}}\right]E_{x}, \ \ \ \ \ \ \ 
OS_{p} =  \exp\left[\frac{2}{3}{S_{p}}\right]E_{p}. 
\end{equation}
And the total quantity is defined by,
\begin{equation}
OS=OS_x \times  OS_p=\exp\left[\frac{2}{3}{S}\right]E. 
\end{equation}
For a QHO, these quantities in position space, assume the form (for four lowest states), 
\begin{eqnarray}
OS_{x}^{0}=\left[\left({\frac{\gamma}{4\pi}}\right)^{\frac{1}{6}}\exp\left[{\frac{1}{3}}\right]
\right], \hspace{10mm}
OS_{x}^{1}=\left[\frac{3}{4}\left({\frac{\gamma}{4\pi}}\right)^{\frac{1}{6}}\exp\left[0.5136\right]
\right],\\ \nonumber 
OS_{x}^{2}=\left[\frac{41}{64}\left({\frac{\gamma}{4\pi}}\right)^{\frac{1}{6}}\exp
\left[0.6175\right]\right], \hspace{10mm}
OS_{x}^{3}=\left[\frac{147}{256}\left({\frac{\gamma}{4\pi}}\right)^{\frac{1}{6}}\exp
\left[0.6916\right]\right], 
\end{eqnarray}
whereas in momentum space, they are simplified as, 
\begin{eqnarray}
OS_{p}^{0}= \left[\frac{1}{2}\left({\frac{4}{\gamma\pi}}\right)^{\frac{1}{6}}\exp
\left[{\frac{1}{3}}\right]\right], \hspace{10mm}
OS_{p}^{1}=\left[\frac{3}{8}\left({\frac{4}{\gamma\pi}}\right)^{\frac{1}{6}}\exp
\left[0.5136\right]\right],\\  \nonumber
OS_{p}^{2}=\left[\frac{41}{128}\left({\frac{4}{\gamma\pi}}\right)^{\frac{1}{6}}\exp
\left[0.6175\right]\right], \hspace{10mm}
OS_{p}^{3}=\left[\frac{147}{512}\left({\frac{4}{\gamma\pi}}\right)^{\frac{1}{6}}\exp
\left[0.6916\right]\right].
\end{eqnarray}
Finally, the corresponding total quantities for QHO can be written as,
\begin{eqnarray}
OS^{0}= \left[\frac{1}{2}\left({\frac{1}{\pi}}\right)^{\frac{1}{3}}\exp\left[{\frac{2}{3}}\right]
\right], \hspace{15mm}
OS^{1}= \left[\frac{9}{32}\left({\frac{1}{\pi}}\right)^{\frac{1}{3}}\exp\left[{1.0271}
\right]\right],\\  \nonumber
OS^{2}=\left[\frac{1681}{8192}\left({\frac{1}{\pi}}\right)^{\frac{1}{3}}\exp\left[{1.235}\right]
\right], \hspace{15mm}
OS^{3}=\left[\frac{21609}{131072}\left({\frac{1}{\pi}}\right)^{\frac{1}{3}}\exp\left[{1.3831}
\right]\right].
\end{eqnarray}

So, $OS_{x}^{n}$ increases with increase of $\gamma$, whereas $OS_{p}^{n}$ decreases. Then, 
as with all previous measures, $OS^{n}$ too is independent of $\gamma$. This occurs because, 
it is a product of $OS_x$ and $OS_p$ (see Eq.~(33)), both of which are individually multiplicative.
The result is a product of exponential of $S$ (an additive measure) time $E$ (a multiplicative
measure). Both $S$, $E$ are independent of $\gamma$; hence the same is true for $OS^n$. Also $OS^n$ 
decreases with increase in $n$. 

\begin{figure}         
\includegraphics[scale=0.45]{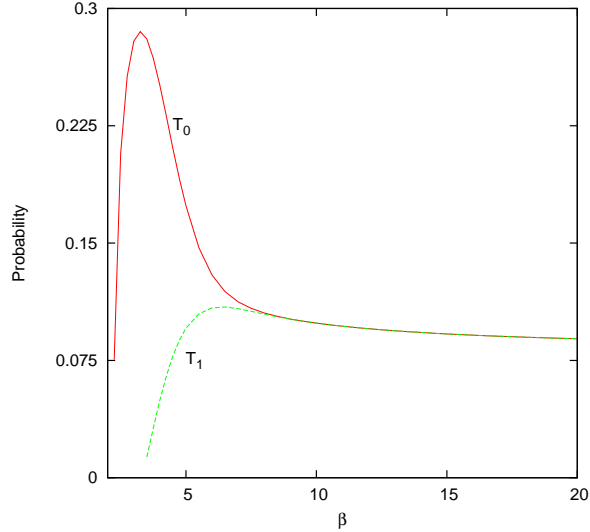}
\caption{Probabilities of finding the particle within the well in ground ($T^0$), first excited 
($T^1$) state, against $\beta$, using Eq.~(3), for DW potential in Eq.~(6).}
\end{figure}

\section{Results and Discussion}
Our present work primarily concerns with understanding the effect of variation of $\beta$ on the 
behavior of a particle in a DW potential. As already mentioned earlier, varying $\beta$ increases 
both delocalization by increasing spacing between classical turning points, and also promotes  
confinement through increase of barrier height. Therefore, an interplay between these two simultaneous 
effects should be felt in the behavior of IE as well--hence one or more extremum is to be expected. 
As observed in the following discussion, this is indeed found to be the case. A proper and complete 
description of such contrasting effects through conventional uncertainty product 
$\triangle{x}\triangle{p} \ $ seems inadequate--no extremal nature is noticed.

Companion calculations are also performed to monitor changes in the behavior of particle with 
parameter $\alpha$. For this, three different values of $\alpha,$ namely, 0.5, 1 and 2 are chosen. 
At the onset, however, it may be noted that, the role of $\alpha$ is relatively more straightforward 
to examine than that of $\beta$. Also, 
differences of $\alpha$ (from unity) can be thought of as a scaling of the parameter $\beta$ to 
$\frac{\beta}{\alpha}$, with the resulting potential being scaled by $\alpha$. Therefore, one would 
expect that as $\alpha$ increases, $\frac{\beta}{\alpha}$ reduces for any $\beta$--therefore any 
points of inflection in IE should shift towards right. Our future discussion would suggest this to
hold true. Same observation also leads one to predict that there should 
not be any qualitative change in the variation of IE--a point which would be validated later as 
well. However, it should be noted that quantitative predictions cannot be made about these shifts 
regarding the position of inflection points--this is because scaling only affects the 
\emph{potential}, whereas wave function (and by extension all aspects of particle's behavior) 
comes out as solution of the \emph{full} Hamiltonian. 

Before proceeding for the IE calculations, we have examined probability of finding the particle 
within the barrier in ground and first excited states, \emph{viz.}, $T^0, T^1$ using Eq.~(3), 
for our DW potential in Eq.~(6). From Fig.~(2), it seems that the particle begins to be trapped 
at values of $\beta$ starting at 2.25 for ground state and 3.5 for first excited state. Tunneling 
sets in when barrier height exceeds energy of the particle such that, $\frac{\beta^{2}}{4\alpha}>E_{n}$. 

\begin{figure}             
\centering
\begin{minipage}[c]{0.20\textwidth}\centering
\includegraphics[scale=0.38]{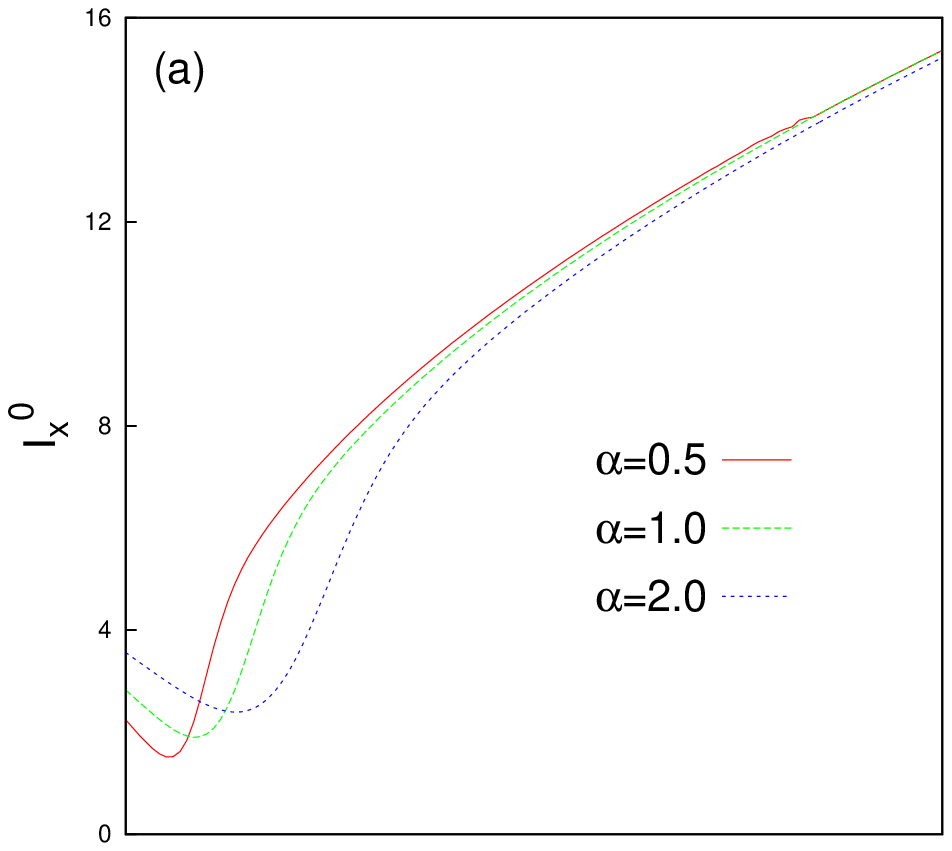}
\end{minipage}\hspace{0.15in}
\begin{minipage}[c]{0.20\textwidth}\centering
\includegraphics[scale=0.38]{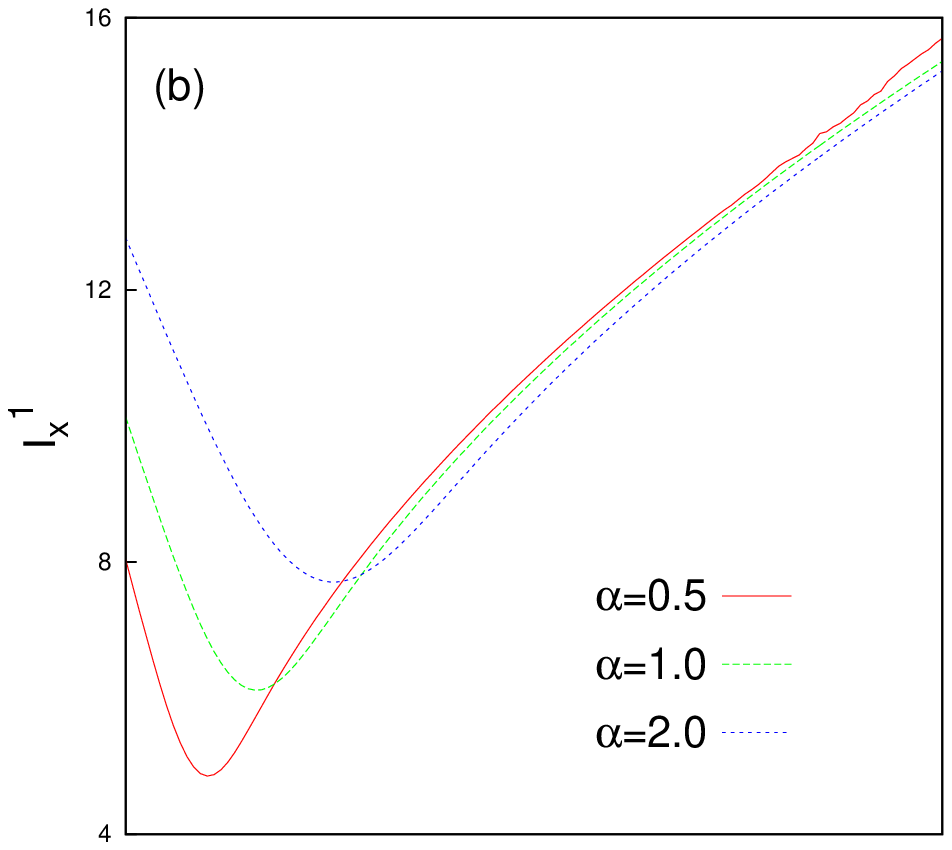}
\end{minipage}\hspace{0.15in}
\begin{minipage}[c]{0.20\textwidth}\centering
\includegraphics[scale=0.38]{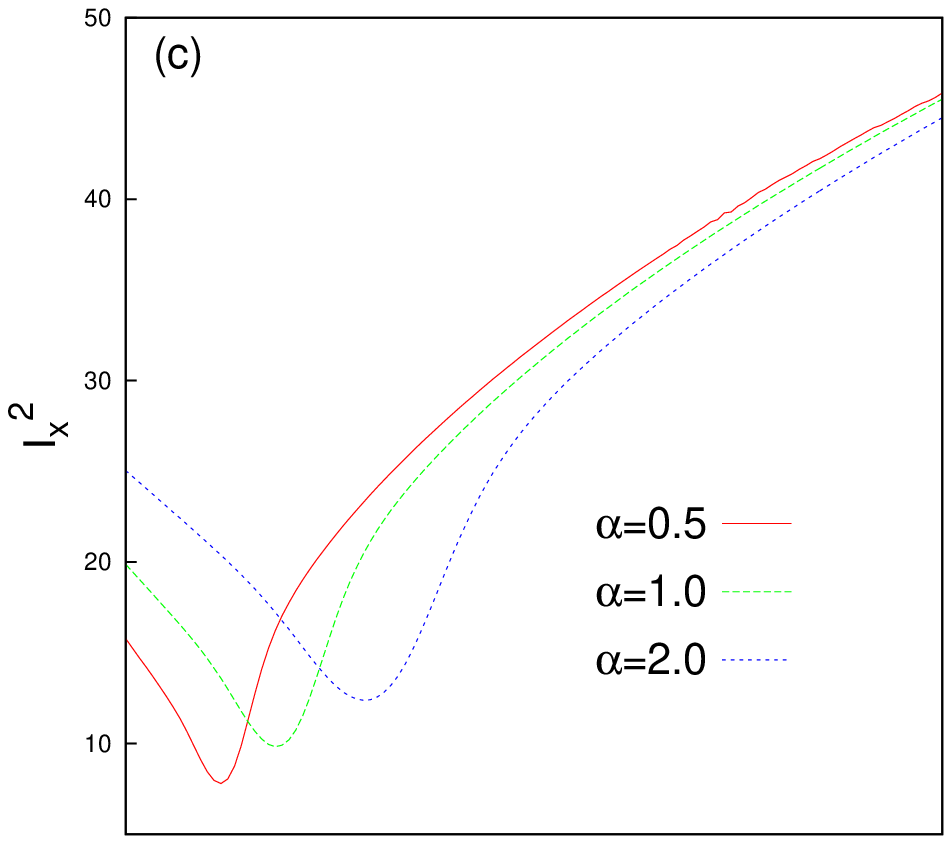}
\end{minipage}\hspace{0.15in}
\begin{minipage}[c]{0.20\textwidth}\centering
\includegraphics[scale=0.38]{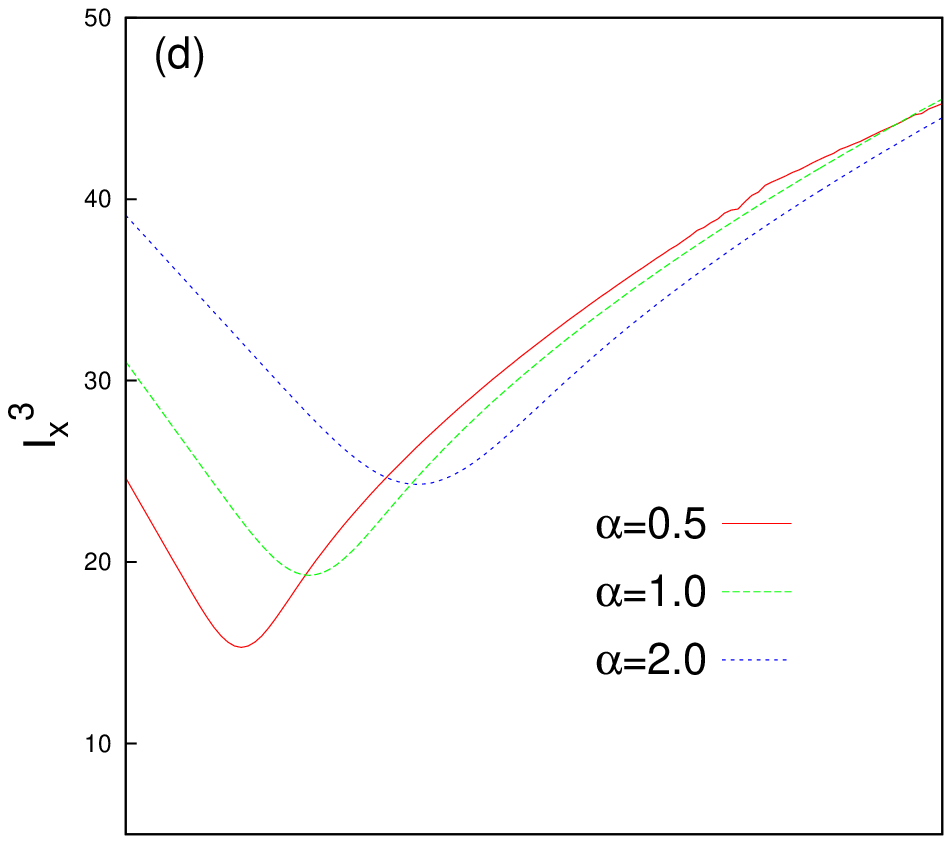}
\end{minipage}
\\[5pt]
\begin{minipage}[c]{0.20\textwidth}\centering
\includegraphics[scale=0.38]{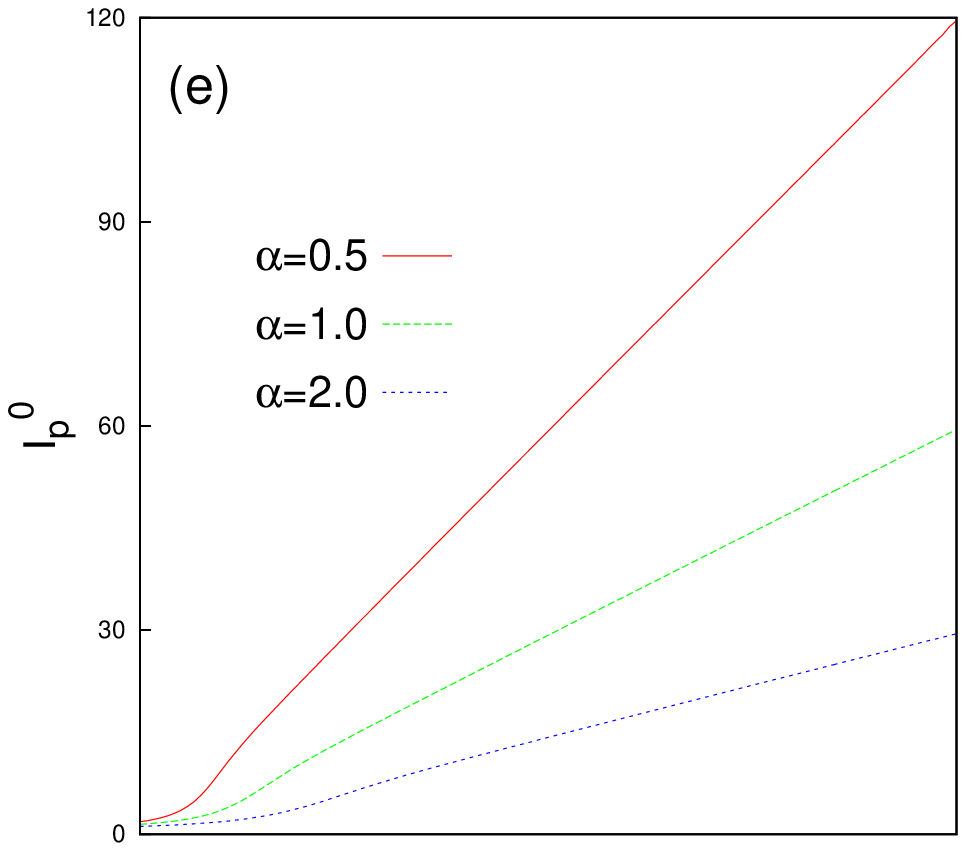}
\end{minipage}\hspace{0.15in}
\begin{minipage}[c]{0.20\textwidth}\centering
\includegraphics[scale=0.38]{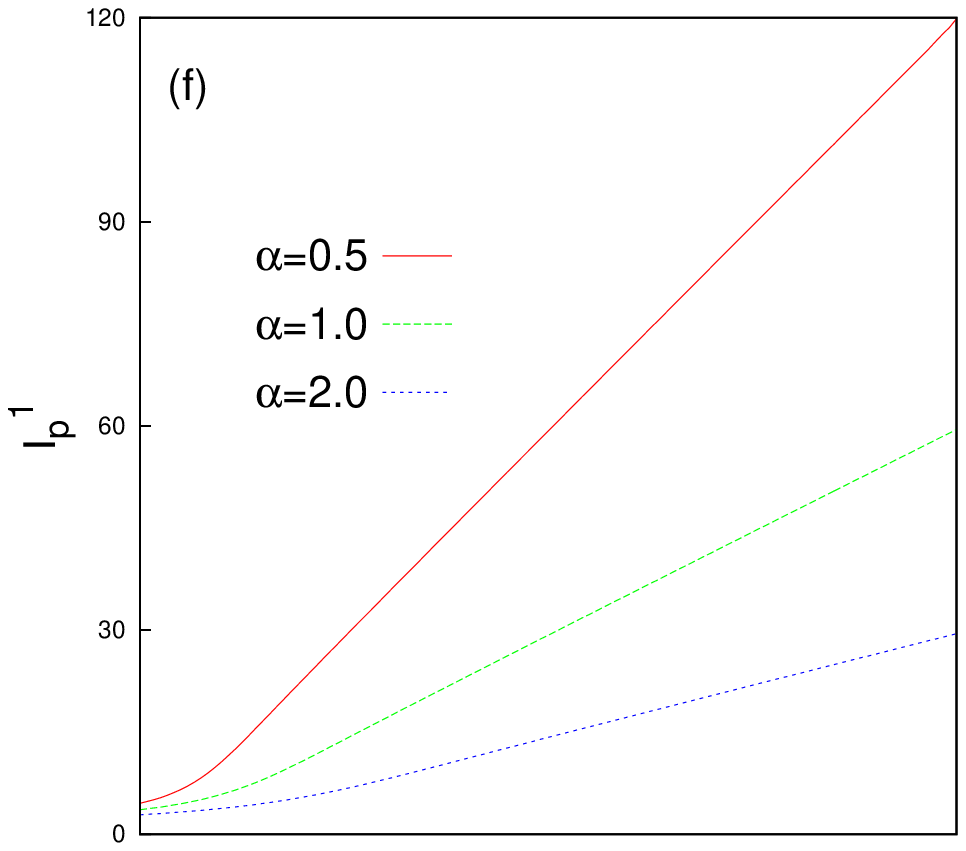}
\end{minipage}\hspace{0.15in}
\begin{minipage}[c]{0.20\textwidth}\centering
\includegraphics[scale=0.38]{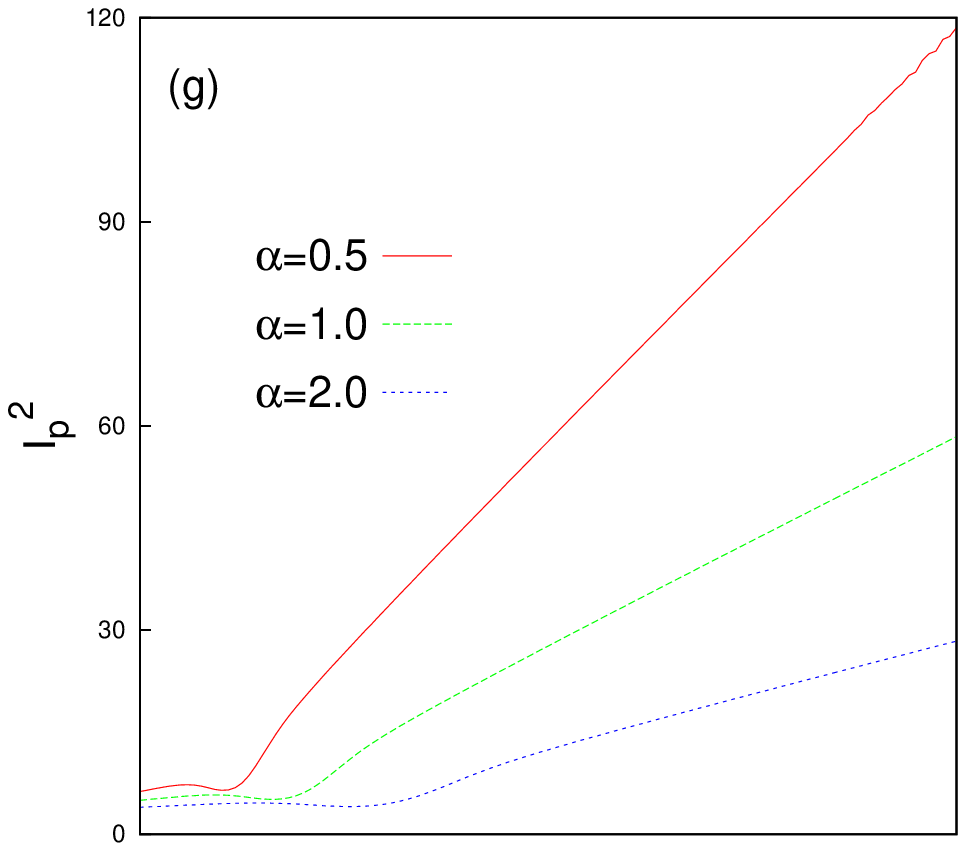}
\end{minipage}\hspace{0.15in}
\begin{minipage}[c]{0.20\textwidth}\centering
\includegraphics[scale=0.38]{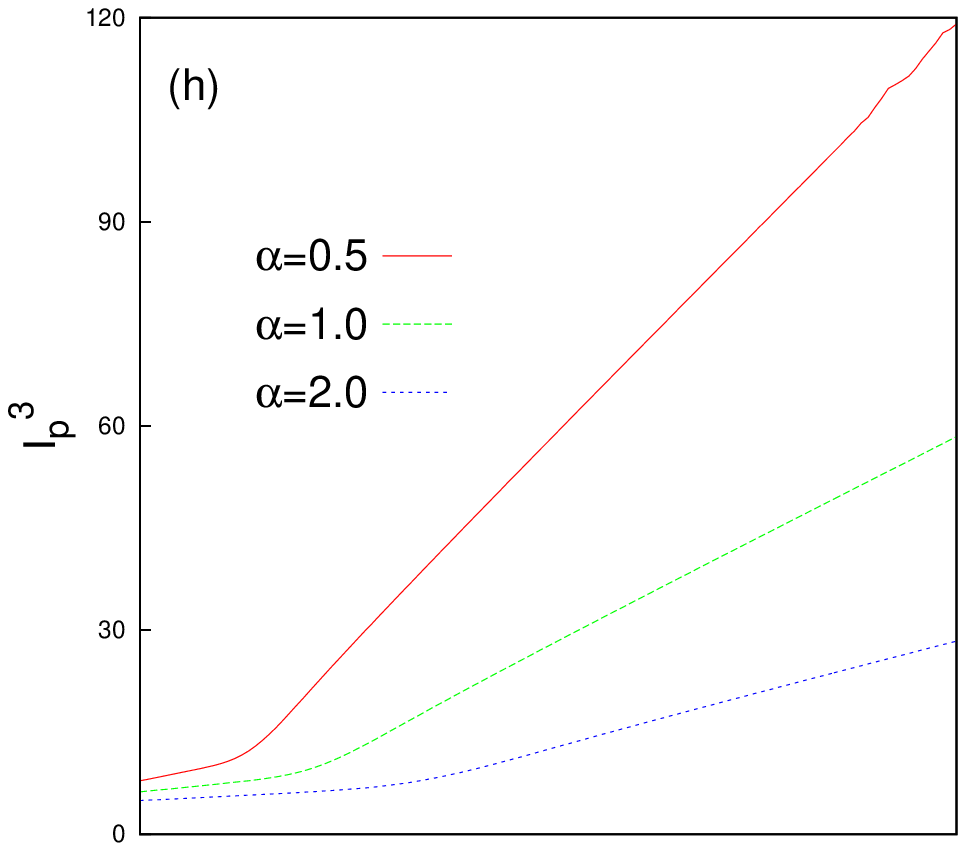}
\end{minipage}
\\[5pt]
\begin{minipage}[c]{0.20\textwidth}\centering
\includegraphics[scale=0.38]{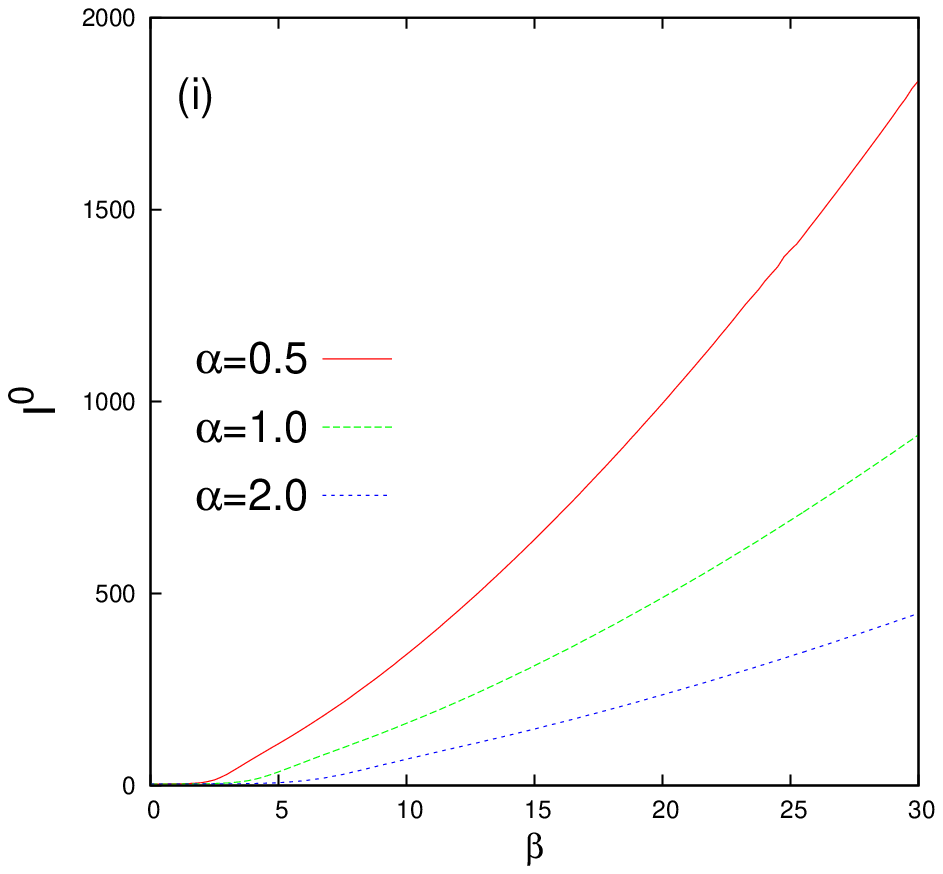}
\end{minipage}\hspace{0.15in}
\begin{minipage}[c]{0.20\textwidth}\centering
\includegraphics[scale=0.38]{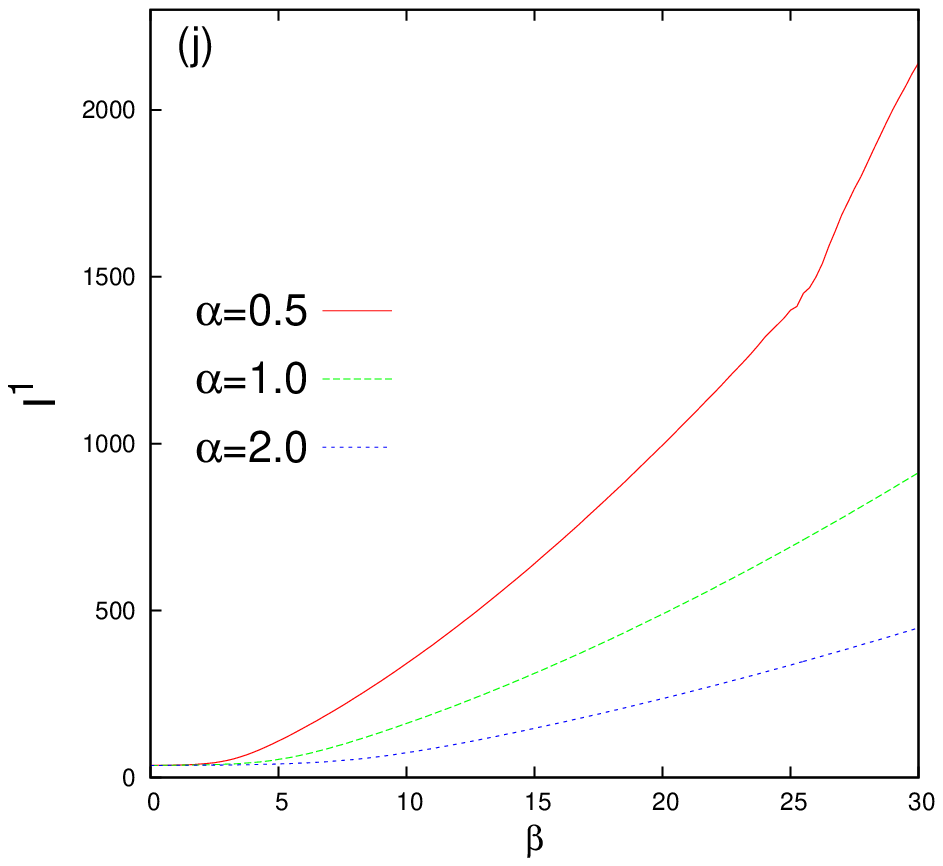}
\end{minipage}\hspace{0.15in}
\begin{minipage}[c]{0.20\textwidth}\centering
\includegraphics[scale=0.38]{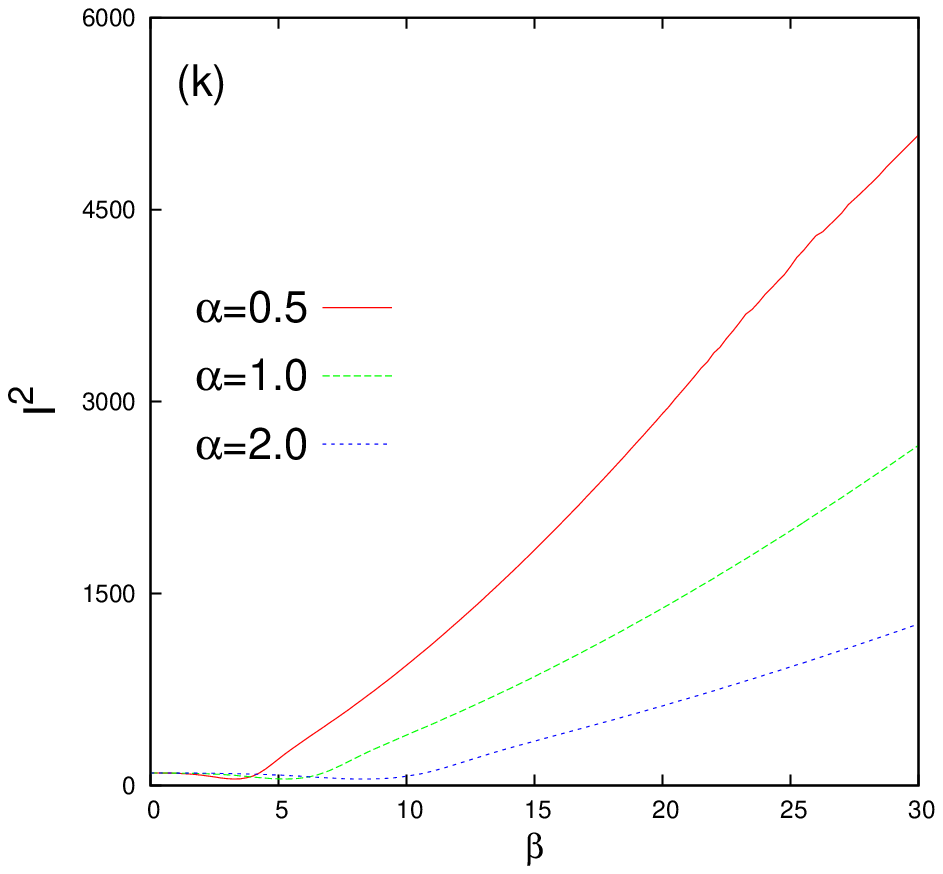}
\end{minipage}\hspace{0.15in}
\begin{minipage}[c]{0.20\textwidth}\centering
\includegraphics[scale=0.38]{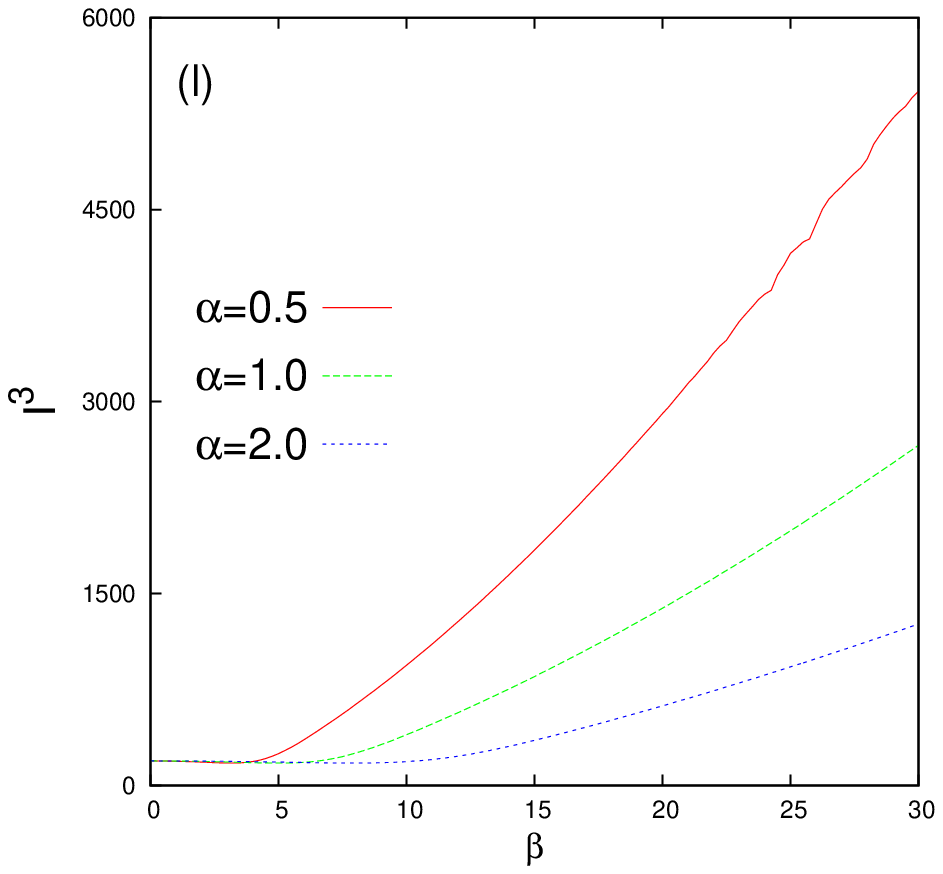}
\end{minipage}
\caption[optional]{Top panels (a)-(d), middle panels (e)-(h) and bottom panels (i)-(l) portray 
$I_{x}$, $I_p$ and $I$ against $\beta$ plots respectively at three different $\alpha \ (0.5, 1, 2)$  
for the DW potential in Eq.~(6). Four figures in each row correspond to lowest four states 
from left to right. See text for details.}
\end{figure}

\subsection{Fisher information}
We start our discussion by first enumerating $I_{x}$, $I_{p}$, $I$ for a particle in a symmetric 
DW potential in Figs.~(3), (4) using Eqs.~(17), (18). Top four segments (a)-(d) in Fig.~(3) display 
variation of $I_{x}$ with $\beta$ for first four energy states at three different selected $\alpha$ 
values, namely, 0.5, 1.0, 2.0 respectively. This indicates that $I_{x}$ for all these four states
are qualitatively analogous; at first it decreases with increase of $\beta$, attains a 
minimum and then progressively increases. For a definite state, these minima shift to right 
and gets more flattened (with a corresponding increase in minimum value) with increase of $\alpha$. 
Appearance of a minimum may be attributed to the balance between two competing effects. 
Study of $I_{p}$ and $I$ may further explain this in a more convincing way, which is also 
presented in Fig.~(3) for easy referencing. Before that, two left panels (a), (d) in Fig.~(4) 
gather plots for $I_{x}^{0}$, $I_{x}^{1}$ and $I_{x}^{2}$, $I_{x}^{3}$ pairs with $\beta$, for a 
specific case of $\alpha=1$. This clearly establishes the merging of two states after a certain 
$\beta$, which is presumably due to a quasi-degeneracy in our DW potential. This convergence 
point of $I_x$ shifts to higher values of $\beta$ with increase of $\alpha$. Note that in Figs~(3) and 
(4), for $I_x$ plots, range of $x$ axis remains same all through while that of $y$ axis changes.  

\begin{figure}             
\centering
\begin{minipage}[c]{0.32\textwidth}\centering
\includegraphics[scale=0.48]{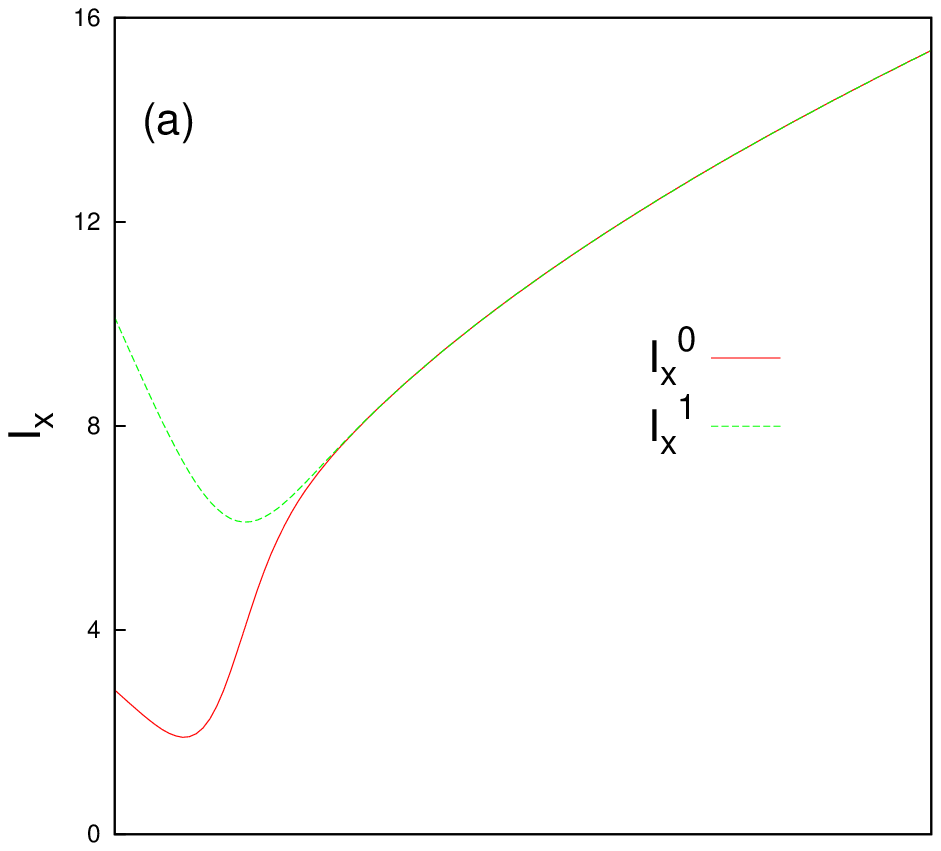}
\end{minipage}\hspace{0.05in}
\begin{minipage}[c]{0.32\textwidth}\centering
\includegraphics[scale=0.48]{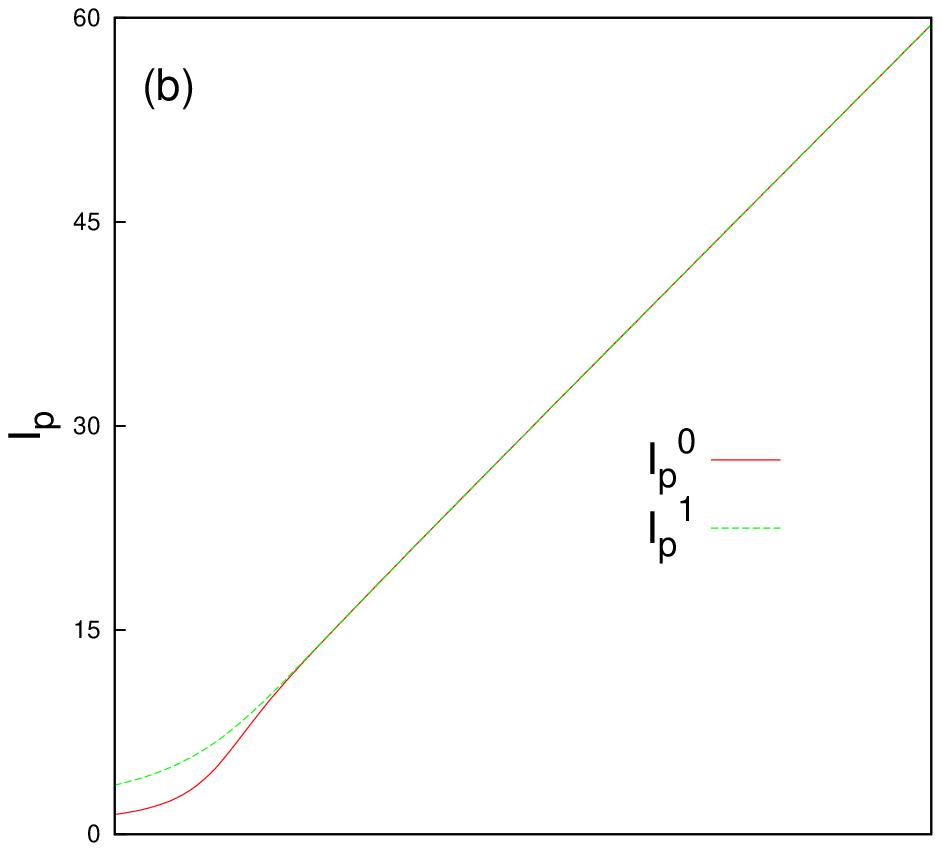}
\end{minipage}\hspace{0.05in}
\begin{minipage}[c]{0.32\textwidth}\centering
\includegraphics[scale=0.48]{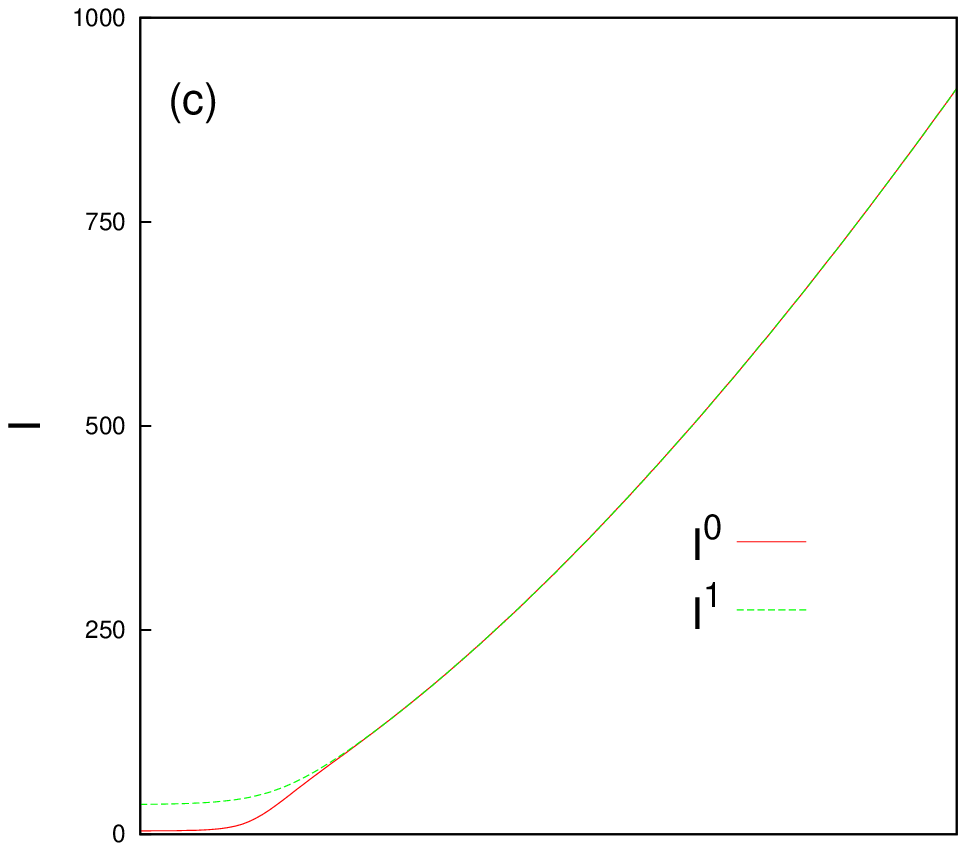}
\end{minipage}
\\[5pt]
\begin{minipage}[c]{0.32\textwidth}\centering
\includegraphics[scale=0.48]{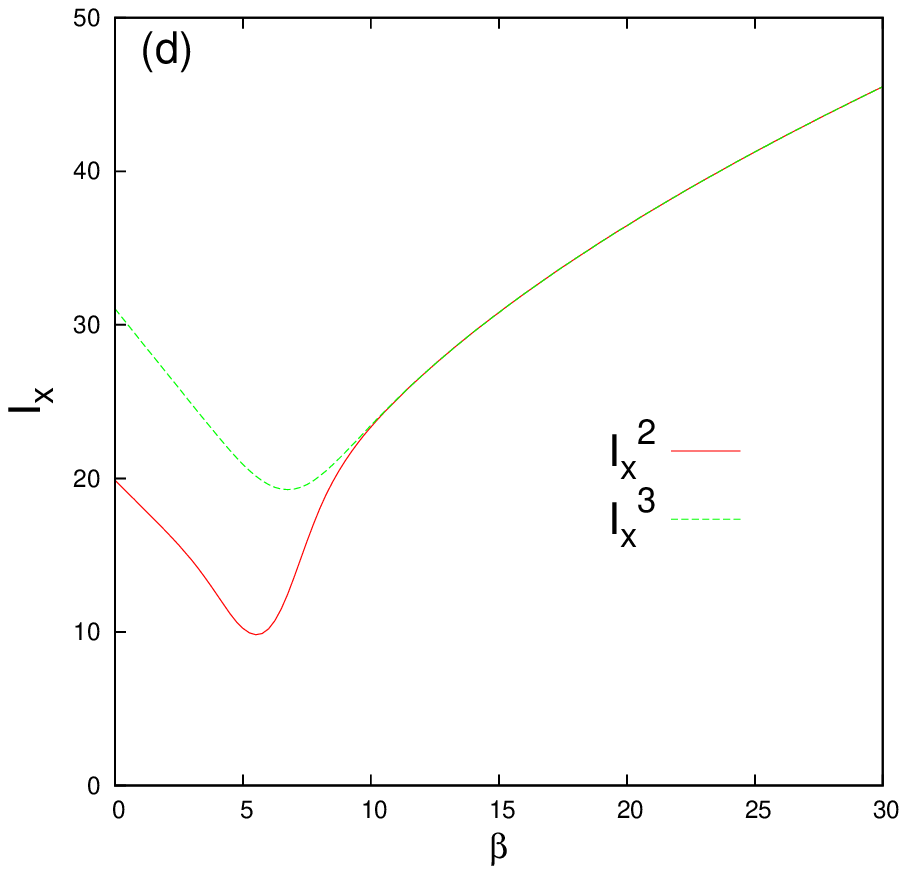}
\end{minipage}\hspace{0.05in}
\begin{minipage}[c]{0.32\textwidth}\centering
\includegraphics[scale=0.48]{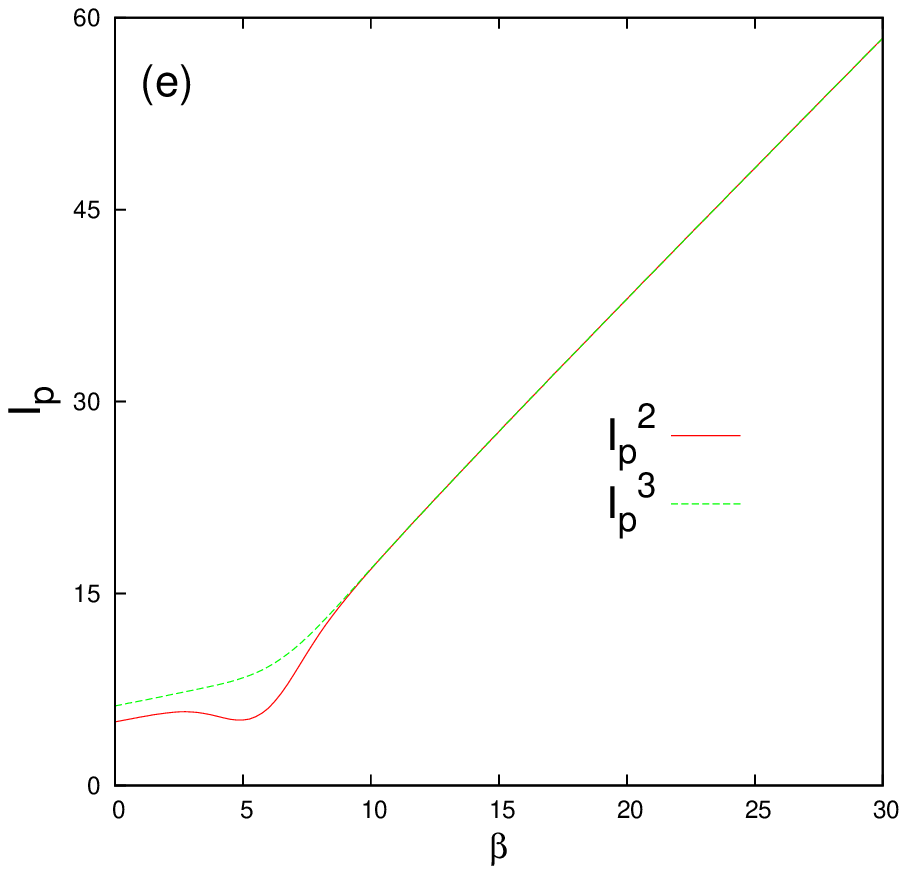}
\end{minipage}\hspace{0.05in}
\begin{minipage}[c]{0.32\textwidth}\centering
\includegraphics[scale=0.48]{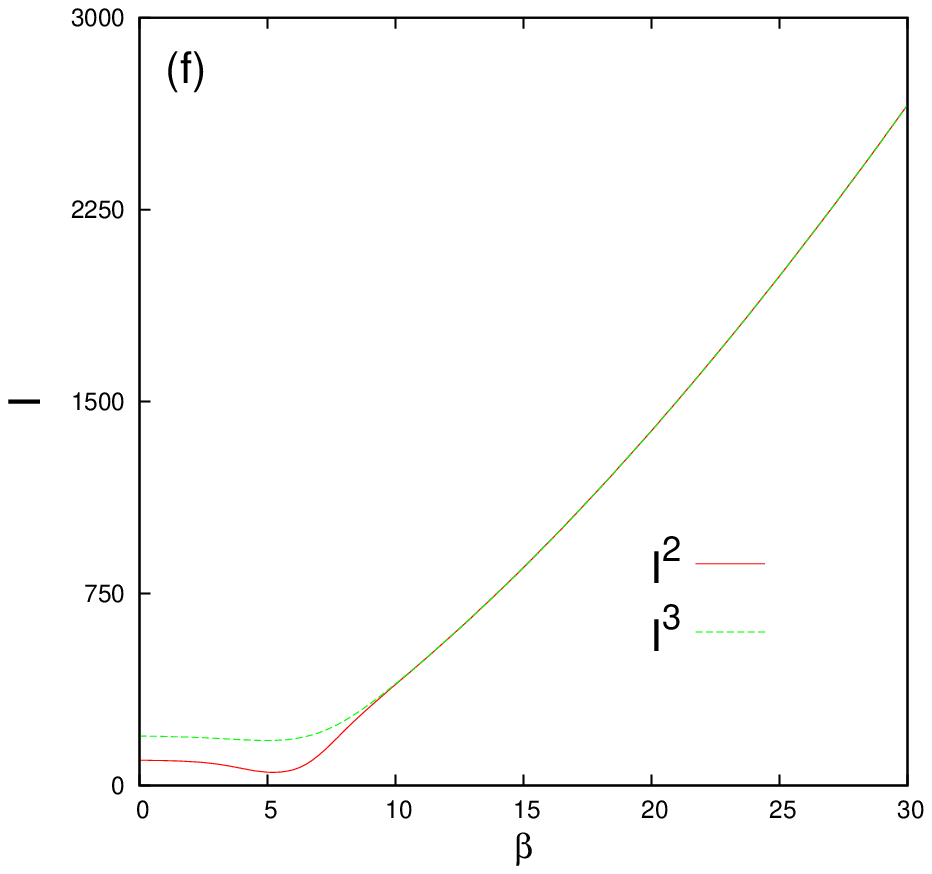}
\end{minipage}
\caption[optional]{Convergence of $I_x, I_p, I$ against $\beta$ at $\alpha=1.0$ for the DW potential in Eq.~(6). Left, 
middle, right columns correspond to $I_x$, $I_p$, $I$ respectively. Top panels (a)-(c) refer to ground and 
first excited states, while bottom panels (d)-(f) refer to third, fourth excited states. See text for details.}
\end{figure}

After observing the extremal nature in $I_x$ plots, it is natural to ask if similar behavior is 
noticed in case of $I_p$ or $I$ plots too. For this, we proceed to four middle panels (e)-(h) in 
Fig.~(3), where $I_p$ variations against $\beta$ are given for four low-lying states of DW potential, 
from left to right. At first glance, general qualitative nature of these plots appear to be quite similar.
It is clear that $I_{p}$ increases as $\beta$ increases, but the rate of increase gets 
considerably slower for higher $\alpha$--a fact that holds true for all states considered. Also, 
one finds that, for all states, at smaller $\beta$, increase of $I_p$ seems rather nominal; 
it remains almost constant until a certain $\beta$ is reached, after which $I_p$ keeps on increasing 
drastically. The $\beta$ value at which this transition occurs normally increases with state 
index. Moreover, for a given state, this $\beta$ is shifted to higher values as $\alpha$ progresses. 
Importantly, however, unlike the case of $I_x$ in panels (a)-(d) of Fig.~(3), no extremal nature is noticed in this
occasion; instead one finds slight flatness for smaller $\beta$. This leads to the conclusion that
like traditional uncertainty measures, $I_{p}$ also is unable to sense the competitive effects 
in a DW. However, as in $I_x$ plots, here also, in two middle panels (b), (e) of Fig.~(4), 
$I_{p}^{0}$, $I_{p}^{1}$ and $I_{p}^{2}$, $I_{p}^{3}$ pairs converge at a particular $\beta$ 
value depending on the value of $\alpha$ (presently 1). This also could be a possible signature 
of quasi-degeneracy in such a potential. Note that range of $y$ axis in (e)-(h) of Fig.~(3) \emph{differs}
from that of (b), (e) of Fig.~(4), to show an enhanced effect of the variation.

Next we discuss total information entropy $I$, recorded in (i)-(l) of Fig.~(3). In general, it increases 
with increase in $\beta$ for all four states. We have adopted a similar representation strategy as 
in $I_x$, $I_p$. Total information initially increases very slowly until a certain $\beta$ is reached and after 
that it shows drastic continuous growth. For a given state, the $\beta$ at which this transition 
occurs, is shifted right as $\alpha$ is increased. Progress of $\alpha$ consistently reduces 
growth rate of $I$ in all four occasions. Like the $I_p$ counterpart, $I$ also fails in bearing 
any characteristic signature indicative of the competing effects in these four states of DW potential.
However once again, as evident from two rightmost panels (c), (f) of Fig.~(4), pairs like 
$I^{0}$, $I^{1}$ and $I^{2}$, $I^{3}$ readily merge at a certain value of $\beta$. Note that, while 
range of $\beta$ is uniform in all these plots, same for $I$ axis varies in Figs.~(3) and (4). 

On the basis of above discussion, Fisher information is not conclusive enough to explain the competing 
effects of delocalization and confinement in a DW potential, as well as tunneling. Now we move on to  
the remaining uncertainty measures, like $S$, $E$ and $OS$.       

\begin{figure}             
\centering
\begin{minipage}[c]{0.20\textwidth}\centering
\includegraphics[scale=0.38]{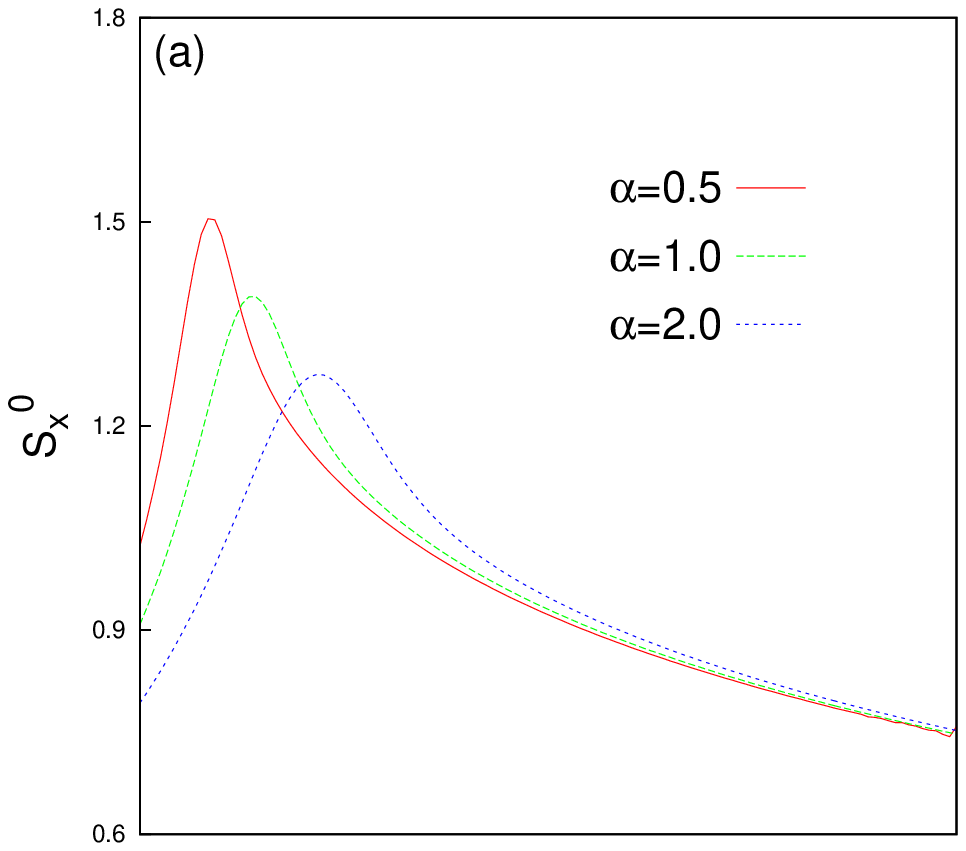}
\end{minipage}\hspace{0.15in}
\begin{minipage}[c]{0.20\textwidth}\centering
\includegraphics[scale=0.38]{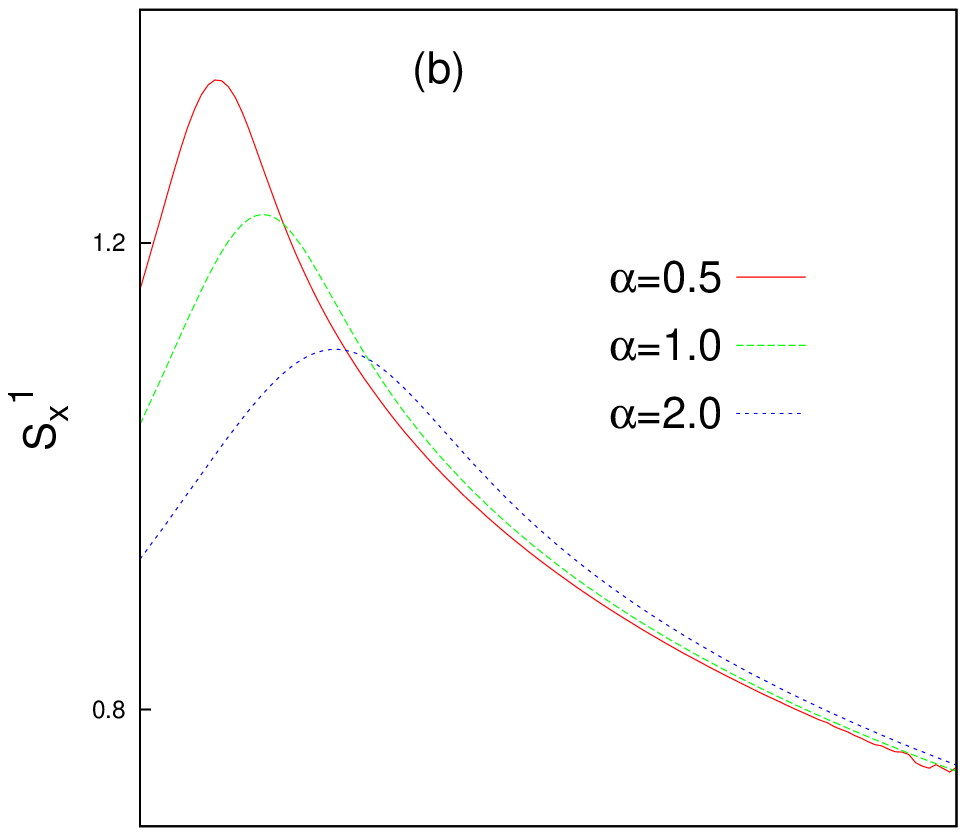}
\end{minipage}\hspace{0.15in}
\begin{minipage}[c]{0.20\textwidth}\centering
\includegraphics[scale=0.38]{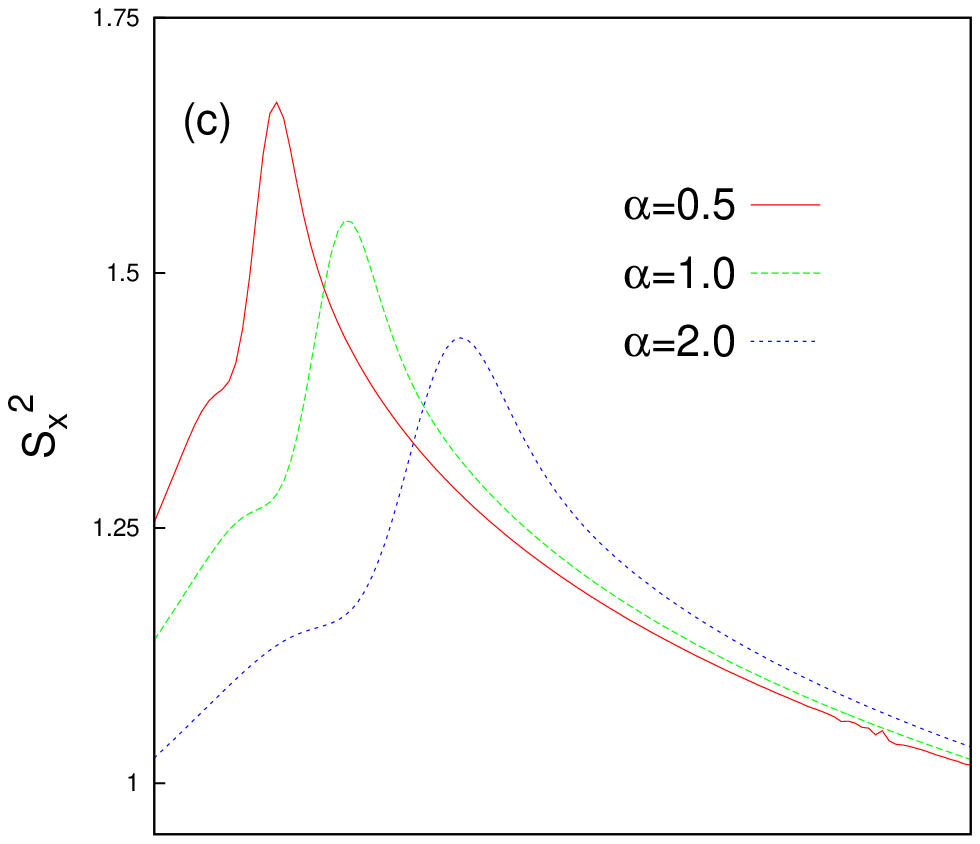}
\end{minipage}\hspace{0.15in}
\begin{minipage}[c]{0.20\textwidth}\centering
\includegraphics[scale=0.38]{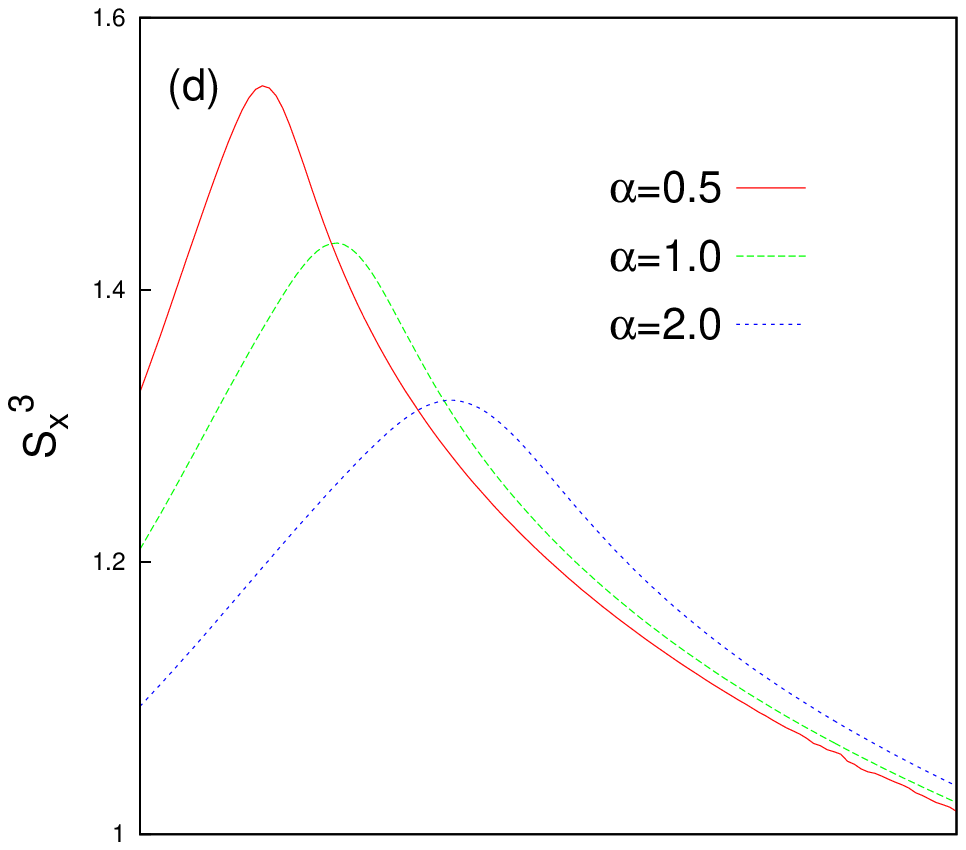}
\end{minipage}
\\[5pt]
\begin{minipage}[c]{0.20\textwidth}\centering
\includegraphics[scale=0.38]{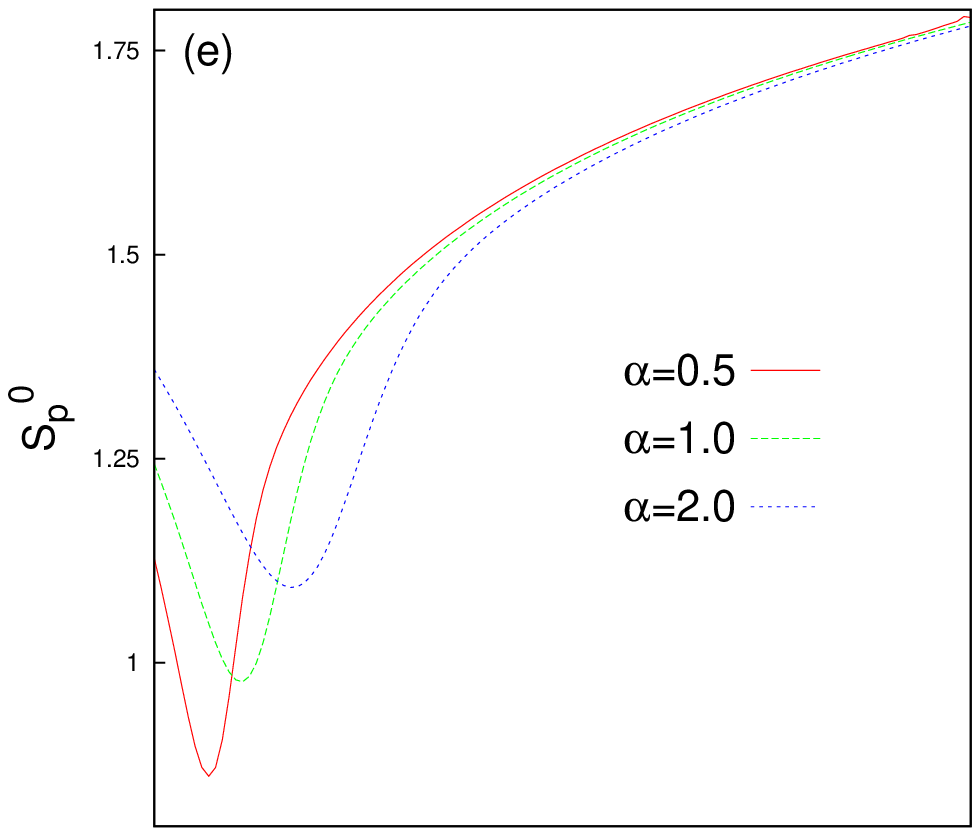}
\end{minipage}\hspace{0.15in}
\begin{minipage}[c]{0.20\textwidth}\centering
\includegraphics[scale=0.38]{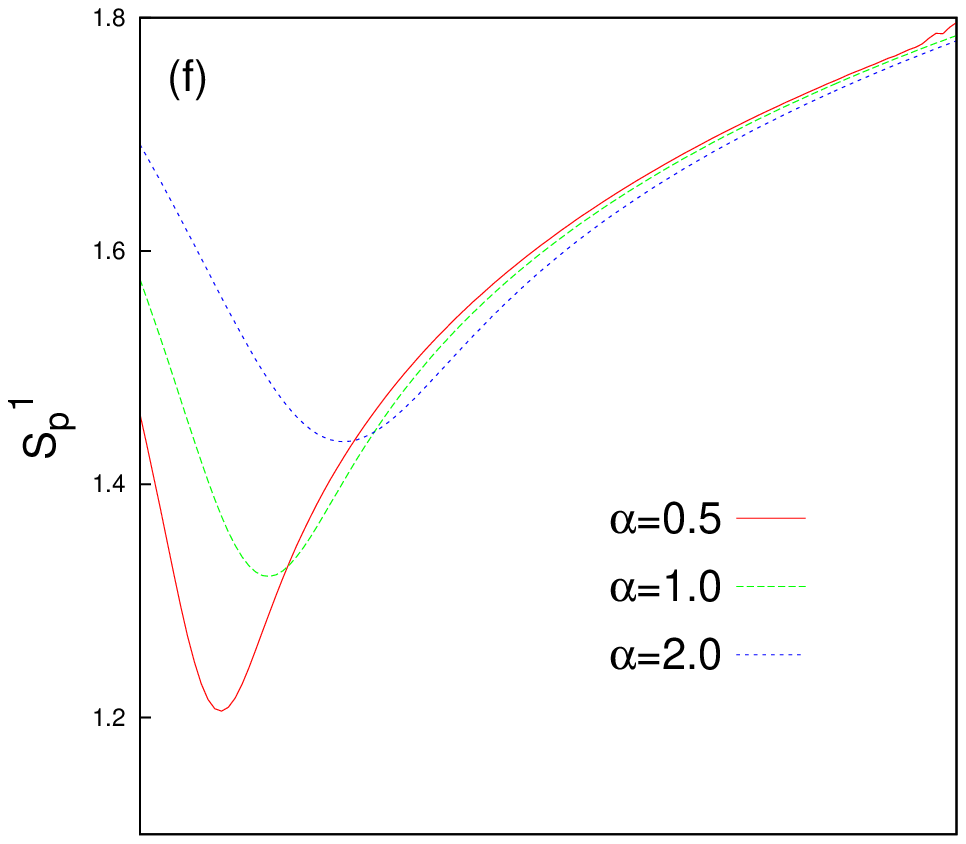}
\end{minipage}\hspace{0.15in}
\begin{minipage}[c]{0.20\textwidth}\centering
\includegraphics[scale=0.38]{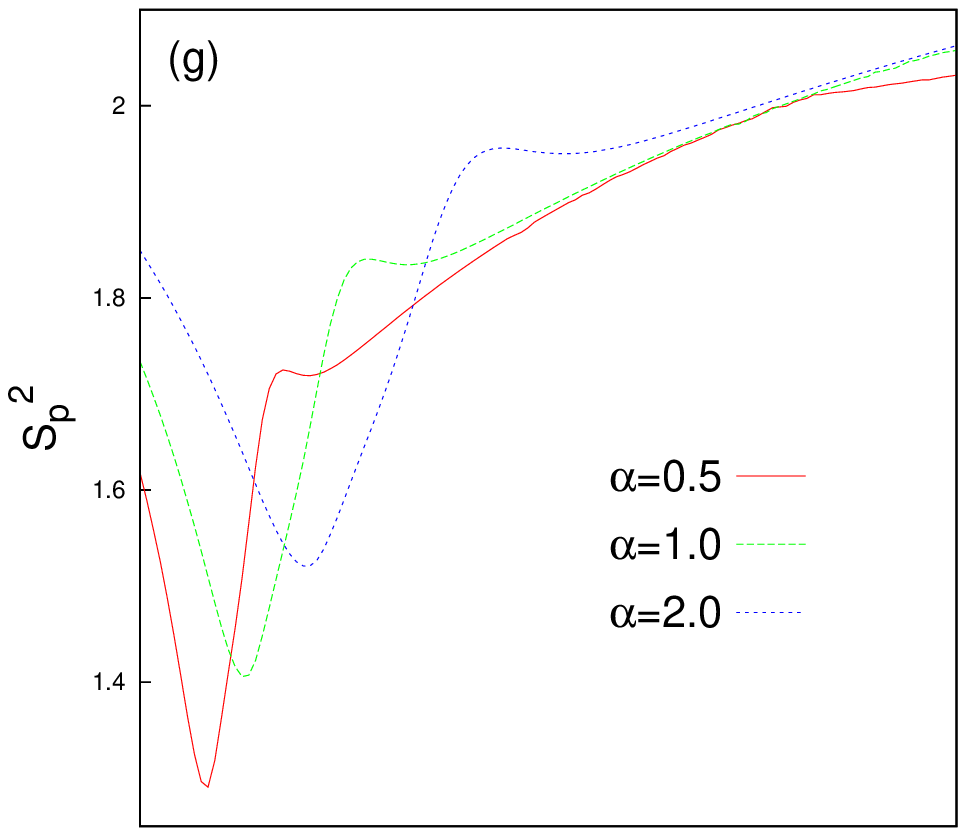}
\end{minipage}\hspace{0.15in}
\begin{minipage}[c]{0.20\textwidth}\centering
\includegraphics[scale=0.38]{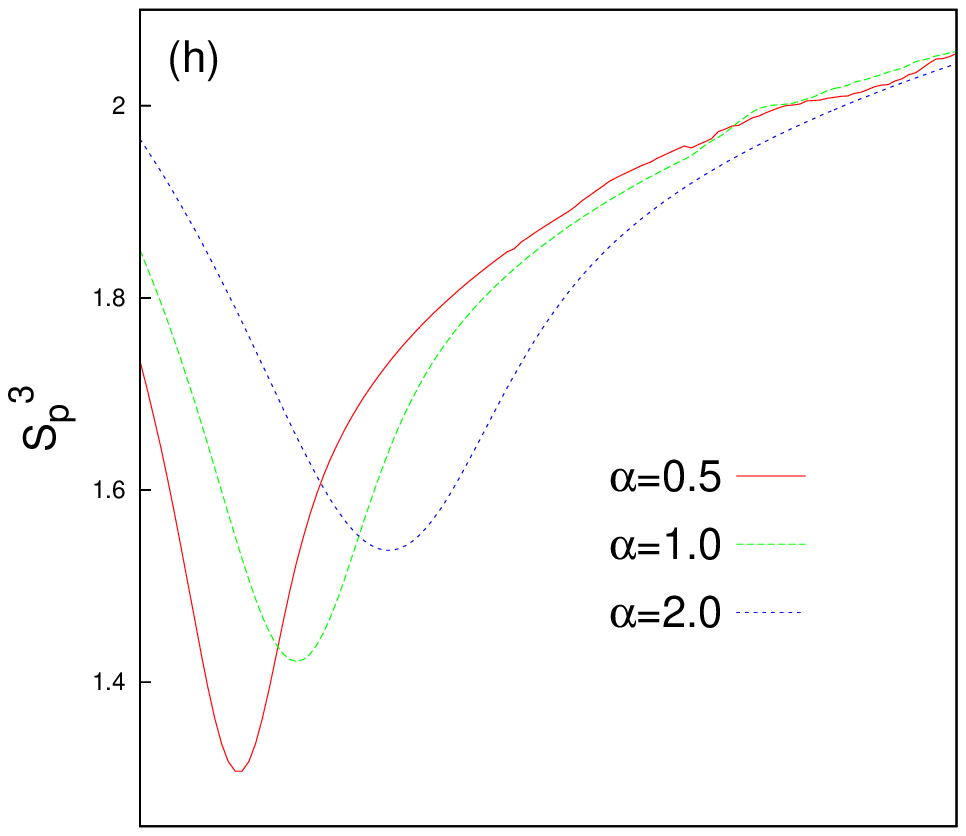}
\end{minipage}
\\[5pt]
\begin{minipage}[c]{0.20\textwidth}\centering
\includegraphics[scale=0.38]{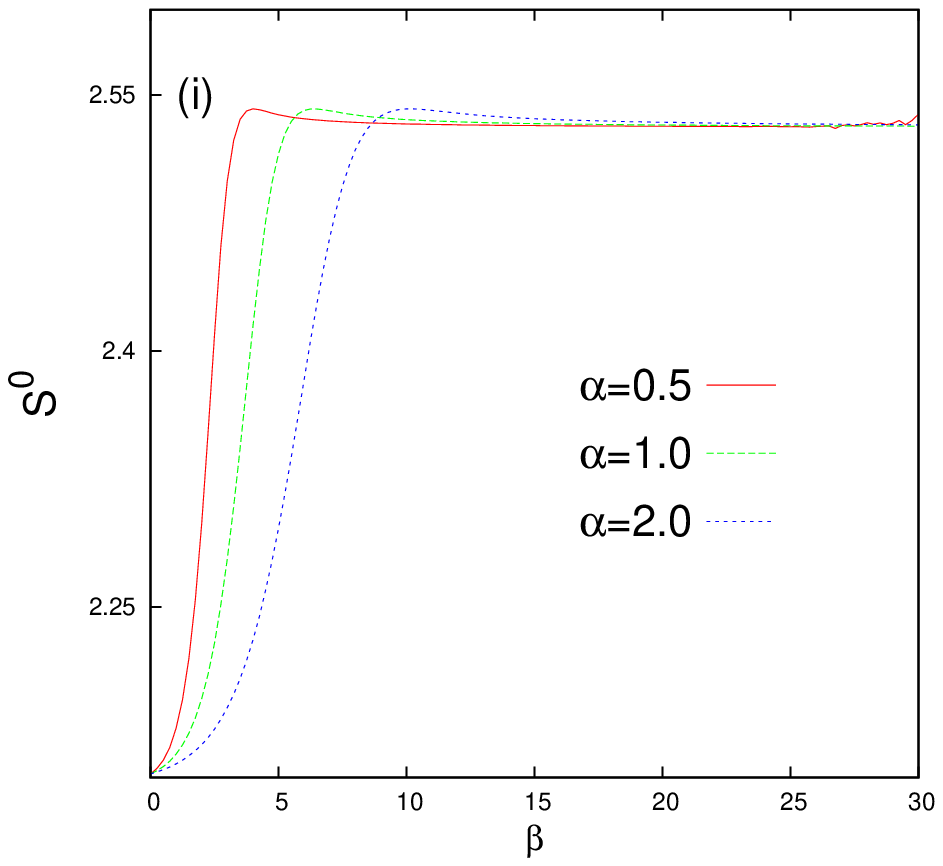}
\end{minipage}\hspace{0.15in}
\begin{minipage}[c]{0.20\textwidth}\centering
\includegraphics[scale=0.38]{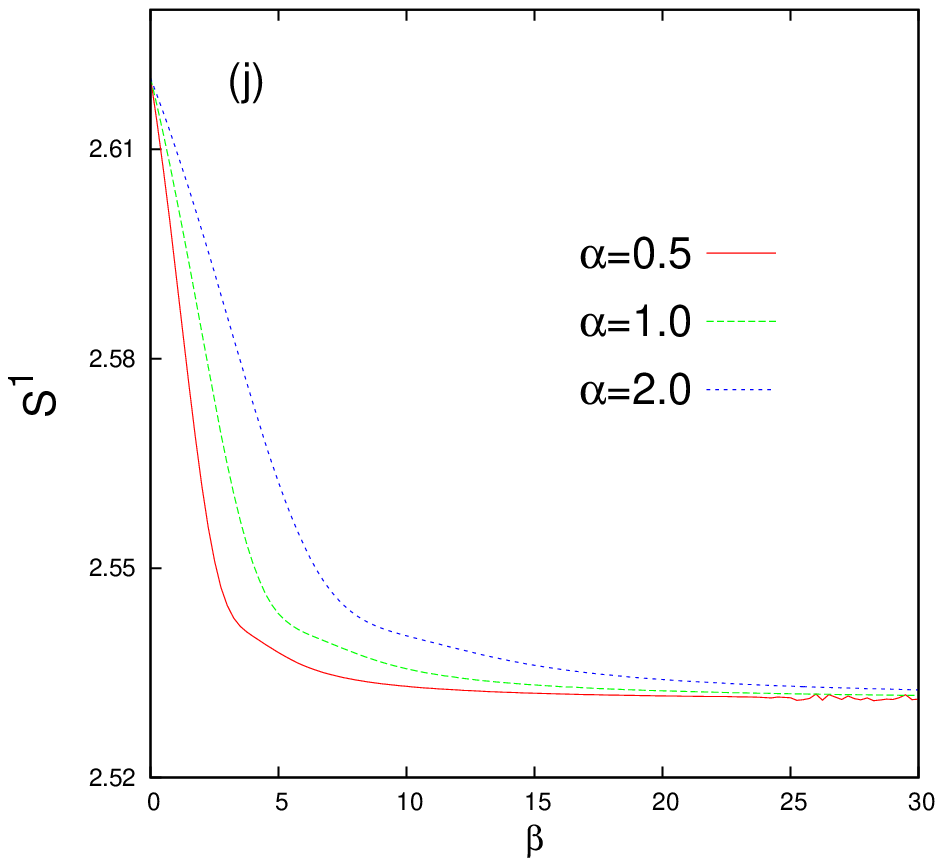}
\end{minipage}\hspace{0.15in}
\begin{minipage}[c]{0.20\textwidth}\centering
\includegraphics[scale=0.38]{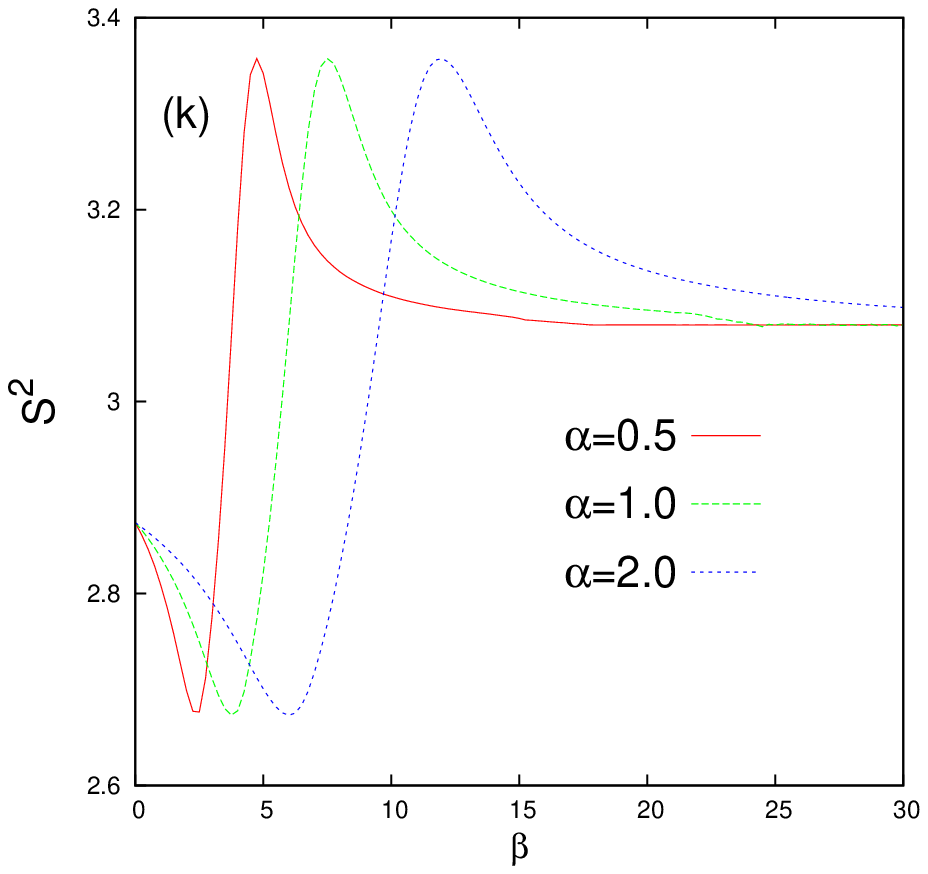}
\end{minipage}\hspace{0.15in}
\begin{minipage}[c]{0.20\textwidth}\centering
\includegraphics[scale=0.38]{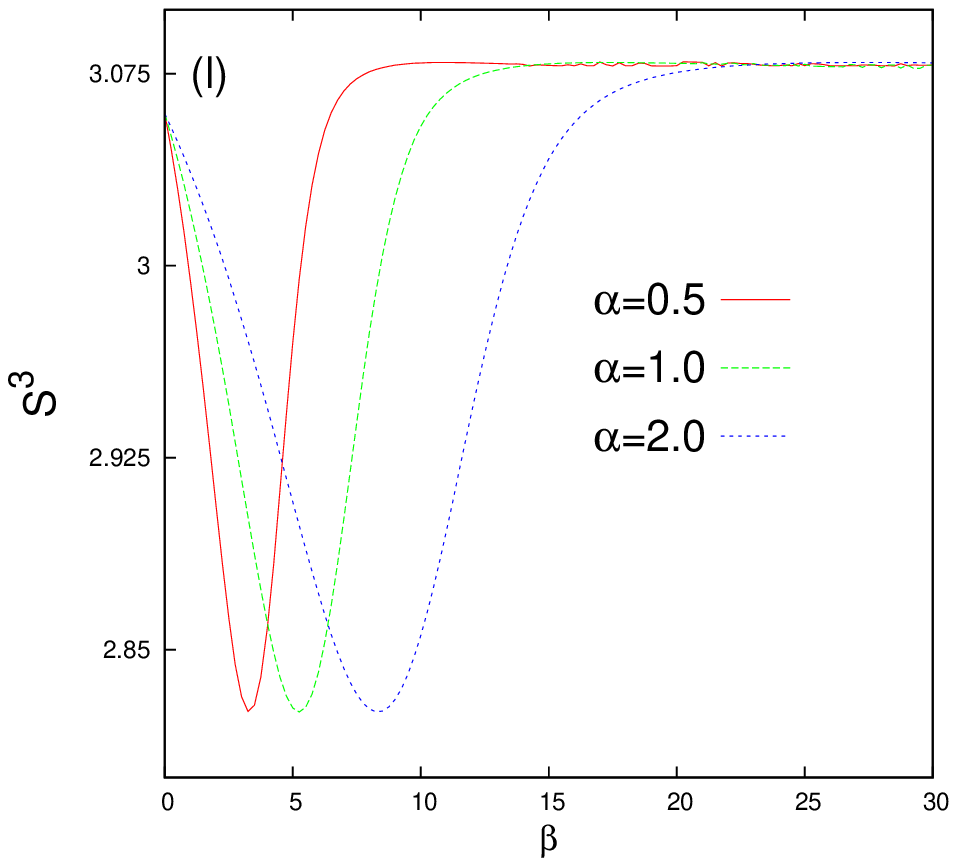}
\end{minipage}
\caption[optional]{Similar plots as in Fig.~(3), for $S_{x}, S_{p}, S$  vs. $\beta$ of DW potential.
See text for details.}
\end{figure} 

\subsection{Shannon entropy}
Next we focus on the Shannon entropy measures for lowest four states in a DW potential, Eq.~(6), 
calculated using Eqs.~(22), (23). Figure~(5) registers effects of $\beta$ on $S_{x}$, for ground 
and first three excited states (top four panels (a)-(d) respectively), at three selected 
values of $\alpha$, namely, 0.5, 1, 2, following same strategy as in previous figures.  
Apparently, there is considerable similarity in general feature amongst these plots--$S_{x}$ 
at first increases steadily, then attains a maximum and finally falls off gradually with increasing 
$\beta$. Also, for a particular state, the positions of these maxima shift to higher values of 
$\beta$ and respective peak heights decrease, making them flatter progressively as $\alpha$ 
increases. This is in keeping with the competing effects of $\beta$ on particle--at small values, 
it seems to predominantly increase delocalization of the particle (reflected in an increase in 
$S_{x}$); however at large values, the prevalent effect seems to be its confinement. At the point 
of maximum, however, these two effects are comparable and seemingly cancel each other out. It is 
worth mentioning here that as predicted earlier, varying $\alpha$ \emph{does not} cause any significant 
change in the qualitative pattern of $S_x$--the maximum shifts to right, and $S_{x}$ values get 
suppressed at maximum with successive increase in $\alpha$. The appearance of a shoulder in 
$S_{x}^{2}$ in sub-figure (c) is, however, noteworthy--this may be caused by the nature of wave function 
of second excited state itself, which has a node inside each of the potential wells along with a 
maximum at center. Position of shoulder also roughly coincides with onset of tunneling. 
Next, from the top leftmost panel (a) of Fig.~(6), it is observed that $S_{x}$ for ground and first excited 
state seem to coalesce at a value of $\beta$ approximately close to 5; while same for second and third 
excited state in bottom leftmost panel (d), occurs at nearly $\beta=9$. Both these are obtained at $\alpha=1.0$. 
This merging of $S_{x}$ is an indication of appearance of quasi-degeneracy in both pair of states, 
and also confinement of particle in either of the wells. Study of $S_{p}$ and $S$ will further 
consolidate this finding, which is presented next in the following paragraph. 

\begin{figure}             
\centering
\begin{minipage}[c]{0.30\textwidth}\centering
\includegraphics[scale=0.48]{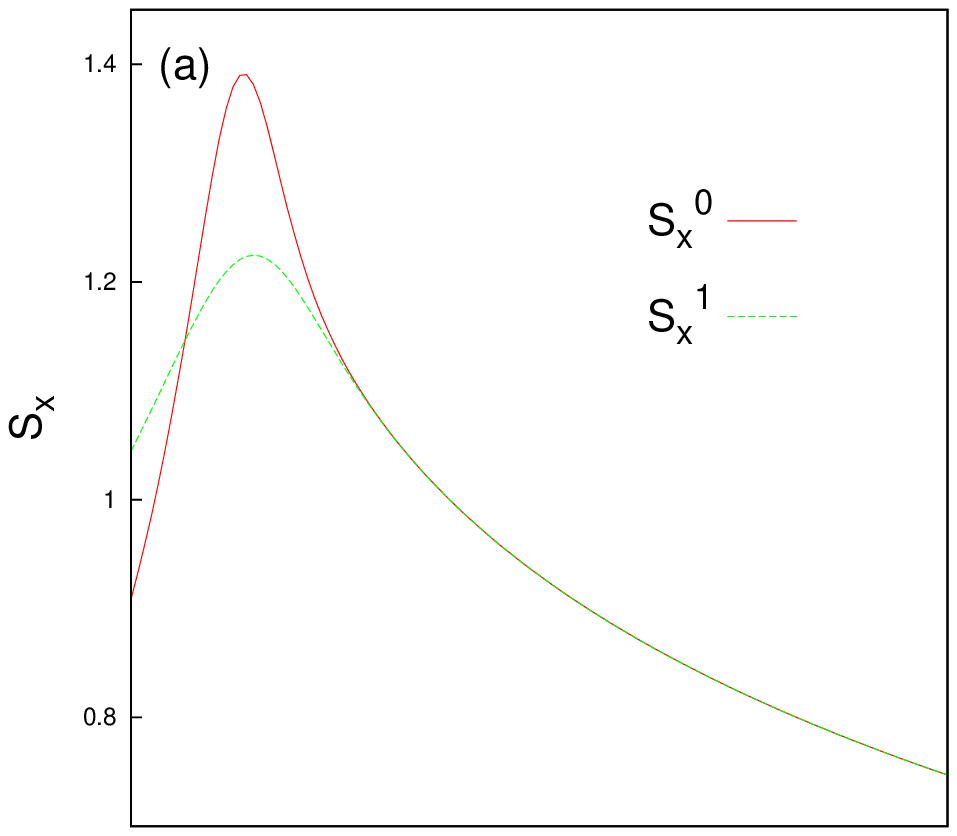}
\end{minipage}\hspace{0.05in}
\begin{minipage}[c]{0.30\textwidth}\centering
\includegraphics[scale=0.48]{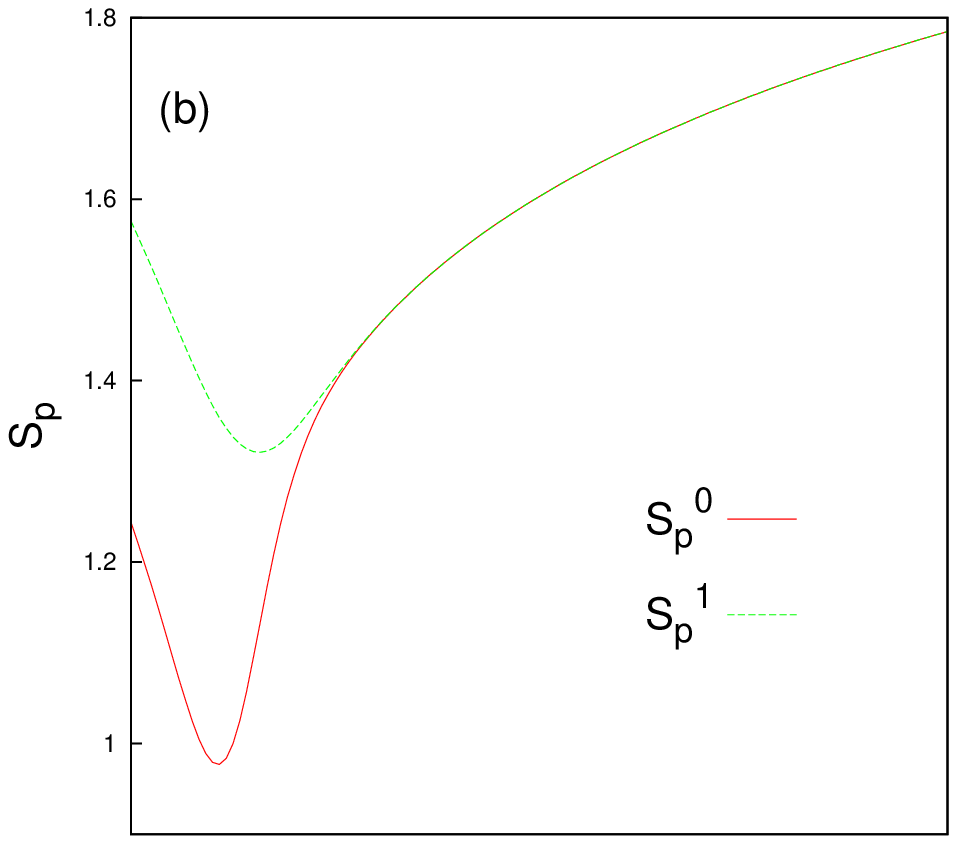}
\end{minipage}\hspace{0.05in}
\begin{minipage}[c]{0.30\textwidth}\centering
\includegraphics[scale=0.48]{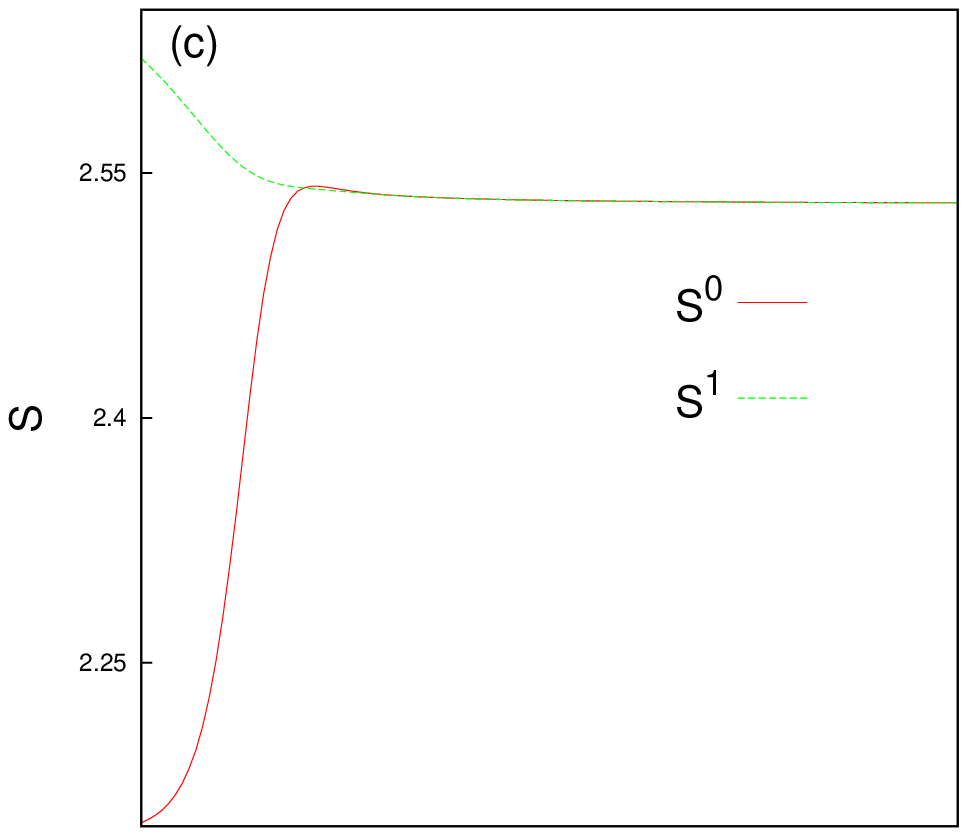}
\end{minipage}
\\[5pt]
\begin{minipage}[c]{0.30\textwidth}\centering
\includegraphics[scale=0.48]{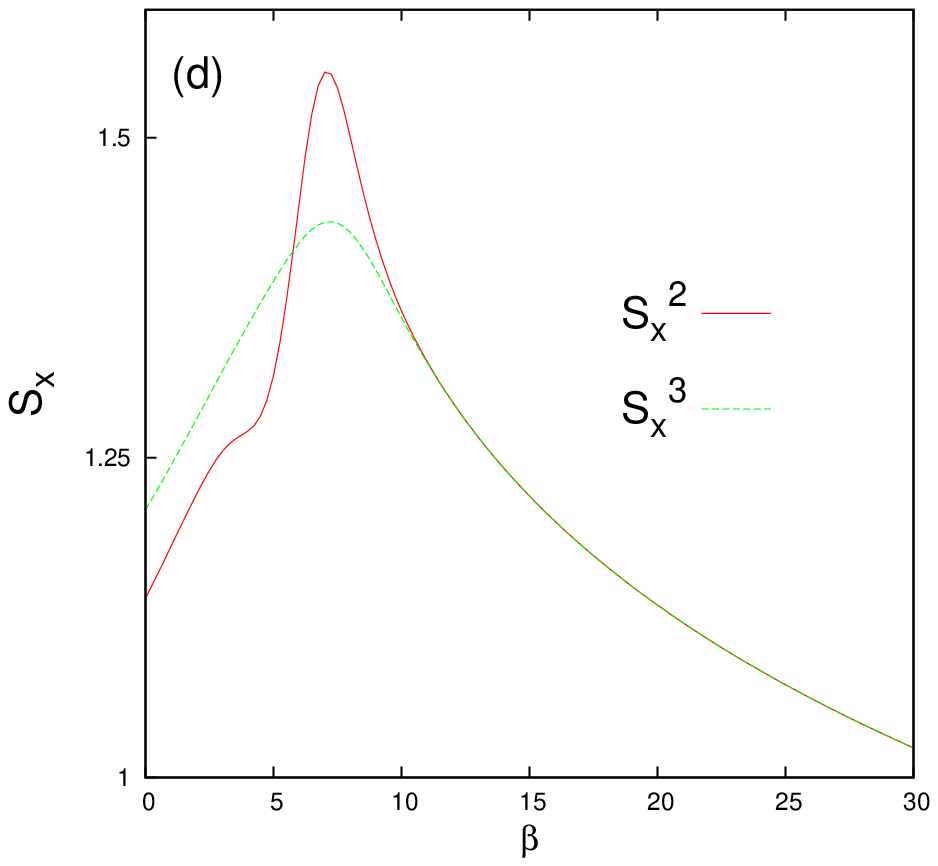}
\end{minipage}\hspace{0.05in}
\begin{minipage}[c]{0.30\textwidth}\centering
\includegraphics[scale=0.48]{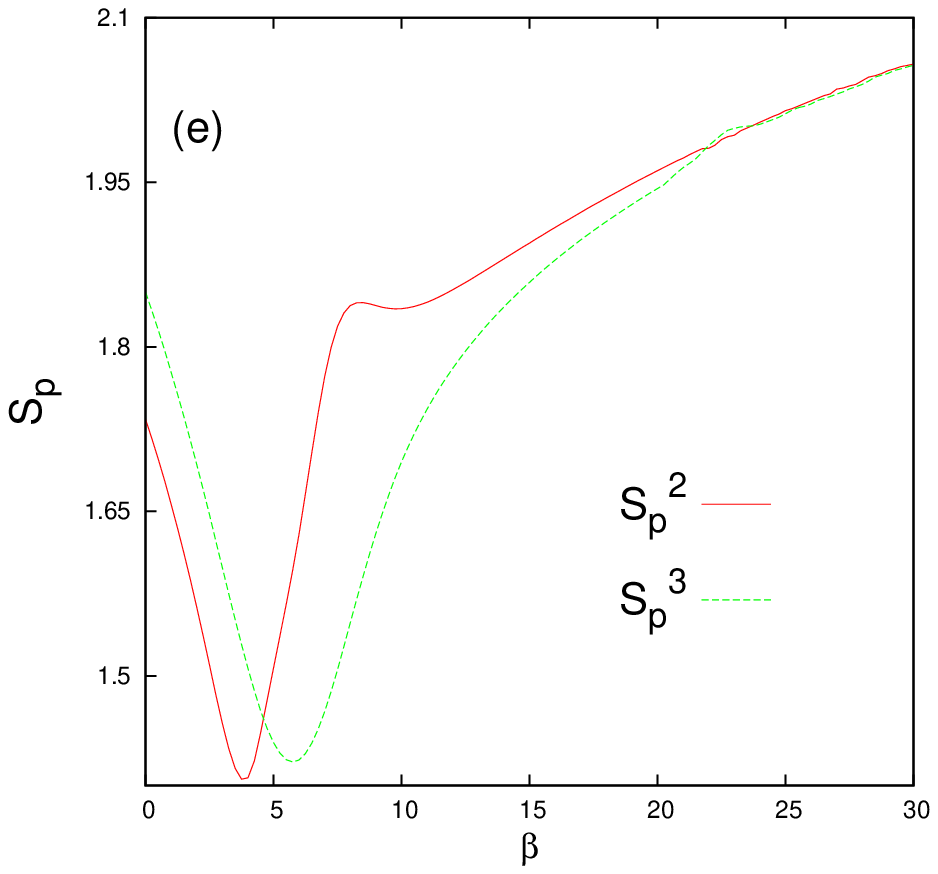}
\end{minipage}\hspace{0.05in}
\begin{minipage}[c]{0.30\textwidth}\centering
\includegraphics[scale=0.48]{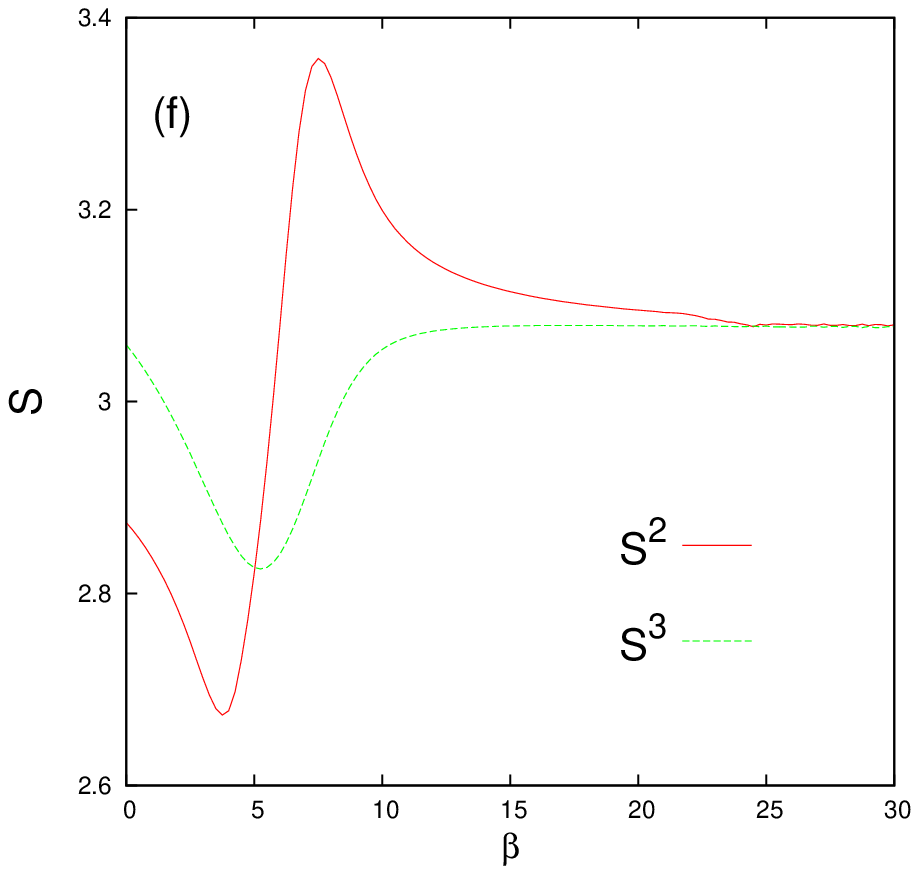}
\end{minipage}
\caption[optional]{Plots of $S_{x}, S_{p}, S$  vs. $\beta$ at $\alpha=1$, of DW potential, as in Fig.~(4).
See text for details.}
\end{figure}

Next, four middle panels (e)-(h) of Fig.~(5) illustrate behavior of $S_{p}$'s for four 
lowest states of DW potential. Once again, general 
trend for ground and first three excited states seems to be quite similar, 
\emph{viz.,} $S_{p}$ initially decreases sharply with increase in $\beta$, then attains a 
minimum and gradually increases thereafter. In this instance also, variation in $\alpha$ shows no 
qualitative change. Positions of these minima shift to higher values of $\beta$ 
with increasing $\alpha$; however this time, values of $S_{p}$ increase overall. A shoulder
again appears for $S_{p}^2$. However, since effects of tunneling in momentum space is not 
clearly understood, a definitive explanation of this phenomenon appears difficult. In top middle segment
(b) of Fig.~(6), $S_p^0$, $S_p^1$ corresponding to two lowest states of DW potential are plotted 
against $\beta$, for a fixed $\alpha=1$; similar plots are given for $S_p^2$, $S_p^3$ in bottom middle panel 
(e). Note that, $S_{p}$'s in case of a DW show a pattern similar to that found in QHO (not shown 
here to save space). Just like the convergence of $S_{x}$ for two lowest pairs at nearly 
$\beta=7$, 9, same in case of $S_p$ occurs at $\beta$ approximately equal to 7, 21 
respectively. While these plots pertain to $\alpha=1$, similar trend is recorded 
for other $\alpha$ as well, with a corresponding shift in position of convergence.

\begin{figure}         
\centering
\begin{minipage}[c]{0.28\textwidth}\centering
\includegraphics[scale=0.30]{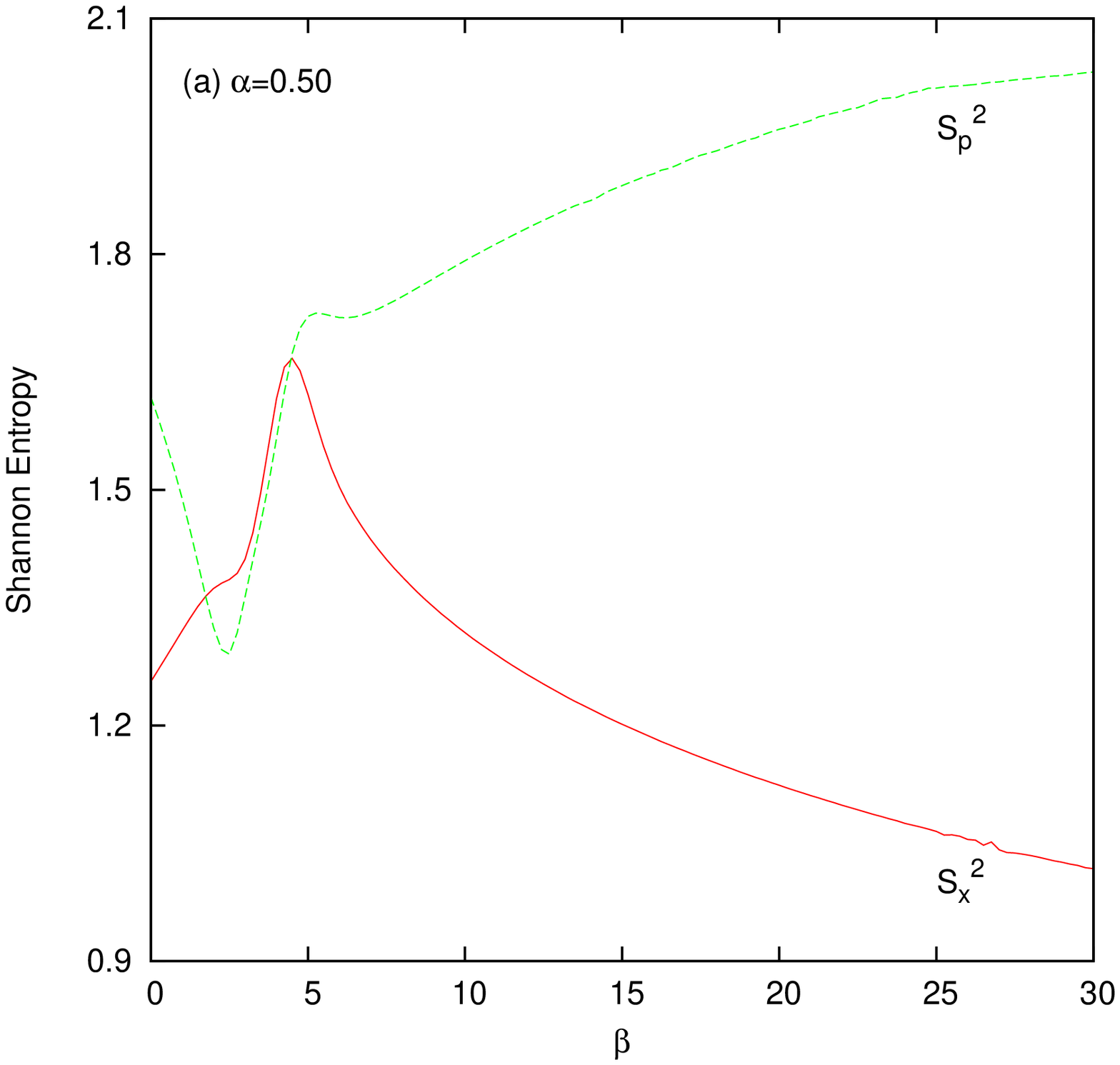}
\end{minipage}
\hspace{0.25in}
\begin{minipage}[c]{0.28\textwidth}\centering
\includegraphics[scale=0.30]{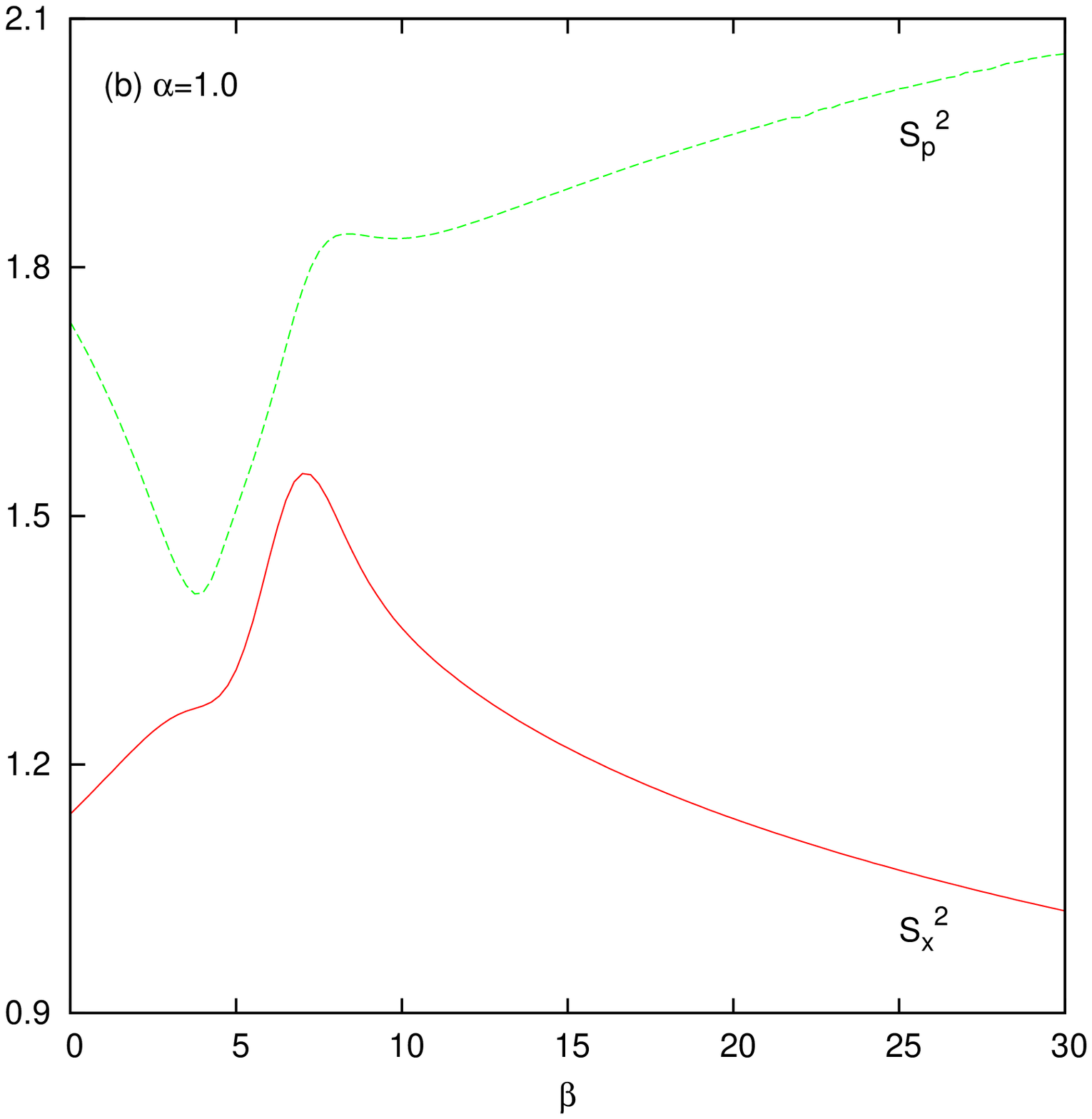}
\end{minipage}
\hspace{0.25in}
\begin{minipage}[c]{0.28\textwidth}\centering
\includegraphics[scale=0.30]{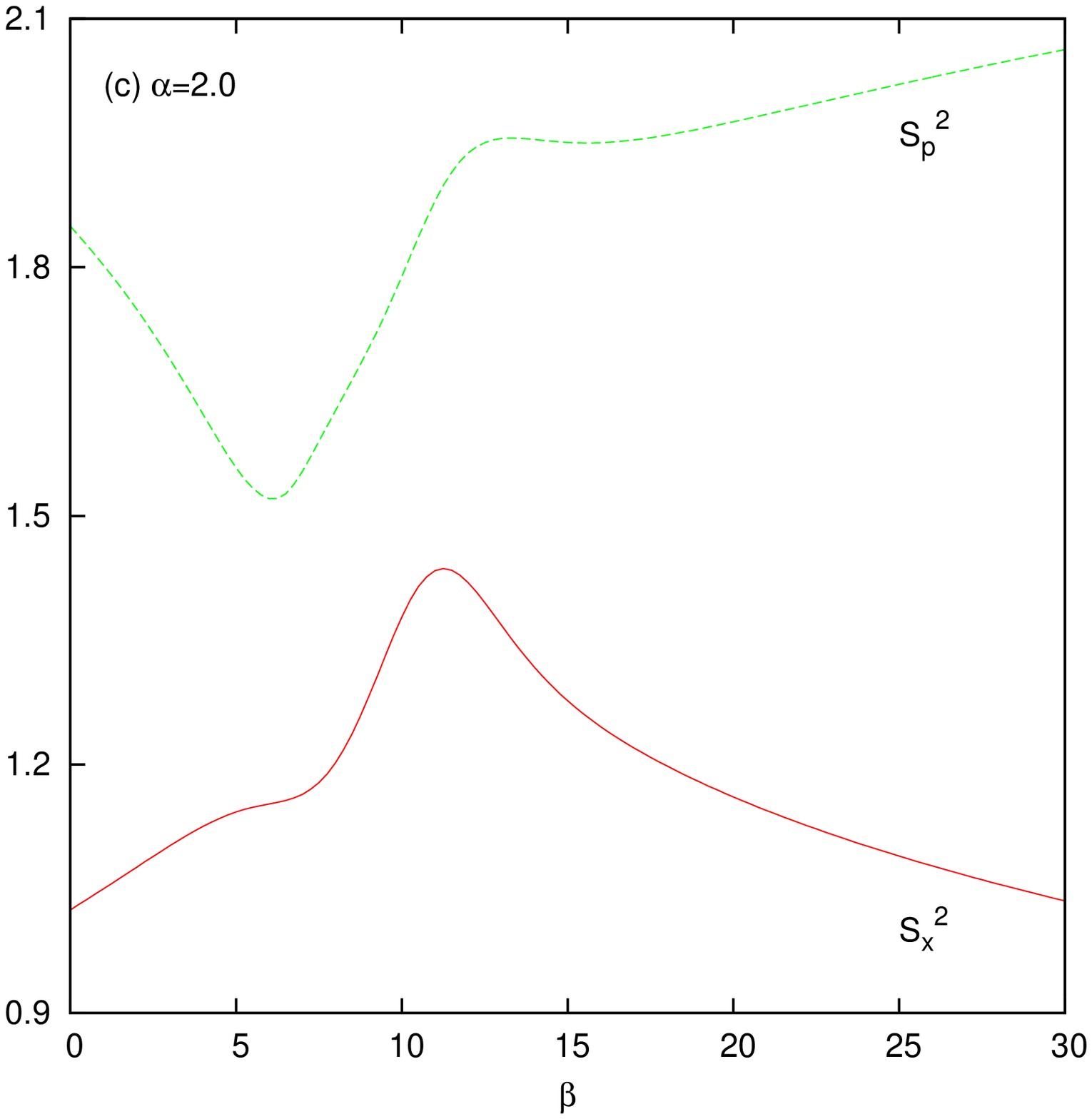}
\end{minipage}
\caption{Shannon entropy in position $(S_{x}^{2})$ and momentum $(S_{p}^{2})$ space, versus 
$\beta$, for second excited state, of DW potential in Eq.~(6). (a)-(c) correspond to three 
$\alpha$ values, namely, 0.5, 1 and 2.}
\end{figure}

To study the cumulative effect of barrier on a given state, one needs to investigate total IE, 
which is offered in bottom four panels (i)-(l) of Fig.~(5). Presence of a barrier at the 
center of well generally has a greater 
impact on ground state than that of first excited state, possibly because the latter possesses a 
node at the maximum of potential. For higher excited states, more complicated behavior is 
anticipated as more nodes start appearing in wave function. From panel (i), one observes
that total IE for ground state quite sharply increases initially, hitting an almost inconspicuous 
maximum and remaining virtually constant thereafter. For first excited state, on the other hand, 
sub-figure (j) shows that same decreases monotonically, until individual $\alpha$ plots merge 
together. Behavior of total IE seems to be unaffected, at least qualitatively, with respect 
to variations in $\alpha$--except for a shift in right. The net IE successfully portrays, 
first delocalization, then tunneling and finally an onset of confinement--initial gradual 
variation gives way to sudden and fast change, followed by a settling down into constancy. This 
could also bear the signature of starting of a quasi-degenerate pair of energy levels. The 
scenario is not as simple for second and third excited states, however. From segment (k), 
it is clear that the maximum in $S^{2}$ (second excited state) is much more distinctly  
defined than its counterpart in ground state. In addition, this maximum is now preceded by the 
appearance of a minimum. Furthermore, the difference in $\beta$ values at which inflection 
points in $S^2$ occur tends to increase with $\alpha$. This is not surprising--since the total IE, 
being a sum, would contain effects from extrema in both $S_{x}^{2}$, $S_{p}^{2}$. $S^{2}$ 
offers a minimum as a consequence of minimum in $S_{p}^{2}$, followed by the effect of a maximum 
in $S_{x}^{2}$ respectively. A similar pattern is however, not registered for ground and first excited 
states, as positions of minimum in $S_{p}$ and maximum in $S_{x}$ are very close to each other
and remain so for all values of $\alpha$ studied--their individual effects are therefore not 
discernible. To address this issue more closely, Fig.~(7) now displays the second excited-state 
quantities, $S_x^2$ and $S_p^2$ against $\beta$ for same three $\alpha$, \emph{viz.,} 0.5, 1 and 2
in segments (a)-(c) respectively. One notices that, overlap between 
$S_{x}^{2}$ and $S_{p}^{2}$ gets reduced with corresponding increase in $\alpha$. Now, the bottom rightmost 
panel (l) of Fig.~(5) reflects changes in total IE for third excited state of DW, with 
respect to $\alpha$. $S^{3}$ initially follows same pathway as its counterpart in $S^1$ 
(panel (j) of Fig.~(5)), falling steeply and monotonically. However, similarities end there; 
for $S^{3}$ then attains a minimum, rises rapidly and eventually decays into constancy. As noted 
previously, values of $S^{3}$ at the minima remain fixed for all $\alpha$ values. This is unlike
the case of $S^2$ (panel (k) of Fig.~(5)) where minima and maxima values change slightly as 
$\alpha$ changes.

\begin{figure}                         
\begin{minipage}[c]{0.40\textwidth}
\centering
\includegraphics[scale=0.45]{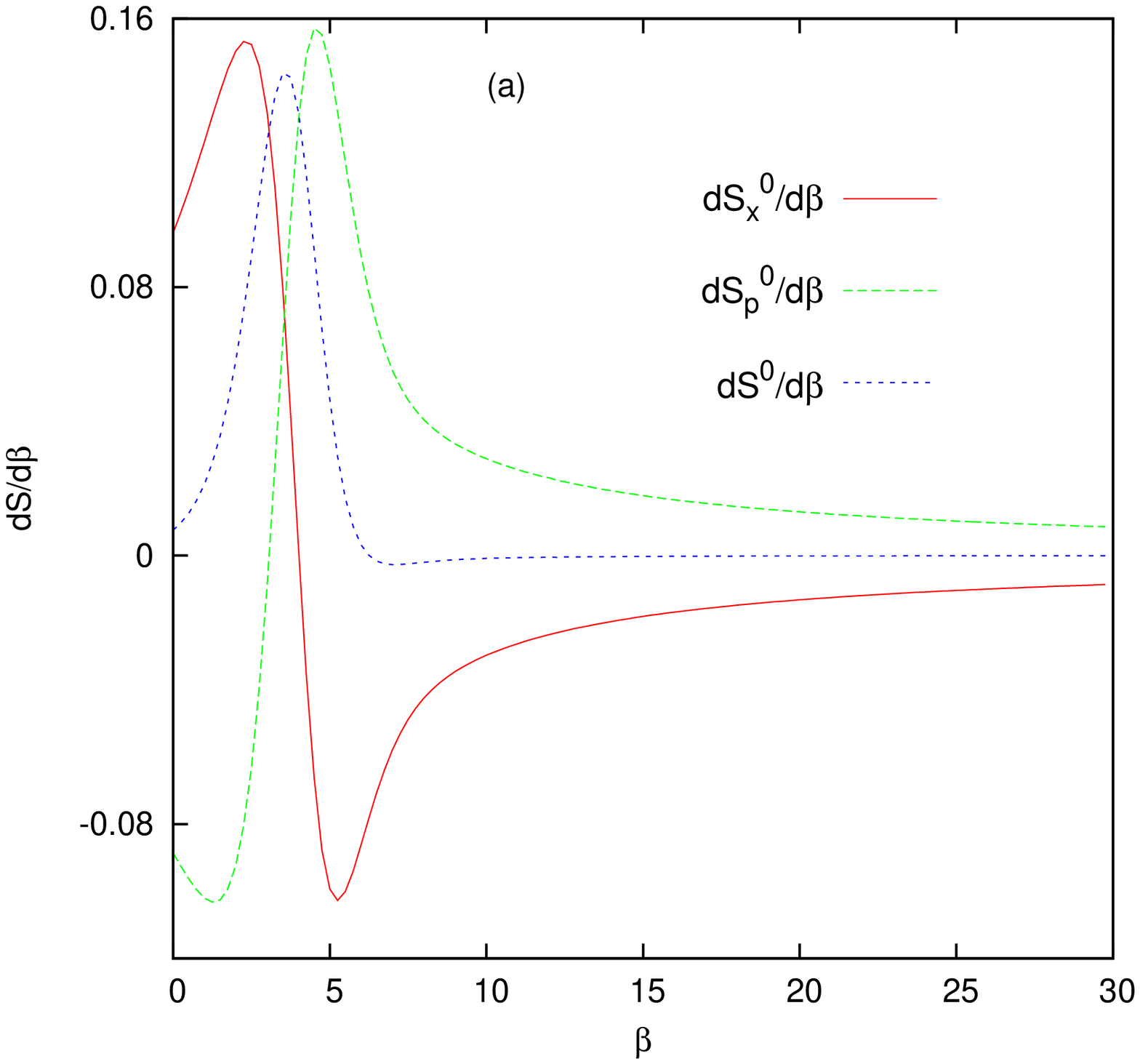}
\end{minipage}%
\hspace{0.9in}
\begin{minipage}[c]{0.40\textwidth}
\centering
\includegraphics[scale=0.45]{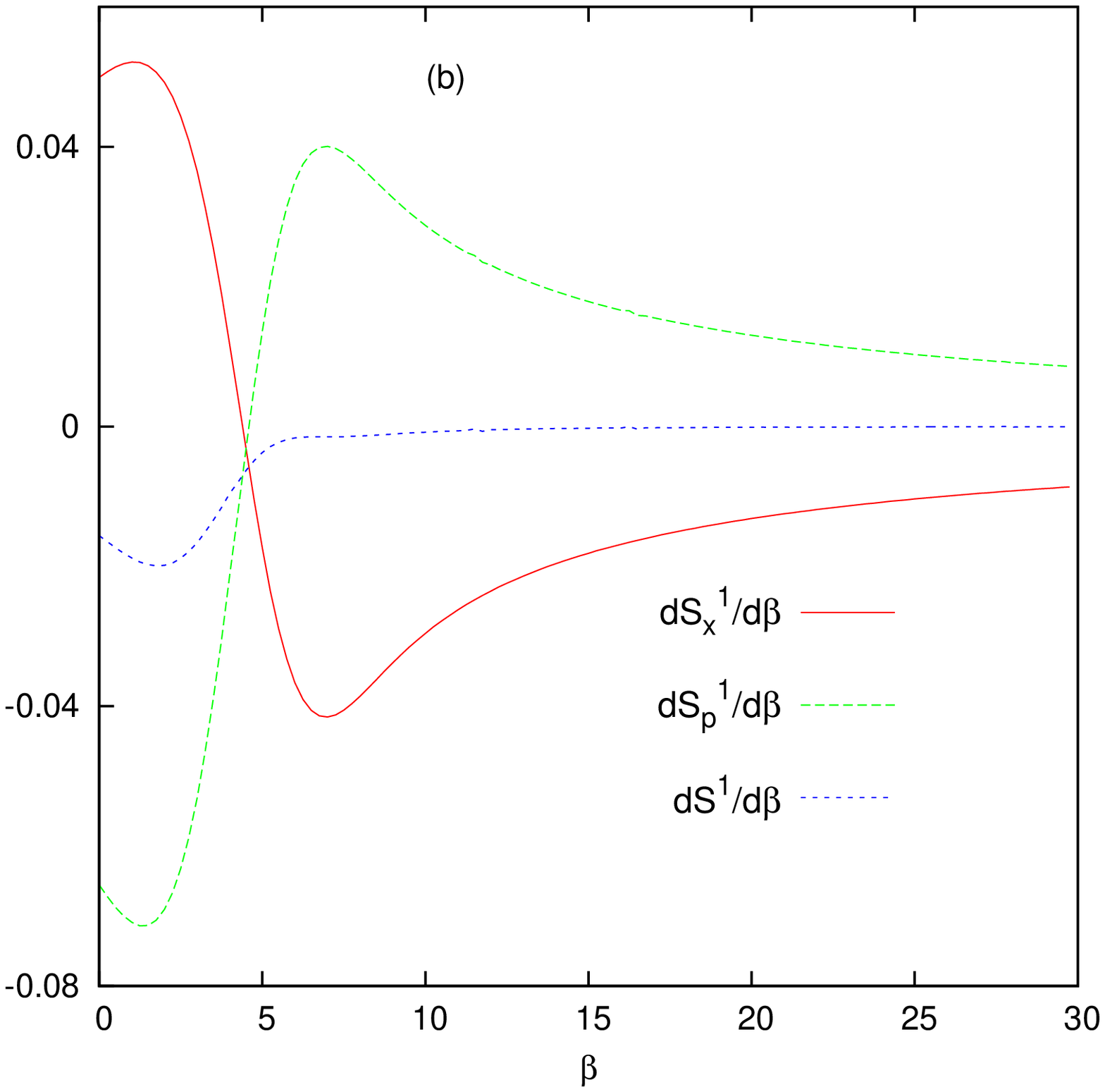}
\end{minipage}%
\caption{$S$ derivatives against $\beta$, for $\alpha=1$, for potential in Eq.~(6). Left panel 
(a) corresponds to $\frac{dS_{x}^{0}}{d\beta}, \frac{dS_{p}^{0}}{d\beta}, \frac{dS^{0}}{d\beta}$, 
in ground state, while (b) refers to $\frac{dS_{x}^{1}}{d\beta}, \frac{dS_{p}^{1}}{d\beta}, 
\frac{dS^{1}}{d\beta}$, in first excited state.}
\end{figure}

Now, two rightmost panels (c), (f) of Fig.~(6) illustrate the convergence of total IE for ground $(S^0)$, first 
excited $(S^1)$ state and second $(S^2)$, third excited $(S^3)$ states respectively, keeping 
$\alpha$ fixed at 1 in both occasions. Total IE becomes stationary at a value of $2.53$ on and 
after $\beta=5.0$ for first pair, while same for latter pair occurs at $\beta=20.0$ with 
$S=3.08$. A parallel exercise for other $\alpha$ values (0.5 and 2) for states under 
consideration, reveals that, these net IE values remain unaltered by variations in $\alpha$. 
Thus, for the family of potentials described by Eq.~(2), it seems that the net 
information attained by any quasi-degenerate pair of states would be a \emph{constant, unique 
to that pair, irrespective of $\alpha$.}

In order to gain further insight on to the effect of $\beta$ on $S$, changes of latter are 
followed in terms of respective derivatives. Thus, Fig.~(8) depicts the first derivatives $S_x$, 
$S_p$ and $S$ of DW potential for ground and first excited state in left (a) and right (b) panels 
respectively. It is found that, for ground state, $\frac{dS_{x}^{0}}{d\beta}$ initially 
increases, reaches a maximum, decreases to attain a negative minimum, and then finally slowly 
grows asymptotically towards zero. It is worth mentioning that these points of inflection 
correspond to $\beta$ values, which mark the \emph{onset of tunneling} (at $\beta=2.25$) and 
\emph{harmonic} trend (at $\beta=5.0$) respectively. Furthermore, presence of a maximum at the 
\emph{exact} location of $\beta=2.25$, which marks the beginning of tunneling is especially 
noteworthy--this could potentially be used as a predictive tool to identify quantum-mechanical 
tunneling in other systems as well. Derivative of $S$ in momentum space for ground state, 
$\frac{dS_{p}^{0}}{d\beta}$, on the other hand, shows a trend exactly complimentary to that of 
$\frac{dS_{x}^{0}}{d\beta}$--it falls off to a minimum, grows rapidly to reach a maximum and 
then slowly decays towards zero. Just like in the previous occasions, points of inflection mark the onset 
and saturation of tunneling respectively. Stated otherwise, this leads to a possibility that 
quantum mechanical tunneling could be characterized by rapid growth of information in momentum 
space and a rapid decay of information in position space--implying that the particle gets lesser 
space to delocalize, in conjunction with a greater range of accessible momentum states. IE 
derivatives in position and momentum space in first excited state, $\frac{dS_{x}^{1}}{d\beta}$ 
and $\frac{dS_{p}^{1}}{d\beta}$ follow similar pattern as their ground-state counterparts, albeit the 
points of inflection are now shifted to match the beginning and decay of tunneling respectively. 
Total $S$ derivative for ground state, $\frac{dS^{0}}{d\beta}$, increases with increase in 
$\beta$, attains a maximum (at a value for which IE derivatives in position and momentum spaces 
cross $\beta$ axis), then slowly falls off towards zero. As it will be shown from a consideration of 
the semi-classical phase-space area as well in following discussion, an initial increase in $\beta$ 
increases phase-space area with a consequent increase in total information available. However, 
as tunneling begins, rate of increase of $S^{0}$ with $\beta$ falls off and gradually $S^{0}$ 
approaches a constant value. This will also be discussed later from the viewpoint of phase-space 
area. However, $\frac{dS^{1}}{d\beta}$ suffers a minimum before leveling off to zero, a trend 
which will be shown to mirror the change in phase-space area in first excited state. 

\begin{figure}             
\centering
\begin{minipage}[c]{0.20\textwidth}\centering
\includegraphics[scale=0.38]{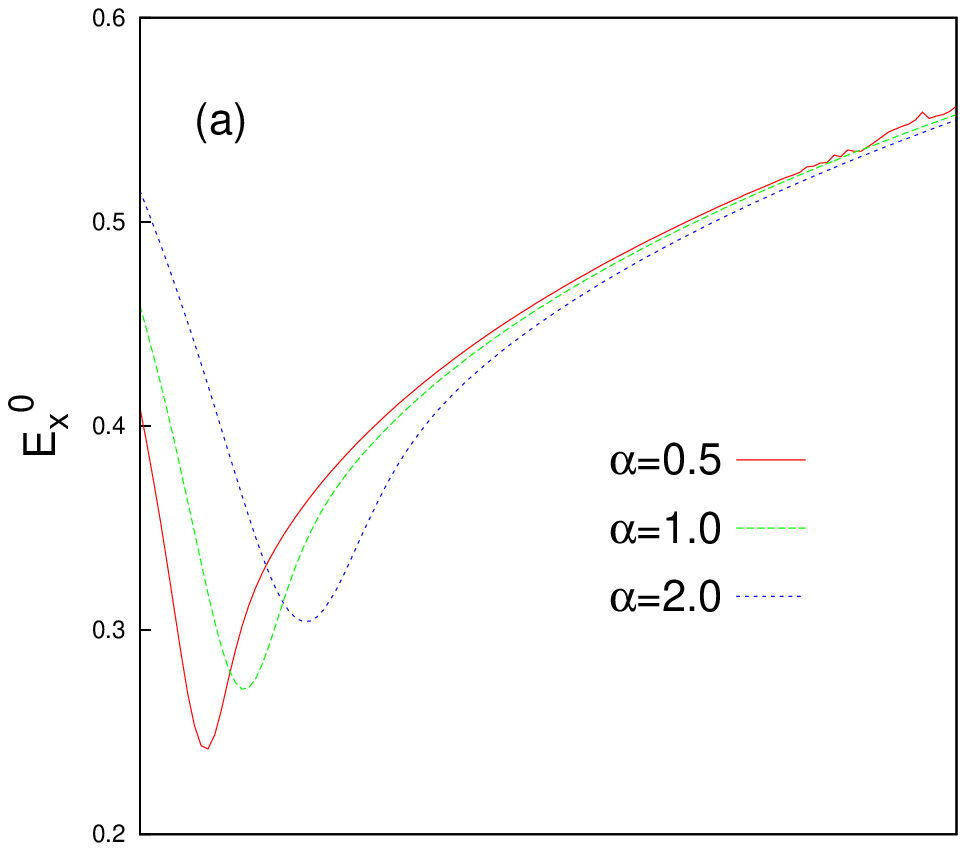}
\end{minipage}\hspace{0.15in}
\begin{minipage}[c]{0.20\textwidth}\centering
\includegraphics[scale=0.38]{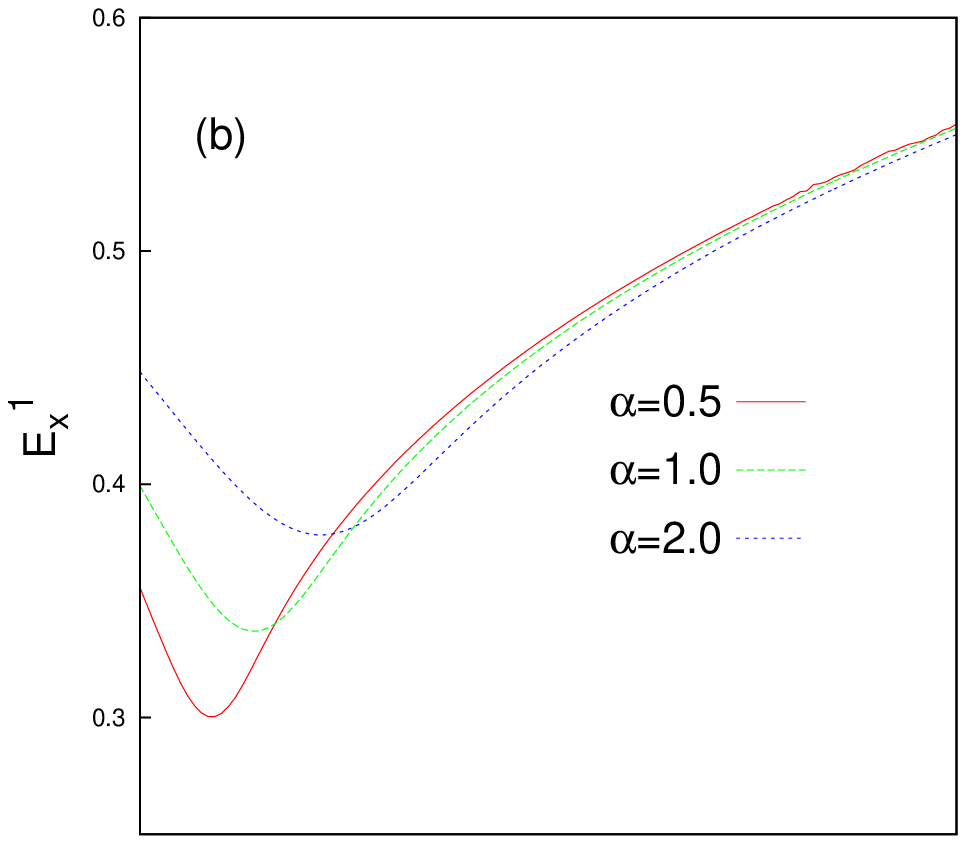}
\end{minipage}\hspace{0.15in}
\begin{minipage}[c]{0.20\textwidth}\centering
\includegraphics[scale=0.38]{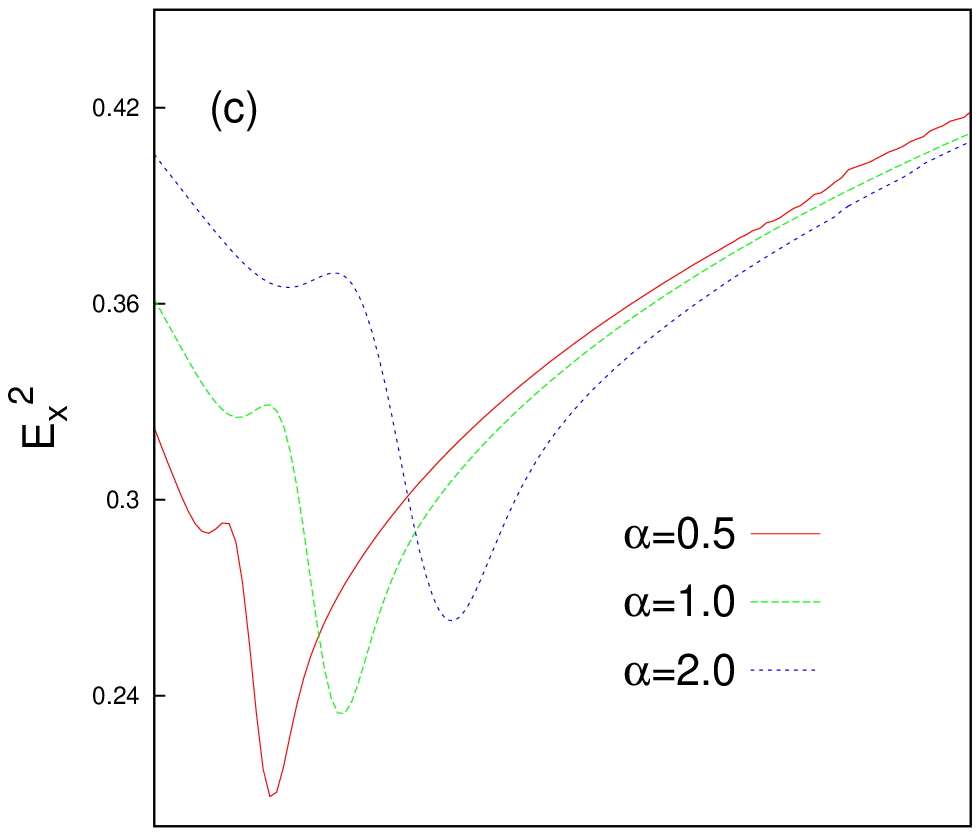}
\end{minipage}\hspace{0.15in}
\begin{minipage}[c]{0.20\textwidth}\centering
\includegraphics[scale=0.38]{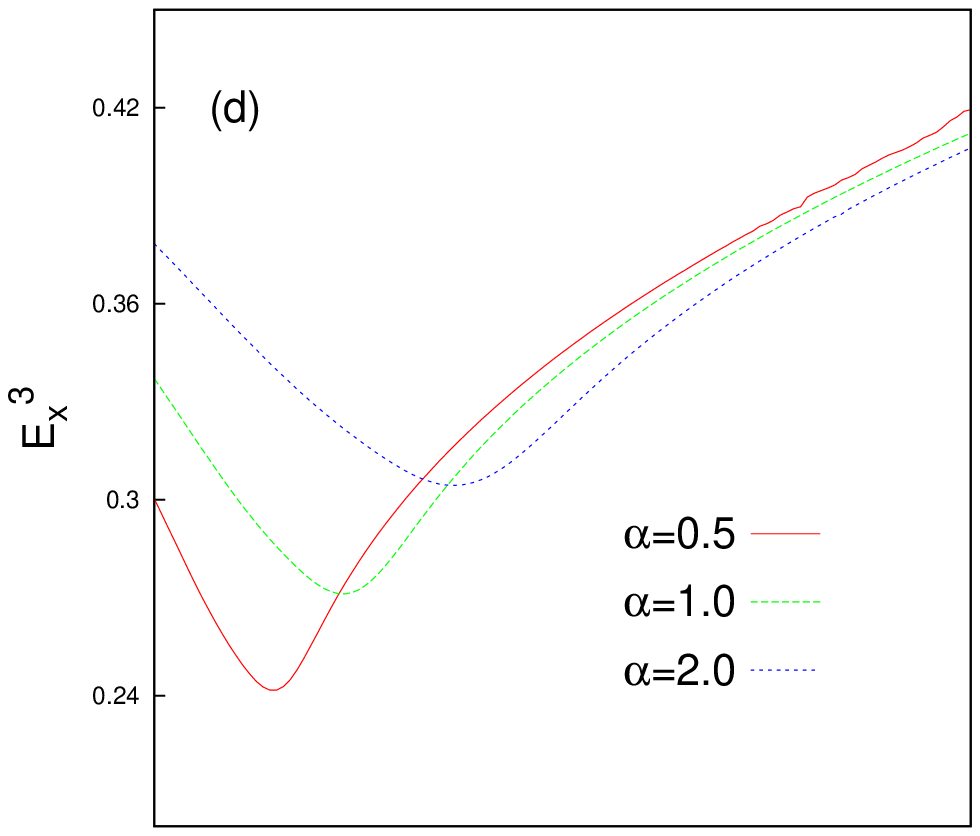}
\end{minipage}
\\[5pt]
\begin{minipage}[c]{0.20\textwidth}\centering
\includegraphics[scale=0.38]{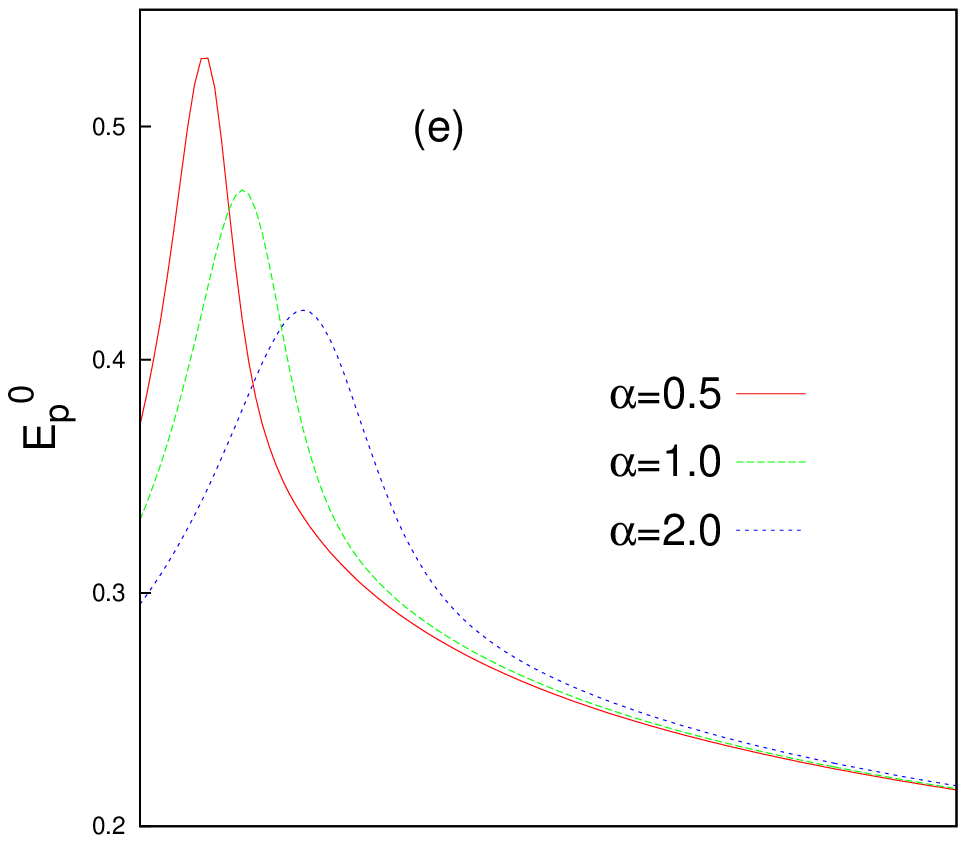}
\end{minipage}\hspace{0.15in}
\begin{minipage}[c]{0.20\textwidth}\centering
\includegraphics[scale=0.38]{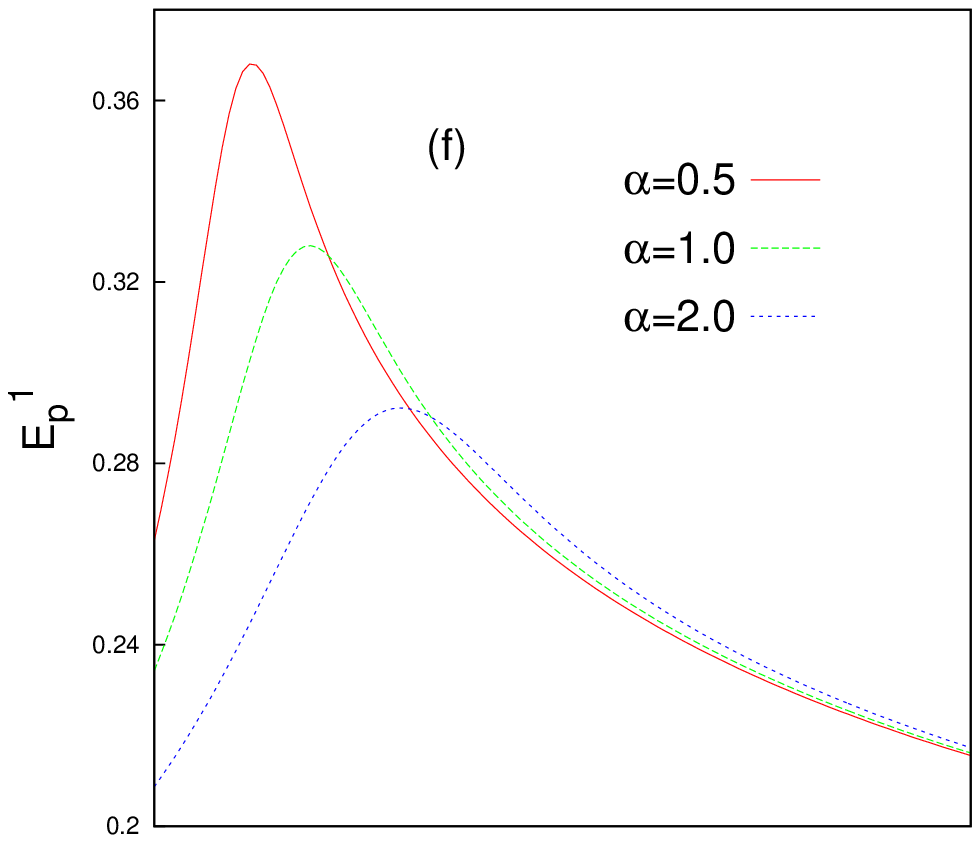}
\end{minipage}\hspace{0.15in}
\begin{minipage}[c]{0.20\textwidth}\centering
\includegraphics[scale=0.38]{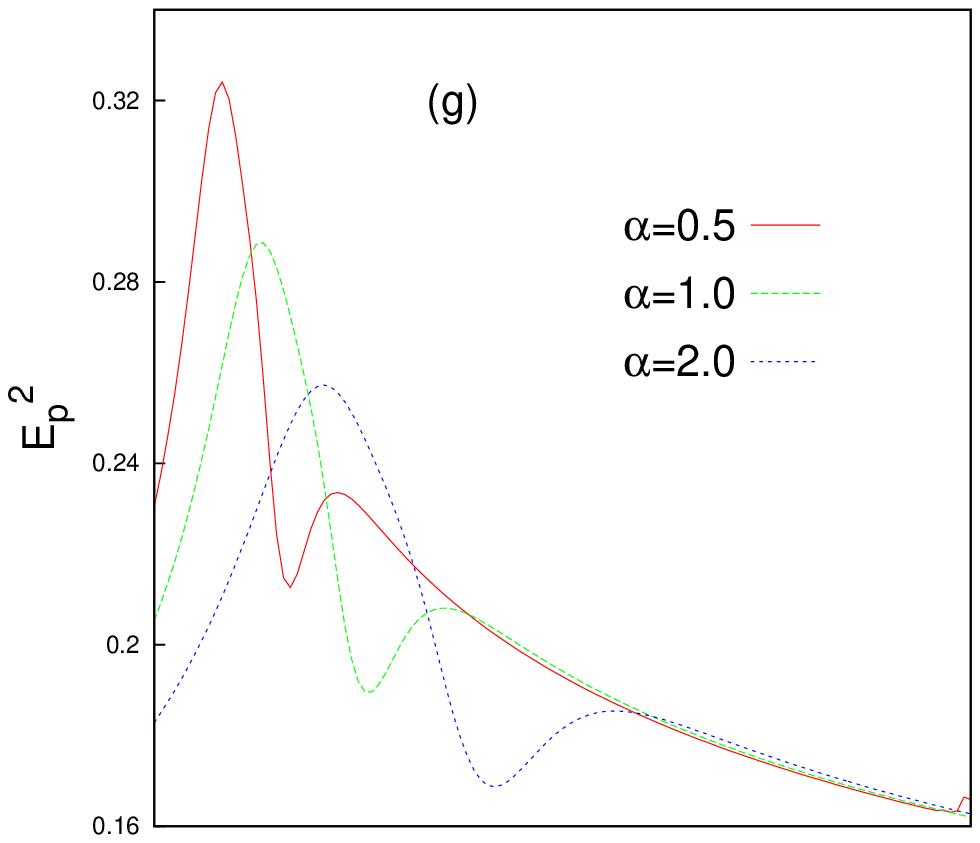}
\end{minipage}\hspace{0.15in}
\begin{minipage}[c]{0.20\textwidth}\centering
\includegraphics[scale=0.38]{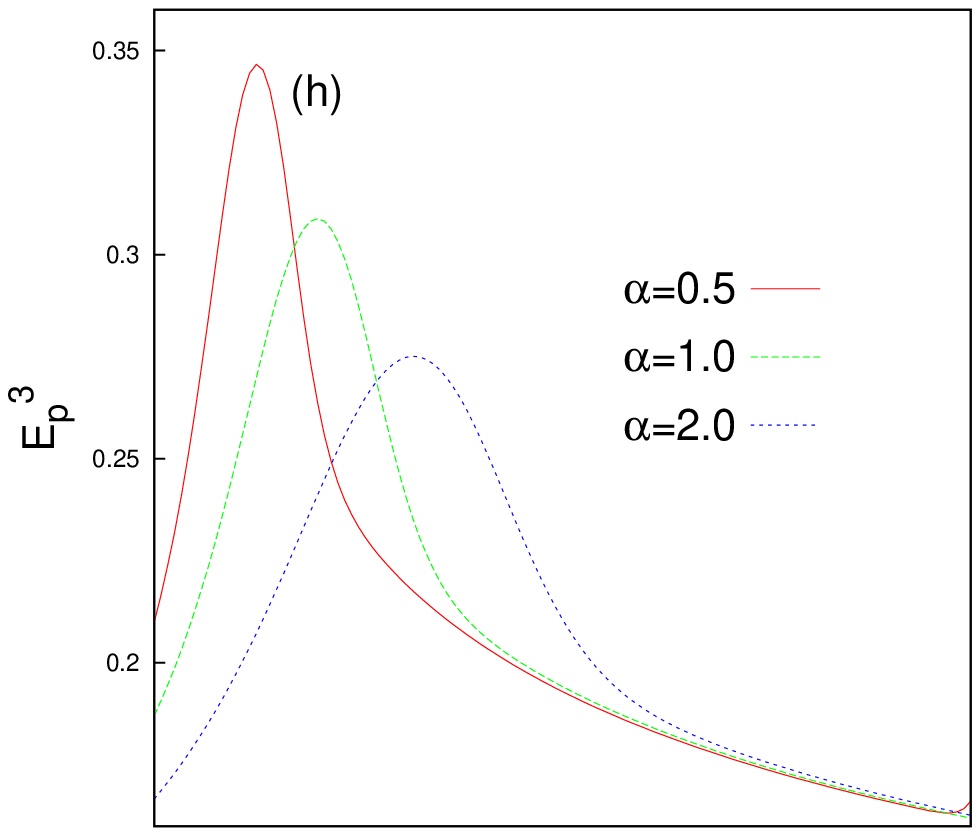}
\end{minipage}
\\[5pt]
\begin{minipage}[c]{0.20\textwidth}\centering
\includegraphics[scale=0.38]{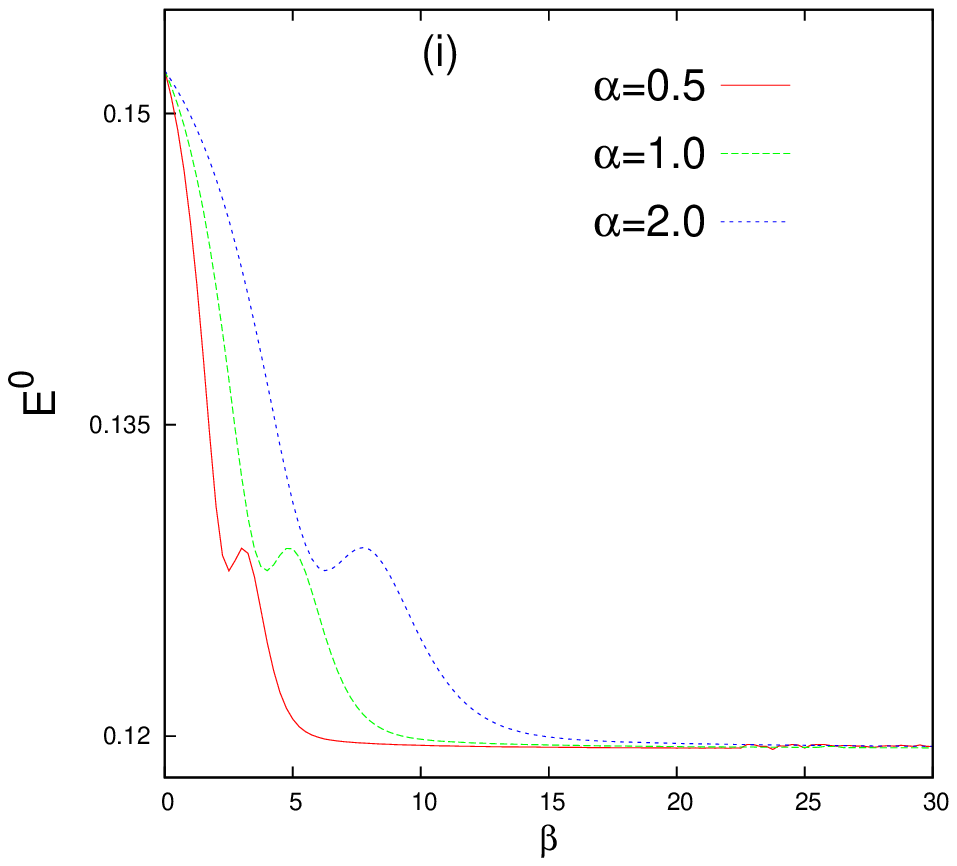}
\end{minipage}\hspace{0.15in}
\begin{minipage}[c]{0.20\textwidth}\centering
\includegraphics[scale=0.38]{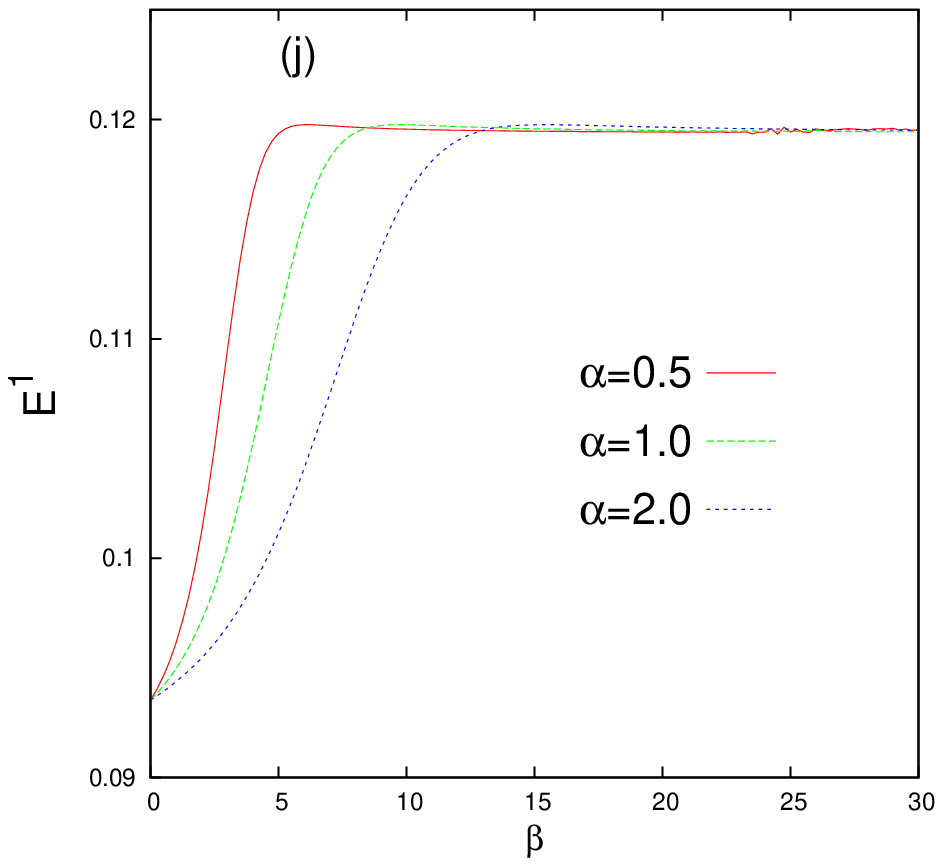}
\end{minipage}\hspace{0.15in}
\begin{minipage}[c]{0.20\textwidth}\centering
\includegraphics[scale=0.38]{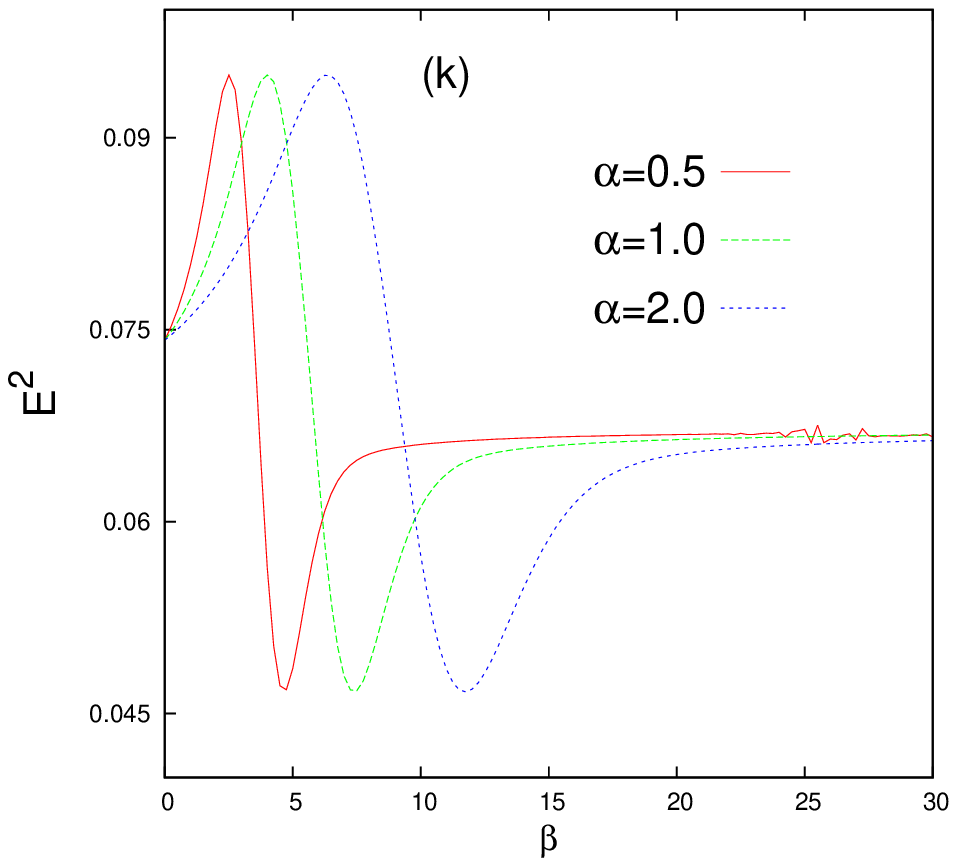}
\end{minipage}\hspace{0.15in}
\begin{minipage}[c]{0.20\textwidth}\centering
\includegraphics[scale=0.38]{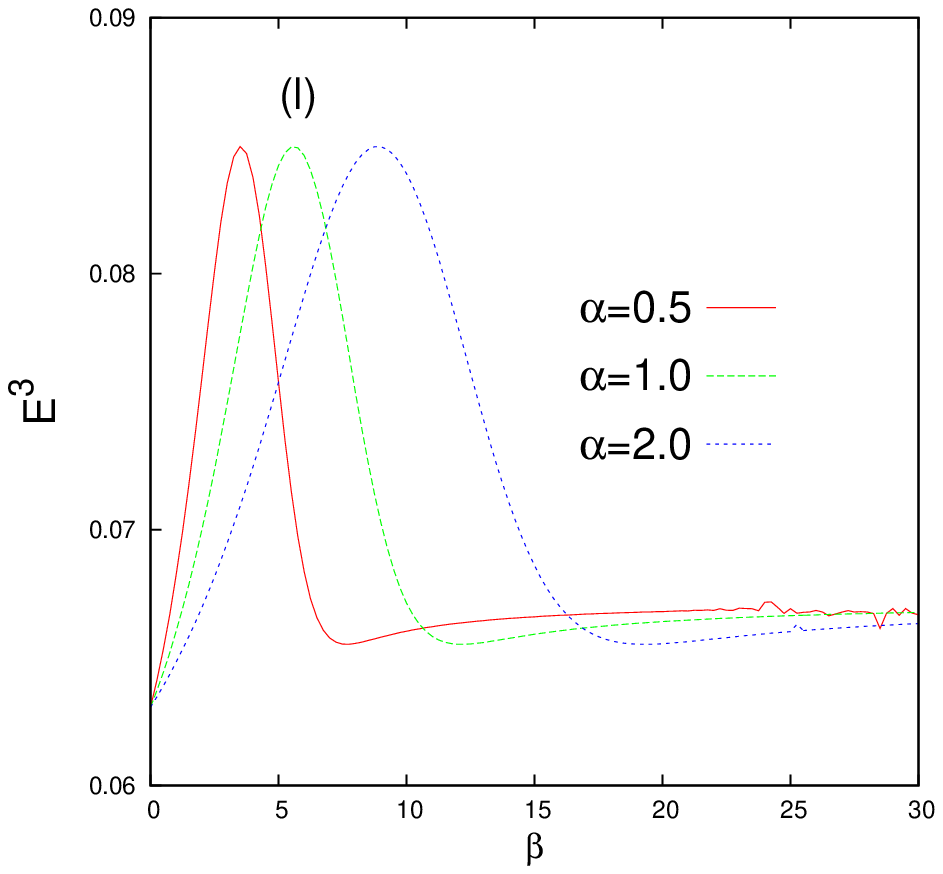}
\end{minipage}
\caption[optional]{Similar plots as in Fig.~(3), for $E_{x}, E_{p}, E$  vs. $\beta$ of DW potential.
See text for details.}
\end{figure}

\subsection{Onicescu energy}  
Now we shift our focus on to Onicescu energy, $E$, a quadratic functional of density, along with its components 
$E_x, E_p$ using Eqs.~(27), (28). At first, variations in $E_x$ with respect to $\beta$ of first two pairs of states
of DW potential are produced in Fig.~(9) at three separate $\alpha$ values, namely, 0.5, 
1, 2. Four columns refer to four lowest states, starting from left to right.
Leaving aside the second excited state in segment (c), other three states in top panels (a), 
(b), (d) maintain qualitative agreement amongst themselves, i.e., $E_{x}$ initially decreases 
from a finite positive value, attains a minimum, then increases monotonically. For a given
state, positions of these minima are shifted to higher values of $\beta$ with progressive 
increase of $\alpha$. Also, the lowest value of $E_{x}$ increases and these minima get flattened
with $\alpha$. The case of second excited state is again different from remaining three states, 
where one sees a shoulder before reaching minimum and then increases continuously. Both
the position of this shoulder and minima get shifted to right with increase in $\alpha$ values. 
Table~III shows respective positions of the extrema for ground and three excited states, at 
three selected $\alpha$ values. It clearly demonstrates the marked contrast between second excited 
state from remaining states; former shows a sequence of extrema while latter three are 
characterized by a minimum. This uniqueness in $n=2$ could be attributed to 
the simultaneous effects of $\beta$ on 
particle--at small values, it indicates the prevailing increase of delocalization of particle 
(reflected in decrease in $E_{x}$); however at large values, the deciding effect seems to be its 
confinement. At minimum, these two effects counter-balance each other and apparently 
cancel. Appearance of a shoulder in $E_{x}^{2}$ is noteworthy--this may again be caused due to the 
nature of wave function of this state itself, which has a node inside each of the 
potential wells along with a maximum at center. Such kink has been observed earlier in 
case of $S_x^2$. Note that, position of shoulder also roughly coincides with onset of 
tunneling. Further, from two left segments (a), (d) of Fig.~(10), it is clear that, $E_x$ for 
lowest two states merge at a value of 
$\beta \approx 5$ for $\alpha=1$, whereas for second, third excited states, convergence
occurs at $\beta \approx 9.5$ for same $\alpha$. This joining of $E_{x}$'s is a clear indication of 
occurrence of quasi-degeneracy in both pairs of states, and also confinement of particle in either 
of the wells.

\begingroup
\squeezetable
\begin{table}
\caption{Positions of the extrema in $E_{x}$ (columns 2--5), $E_p$ (columns 6--9) of the DW 
potential in Eq.~(6), in terms of $\beta$, for four lowest states, at three different 
$\alpha$ values given in column 1. Here ``$ext$" implies ``extrema" and identifying the sequences 
\emph{min-max-min}, \emph{max-min-max} in case of $E_x$ and $E_p$ for second excited states 
respectively. See text for details.}
\centering
\begin{ruledtabular}
\begin{tabular}{c|cccc|cccc}
$\alpha$ &  $\beta_{min}^{0}$  &  $\beta_{min}^{1}$    & $\beta_{ext}^{2}$    & $\beta_{min}^{3}$  
         &  $\beta_{max}^{0}$  &  $\beta_{max}^{1}$    & $\beta_{ext}^{2}$    & $\beta_{max}^{3}$  \\ 
\hline 
0.5      &   2.5  &  2.75   &  2.0,2.5,4.25   &  4.25  &   2.5    &   3.5   &  2.5,5.0,6.75     &  3.5    \\
1.0      &  3.75  &  4.25   &  3.0,4.25,7.0   &  7.0   &   3.75   &  5.75   &  4.0,7.75,10.75   &  5.75    \\
2.0      &  6.0   &  6.75   &  5.0,6.5,11.0   & 11.0   &   6.0    &  9.0    &  6.25,12.5,17.0   &  9.0     \\
\end{tabular}
\end{ruledtabular}
\end{table}
\endgroup

Next, four middle panels (e)-(h) of Fig.~(9) illustrate behavior of $E_{p}$'s of lowest two pairs 
of states of DW. Like $E_x$, in this case also, general trend for ground, first and third 
excited state maintains a qualitative similarity amongst themselves; i.e., $E_{p}$ initially 
increases sharply with increase 
in $\beta$, attains a maximum and gradually decreases afterwords. Once again, variation in $\alpha$ 
shows no qualitative change in these plots. Positions of maxima shift to higher values of 
$\beta$ and get flatter with increase in $\alpha$. Changes of $E_{p}^2$ with $\beta$ is, 
however, substantially different and more interesting; one notices two maxima sandwiched by a 
minimum in $E_{p}^2$. At first, $E_{p}^2$ increases attaining a maximum, then fall down to a 
minimum, again increases to a maximum and finally decreases. Positions of these maxima have 
been tabulated in columns 6-9 of Table~III for three $\alpha$. The table is self-explanatory. 
Clearly, for increasing $\alpha$, 
they get right-shifted and individual plots flatten. Again, this effect may be due to 
quasi-degeneracy in this DW potential. However, effects of tunneling in momentum space is not 
clearly understood; therefore a clear-cut explanation of this phenomenon is not forthcoming. 
Next, two middle panels (b), (e) of Fig.~(10) depict momentum-space IEs for two lowest pair of 
states of DW potential against $\beta$. $E_{p}$ for two lowest pairs converge at nearly 
$\beta=7.5$ and 11.75. Note that while these plots pertain to $\alpha=1$, a similar trend is 
observed for other $\alpha$ as well, with a corresponding shift in location of convergence point.

\begin{figure}             
\centering
\begin{minipage}[c]{0.30\textwidth}\centering
\includegraphics[scale=0.48]{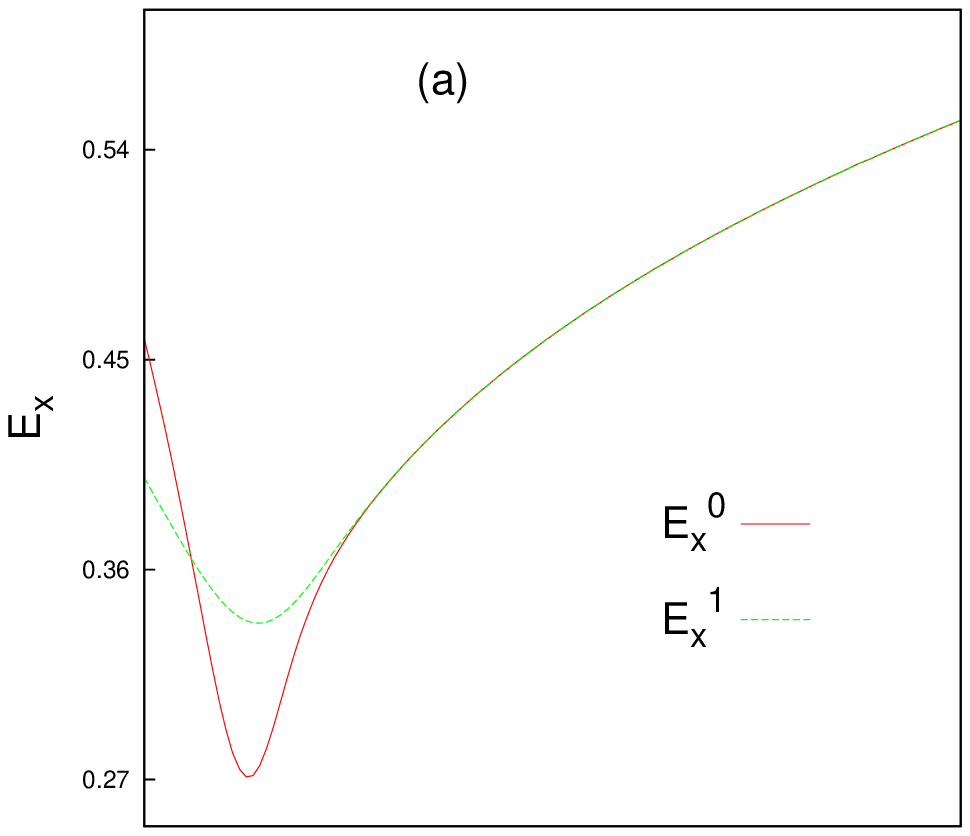}
\end{minipage}\hspace{0.05in}
\begin{minipage}[c]{0.30\textwidth}\centering
\includegraphics[scale=0.48]{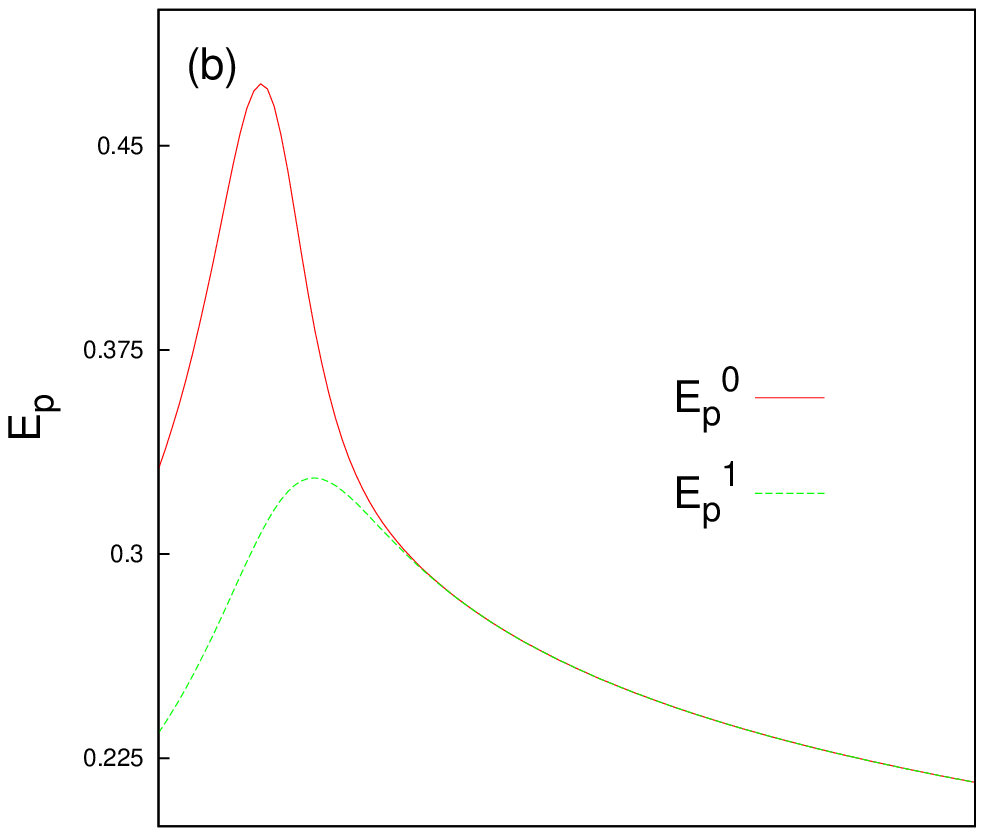}
\end{minipage}\hspace{0.05in}
\begin{minipage}[c]{0.30\textwidth}\centering
\includegraphics[scale=0.48]{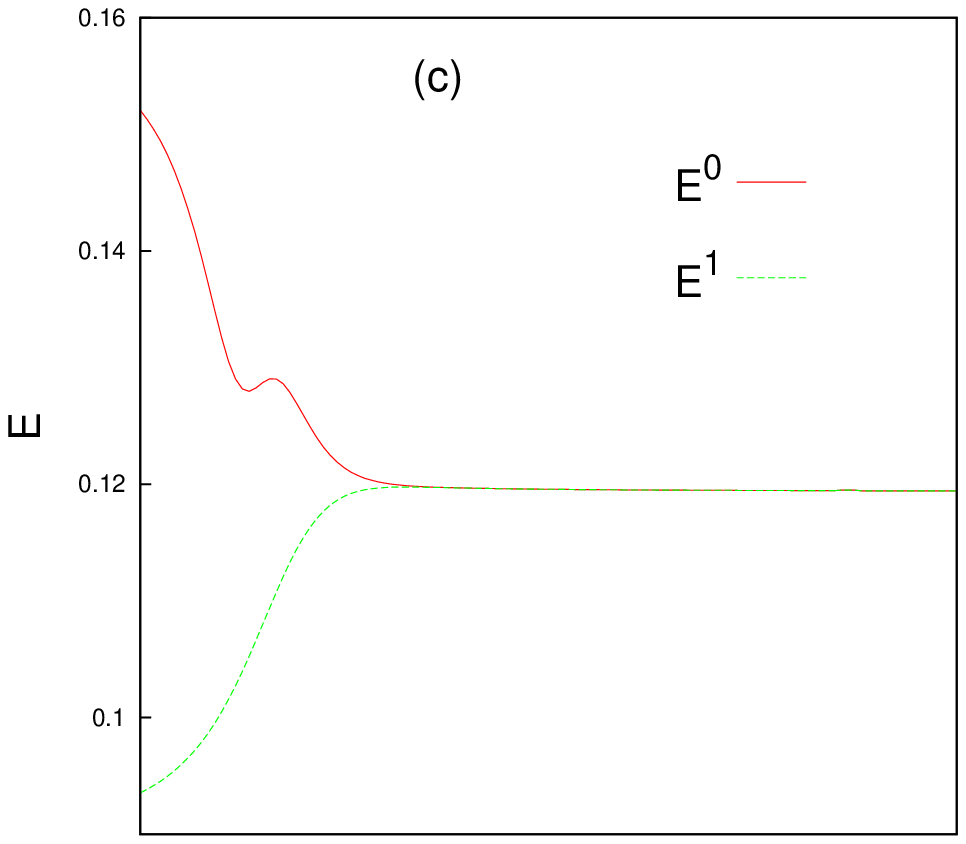}
\end{minipage}
\\[5pt]
\begin{minipage}[c]{0.30\textwidth}\centering
\includegraphics[scale=0.48]{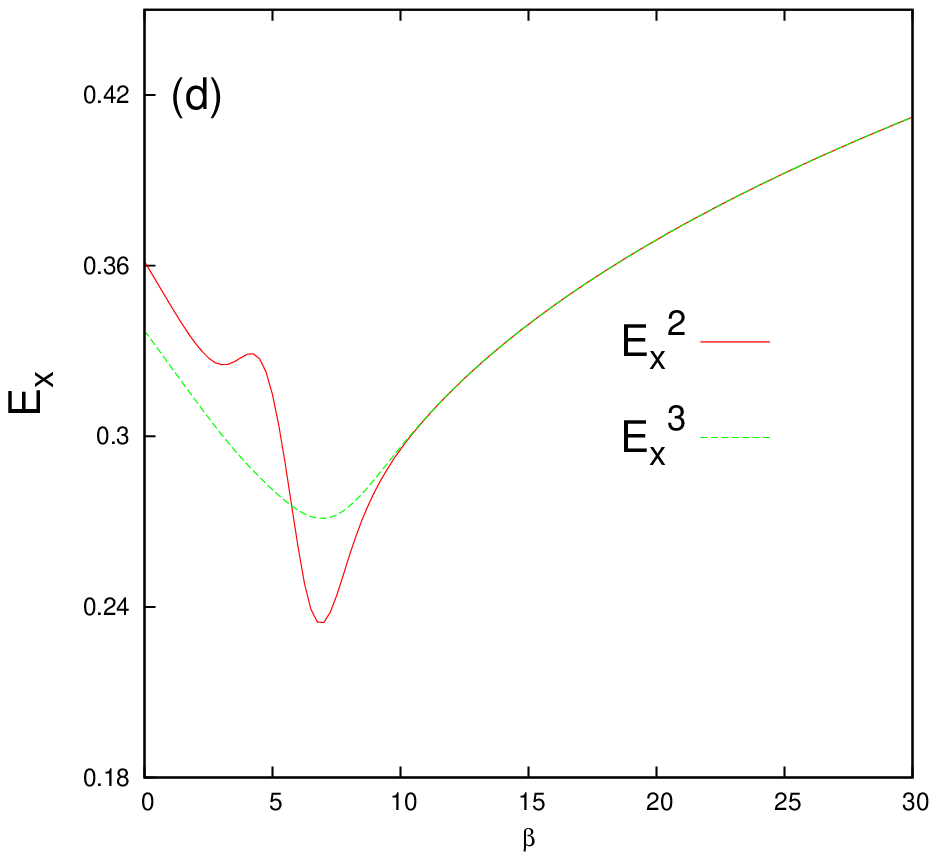}
\end{minipage}\hspace{0.05in}
\begin{minipage}[c]{0.30\textwidth}\centering
\includegraphics[scale=0.48]{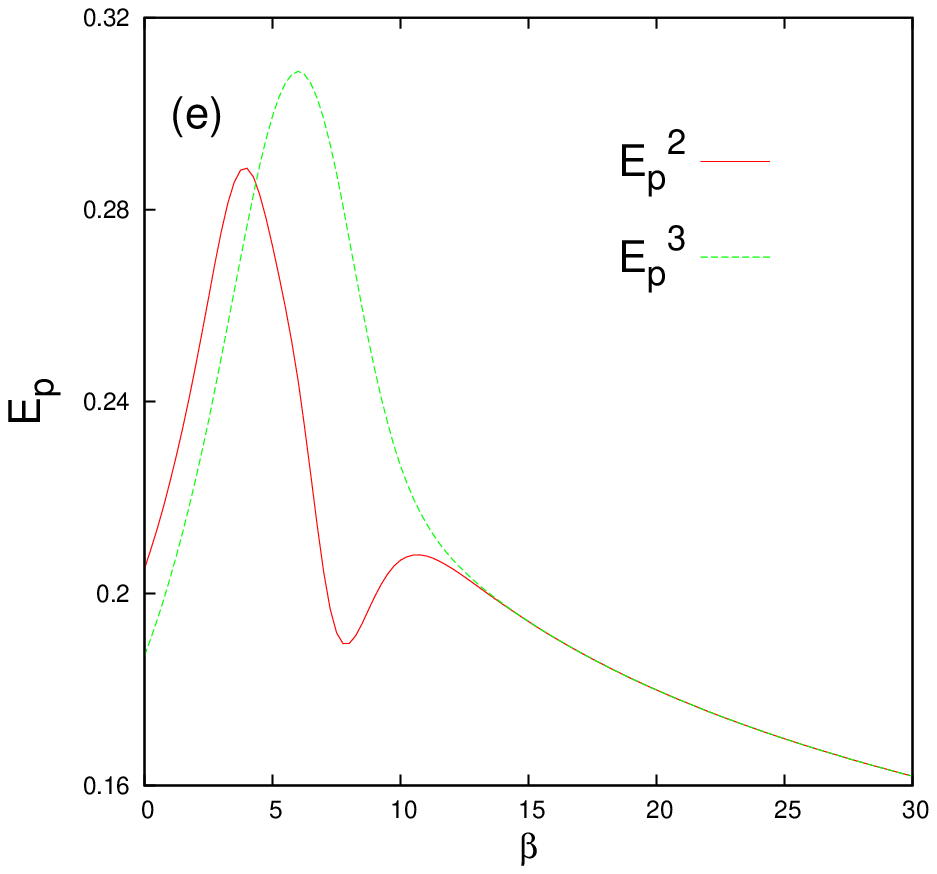}
\end{minipage}\hspace{0.05in}
\begin{minipage}[c]{0.30\textwidth}\centering
\includegraphics[scale=0.48]{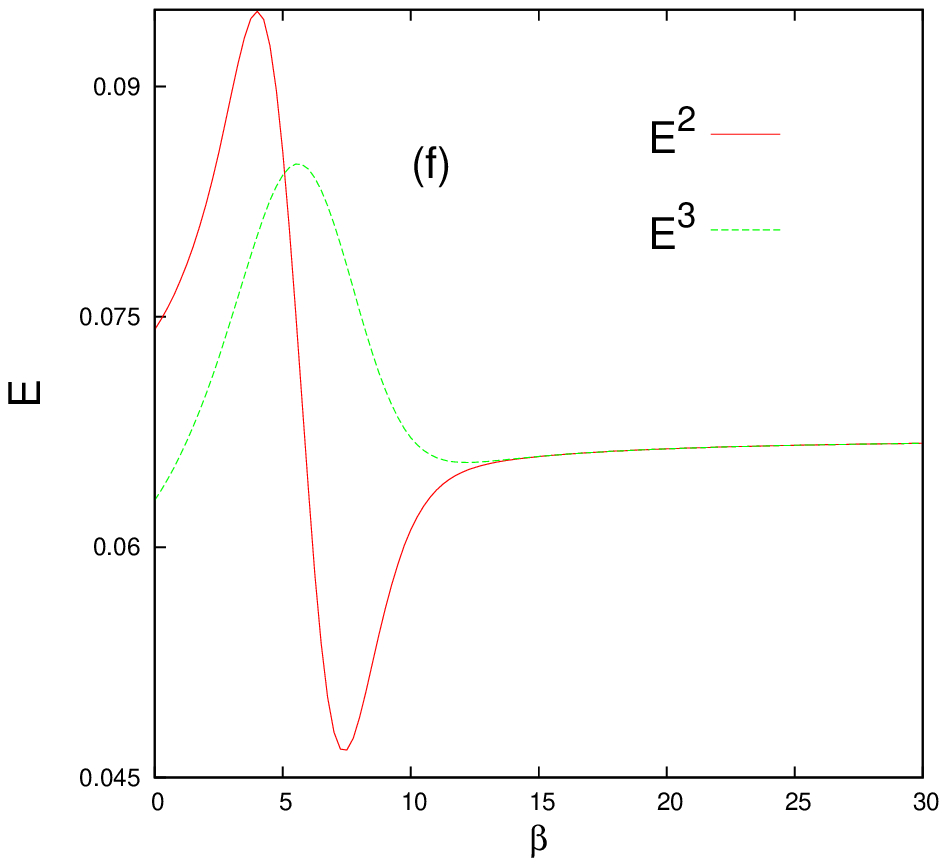}
\end{minipage}
\caption[optional]{Plots of $E_{x}, E_{p}, E$  vs. $\beta$ at $\alpha=1$, of DW potential, as in Fig.~(4).
See text for details.}
\end{figure}

To study the effect of barrier on a given state more fully, we need to look into $E$, which is provided 
now in four bottom panels (i)-(l) of Fig.~(9). Appearance of a barrier at the center of well generally has 
greater effect on even states of panels (i), (k) than odd states of panels (j), (l), possibly 
because latter states possess a node at the maximum of potential. Panel (i) reveals that
$E^{0}$ quite sharply decreases initially to attain a minimum, then almost 
immediately hits a maximum, again starts decreasing to reach a constant value of 0.1195, for all
$\alpha$. Positions of both minima and maxima shift towards right for higher $\alpha$.
Moreover, distance (in values of $\beta$) between these two extrema increases 
with increase of $\alpha$. In contrast, in panel (j), $E^{1}$ increases monotonically, until 
individual $\alpha$ plots converge together at a value of 0.1195. This situation changes for second, 
third excited states, however. Panel (k) indicates that $E^{2}$ initially increases 
to reach a maximum, then follows through a minimum and again increases until attains a constant 
value of 0.0667. The extrema, in this case, are much more clearly defined and well separated than 
its counterpart in ground state. Values of $E$ at extrema are quite comparable for all 
$\alpha$ values studied. Increase of $\alpha$ shifts the extrema towards right but all of them 
eventually converge to same constant value of 0.0667. For third excited state in panel (l), 
$E^{3}$ increases to reach a maximum, then gradually decreases monotonically towards a constant 
value of $0.0667$. Now, top right panel (c) of Fig.~(10) visualizes convergence of 
$E^0$, $E^1$; bottom right panel (f) does 
same for $E^2$, $E^3$, keeping $\alpha$ fixed at 1 in both occasions. $E$ becomes 
stationary at a value of $0.1195$ on and after $\beta=5$ for first pair, while same for 
latter pair occurs at $\beta=20$ with $E=0.0667$. A similar exercise for other $\alpha$ values 
(0.5, 2) for same states, reveals that, these $E$ values remain unchanged 
by variation in $\alpha$. Thus, for the DW potential under consideration, it seems 
that total $E$, like total $S$, possessed by any quasi-degenerate pair of states would also
be a \emph{constant, unique to that pair, independent of $\alpha$} (see discussion on Shannon
entropy earlier). 

\begin{figure}                         
\begin{minipage}[c]{0.40\textwidth}
\centering
\includegraphics[scale=0.45]{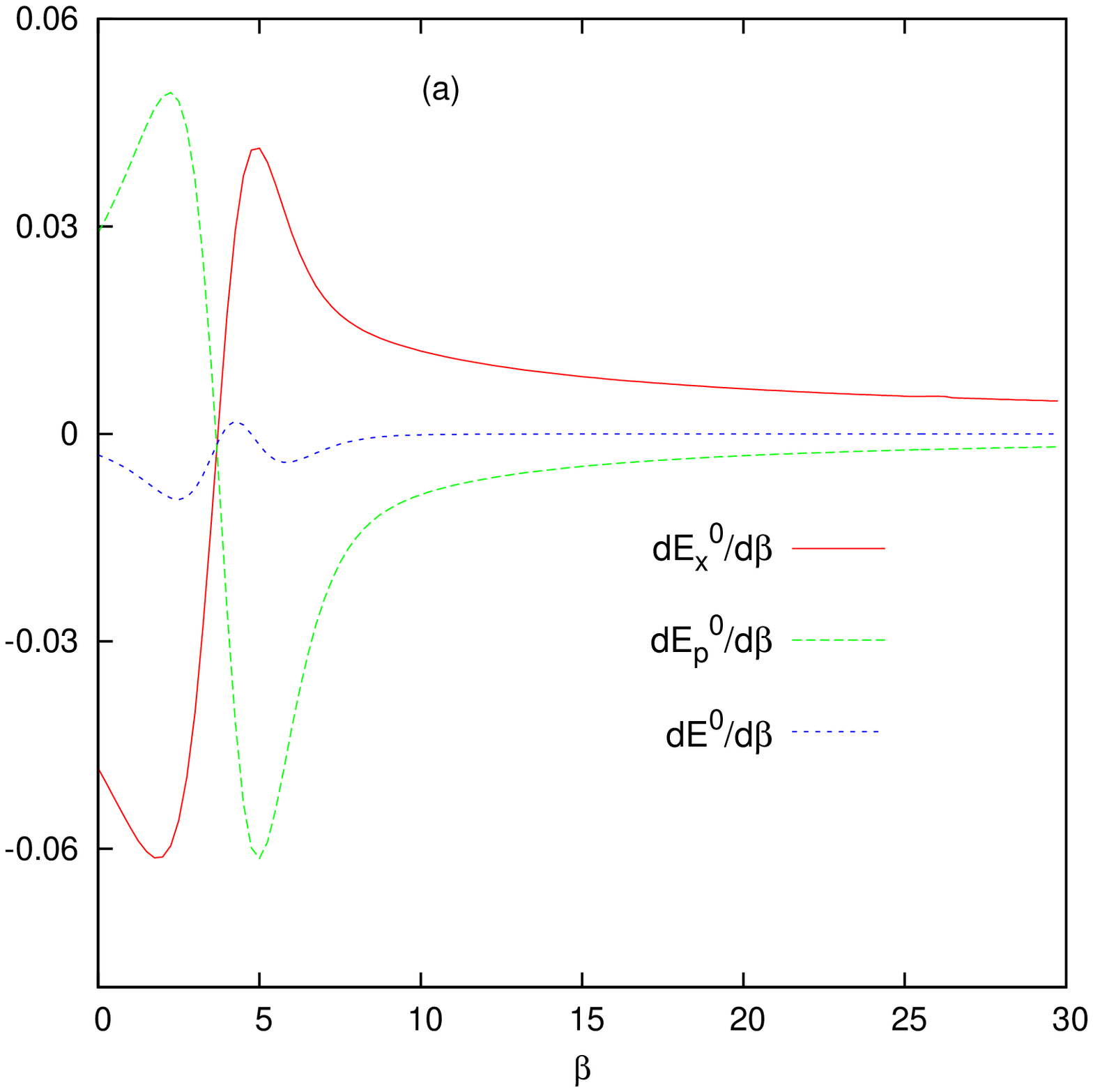}
\end{minipage}%
\hspace{0.9in}
\begin{minipage}[c]{0.40\textwidth}
\centering
\includegraphics[scale=0.45]{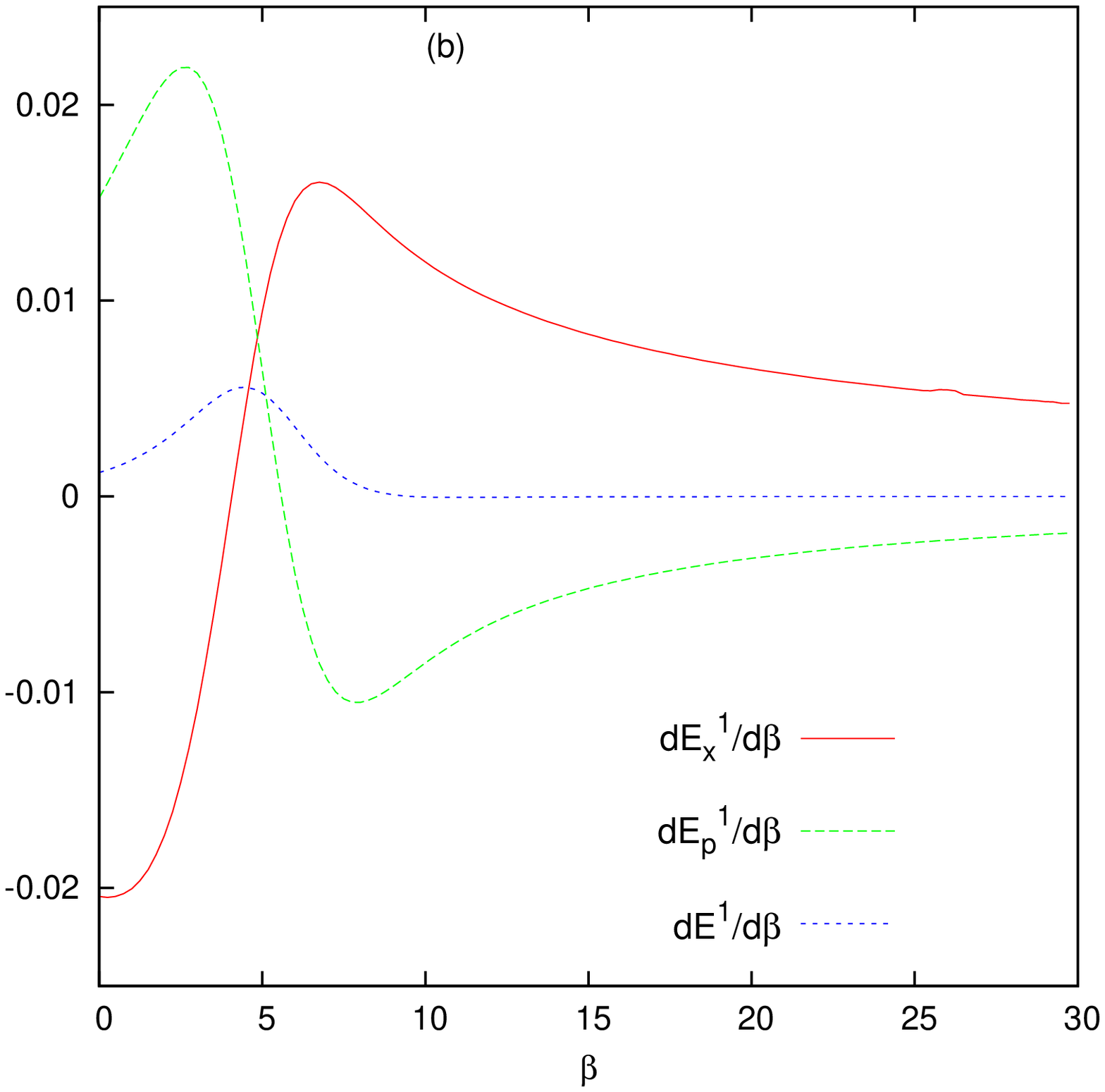}
\end{minipage}%
\caption{$E$ derivatives (first) against $\beta$, for $\alpha=1$, for potential in Eq.~(6). Left panel 
(a) corresponds to $\frac{dE_{x}^{0}}{d\beta}, \frac{dE_{p}^{0}}{d\beta}, \frac{dE^{0}}{d\beta}$, 
in ground state; (b) refers to $\frac{dE_{x}^{1}}{d\beta}, \frac{dE_{p}^{1}}{d\beta}, 
\frac{dE^{1}}{d\beta}$, in first excited state.}
\end{figure}

To throw more light on to the effect of $\beta$ on $E$'s, changes of latter are pursued in terms 
of respective first derivatives. Thus, Fig.~(11) displays 1st derivatives $E_x$, $E_p$, $E$ of 
DW potential for ground (a) and first excited (b) state respectively. One notices that, 
for ground state, $\frac{dE_{x}^{0}}{d\beta}$ initially falls down to reach a negative 
minimum, then rises to attain a positive maximum, and finally proceeds slowly asymptotically 
towards zero. The $E$ derivative in momentum space for same, $\frac{dE_{p}^{0}}{d\beta}$, 
on the other hand, shows a trend exactly complimentary to that in position space--initially it 
rises to a maximum then falls down to a negative minimum and gradually approaches towards zero. 
In position and momentum space, the $E$ derivatives in first excited state, \emph{viz.,}
$\frac{dE_{x}^{1}}{d\beta}$, $\frac{dE_{p}^{1}}{d\beta}$ follow similar trend as their 
corresponding ground-state counterparts; however the points of inflection are now shifted. Total 
$E$ derivative for ground state, $\frac{dE^{0}}{d\beta}$, decreases with increase in $\beta$
until it attains a negative minimum and then goes through a sequence of positive maximum and 
negative minimum. Finally it decays towards zero. However, $\frac{dE^{1}}{d\beta}$ offers only
a maximum before leveling off to zero.

\begin{figure}             
\centering
\begin{minipage}[c]{0.20\textwidth}\centering
\includegraphics[scale=0.38]{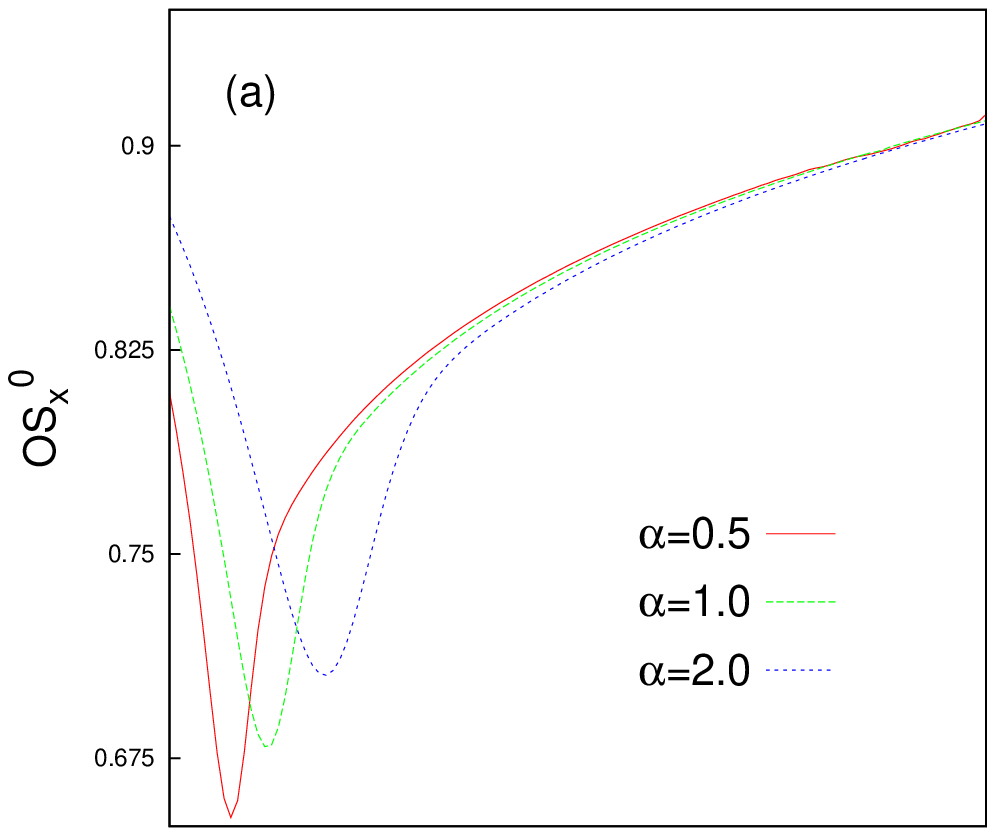}
\end{minipage}\hspace{0.15in}
\begin{minipage}[c]{0.20\textwidth}\centering
\includegraphics[scale=0.38]{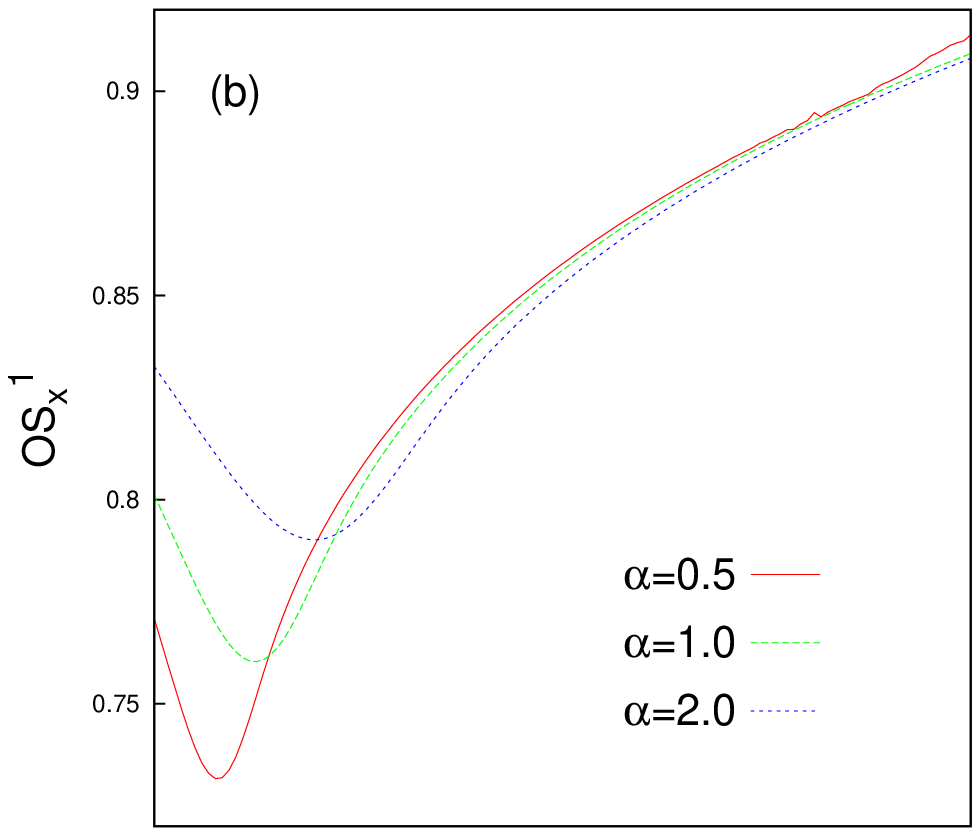}
\end{minipage}\hspace{0.15in}
\begin{minipage}[c]{0.20\textwidth}\centering
\includegraphics[scale=0.38]{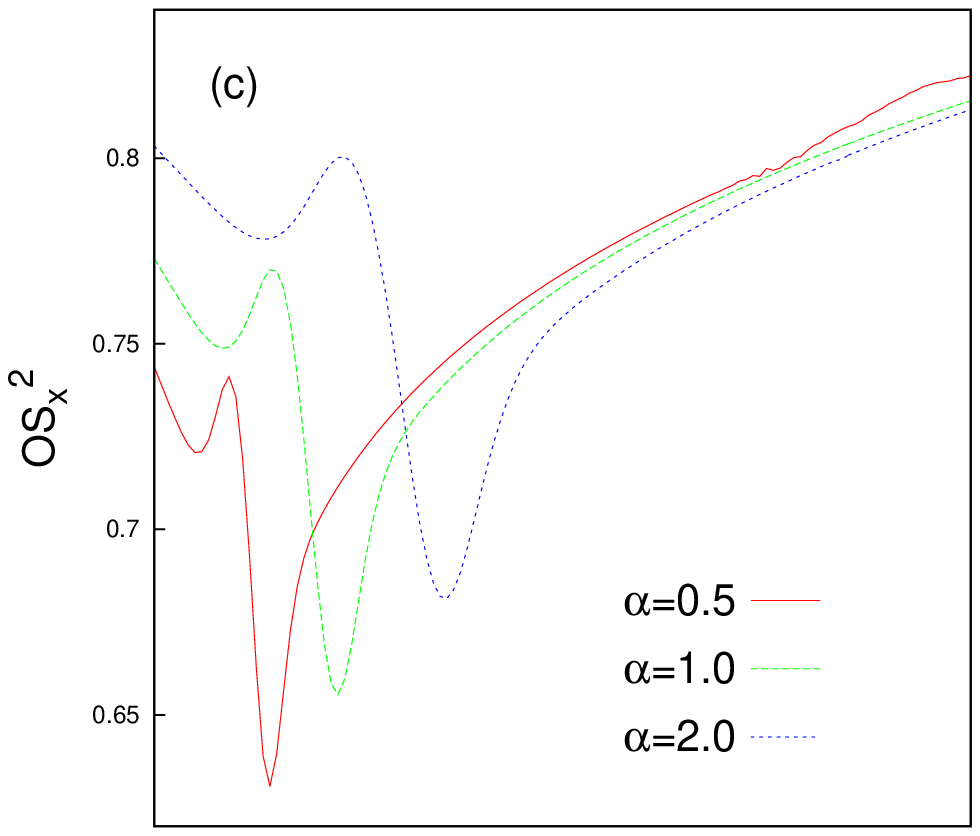}
\end{minipage}\hspace{0.15in}
\begin{minipage}[c]{0.20\textwidth}\centering
\includegraphics[scale=0.38]{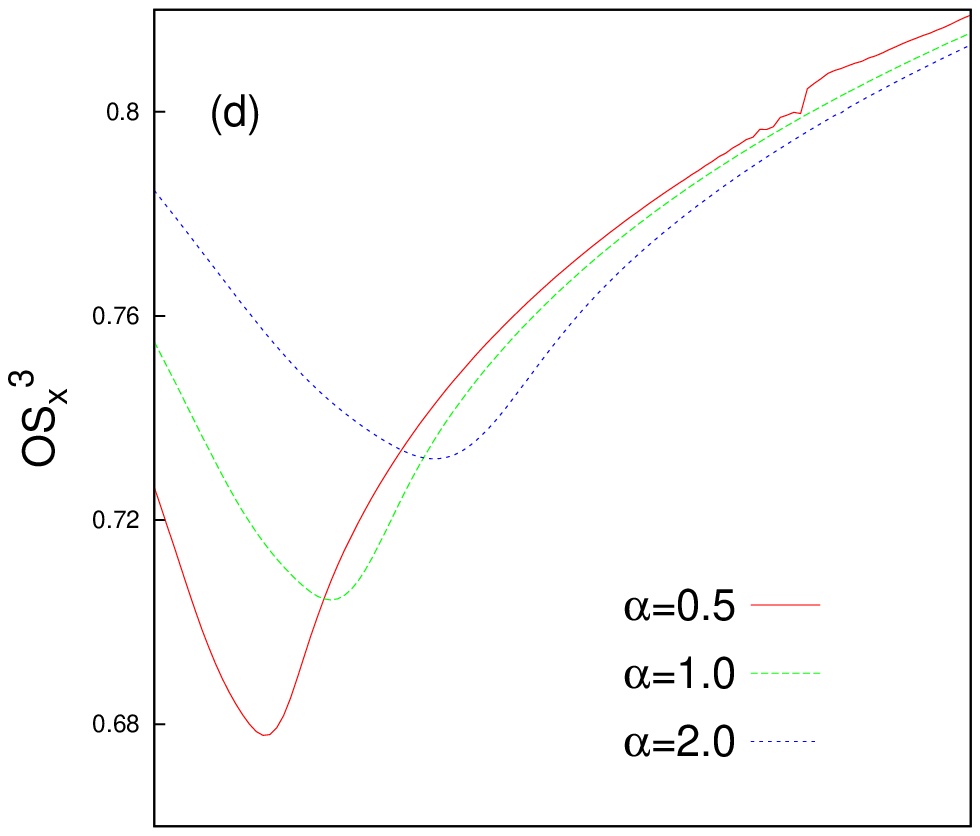}
\end{minipage}
\\[5pt]
\begin{minipage}[c]{0.20\textwidth}\centering
\includegraphics[scale=0.38]{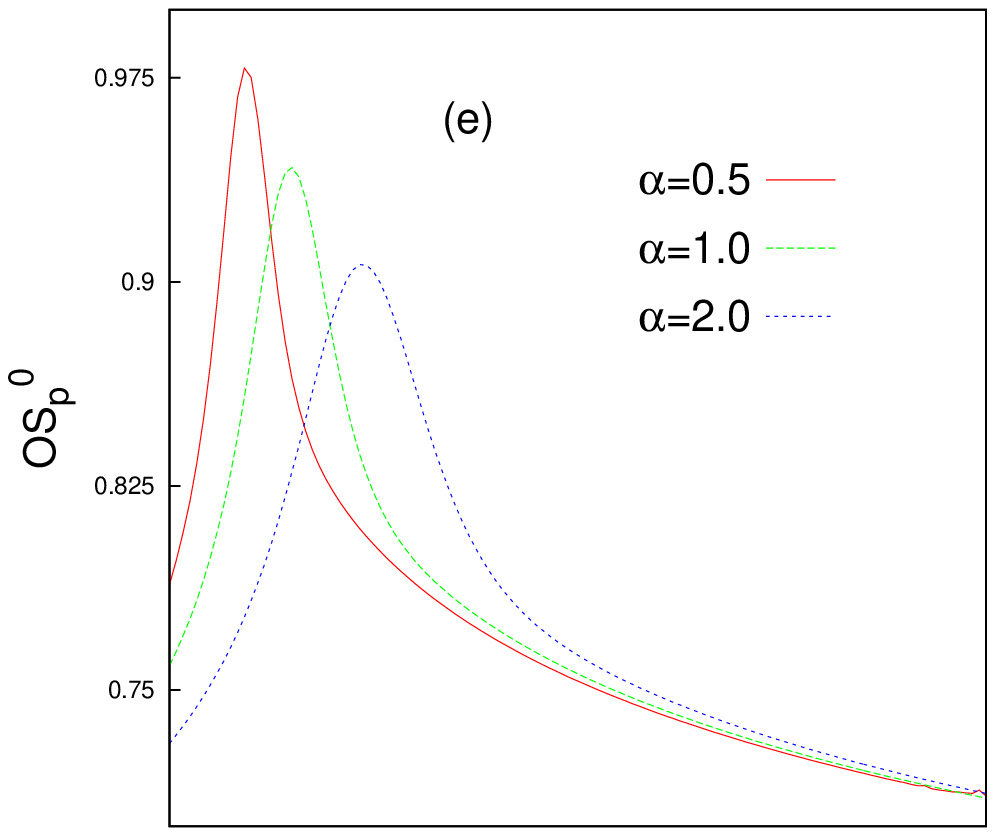}
\end{minipage}\hspace{0.15in}
\begin{minipage}[c]{0.20\textwidth}\centering
\includegraphics[scale=0.38]{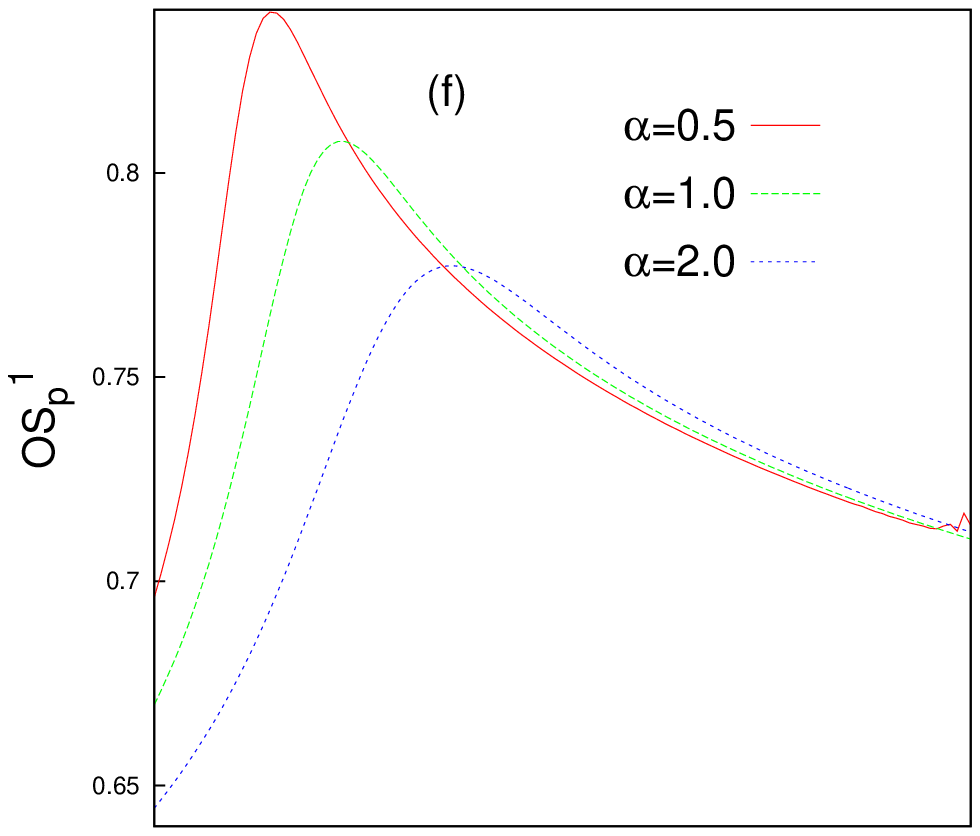}
\end{minipage}\hspace{0.15in}
\begin{minipage}[c]{0.20\textwidth}\centering
\includegraphics[scale=0.38]{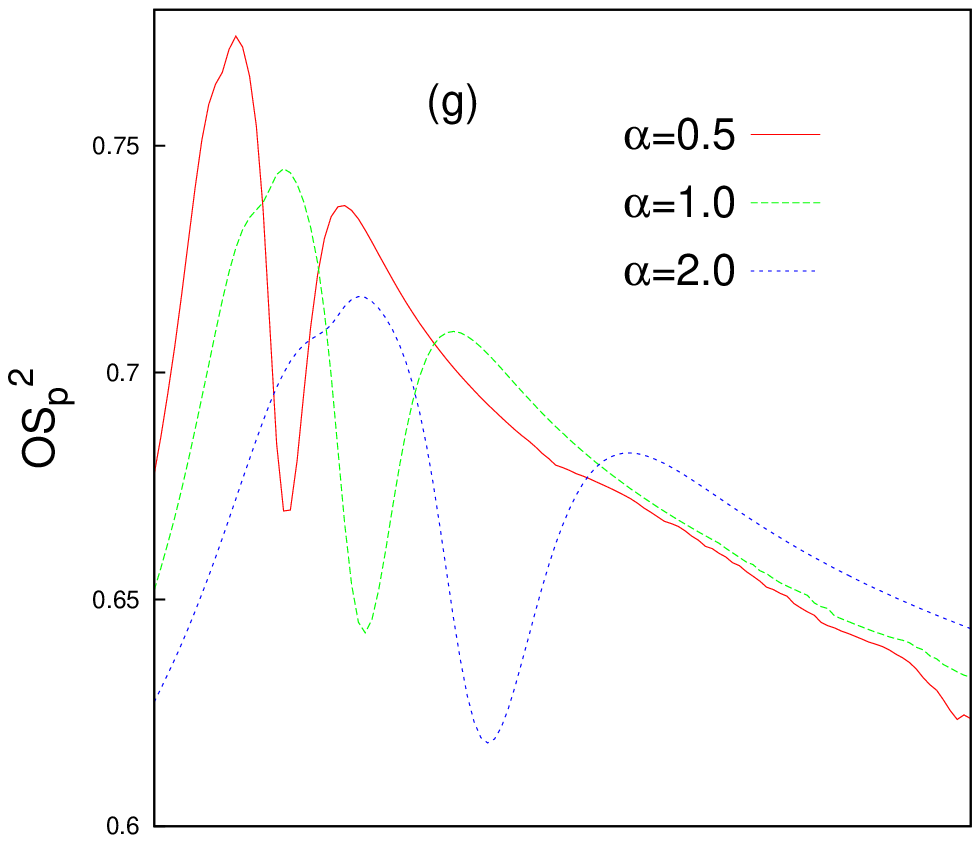}
\end{minipage}\hspace{0.15in}
\begin{minipage}[c]{0.20\textwidth}\centering
\includegraphics[scale=0.38]{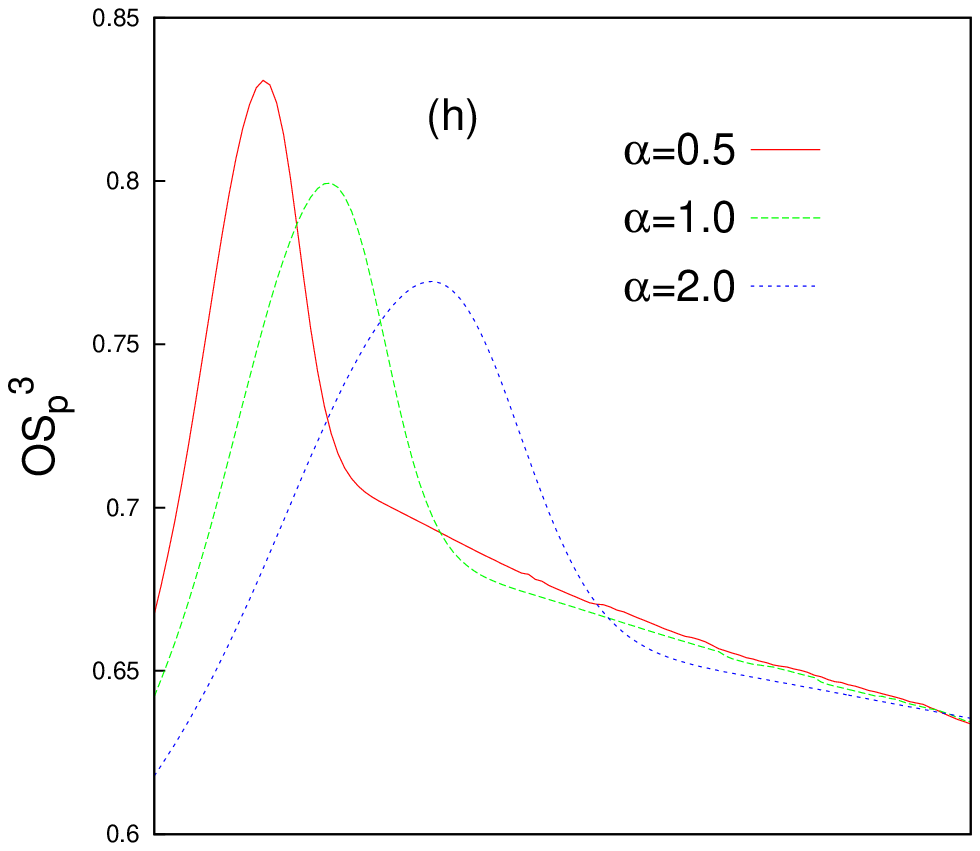}
\end{minipage}
\\[5pt]
\begin{minipage}[c]{0.20\textwidth}\centering
\includegraphics[scale=0.38]{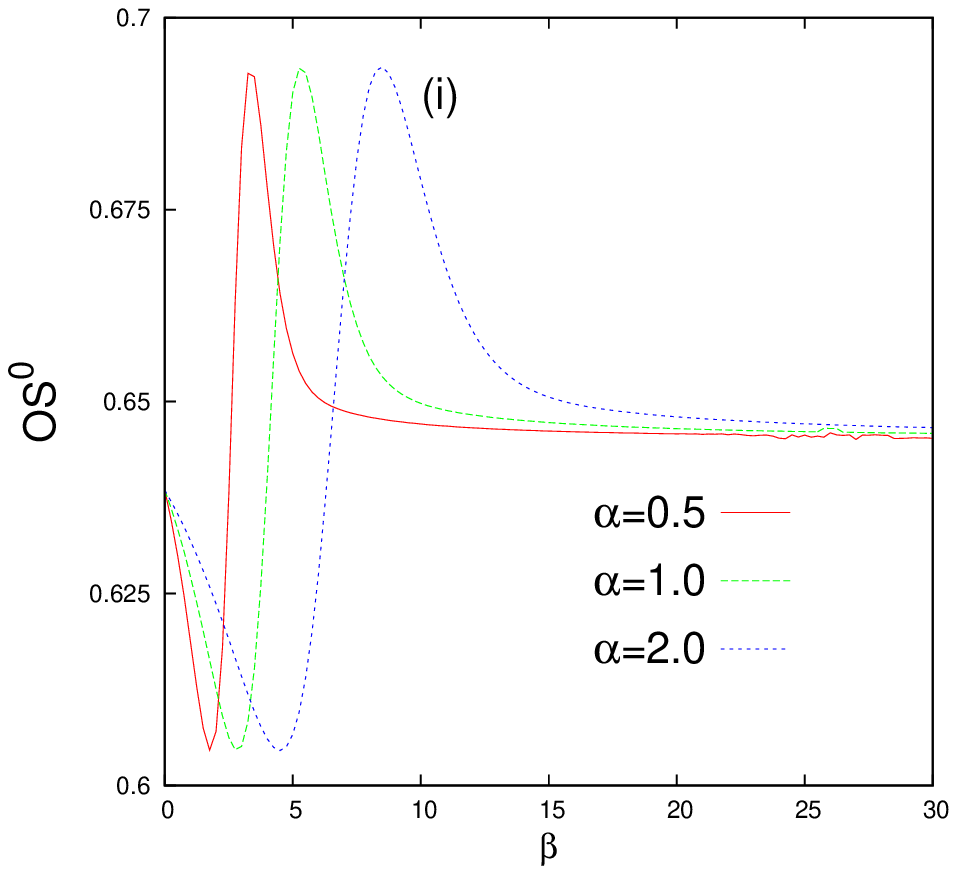}
\end{minipage}\hspace{0.15in}
\begin{minipage}[c]{0.20\textwidth}\centering
\includegraphics[scale=0.38]{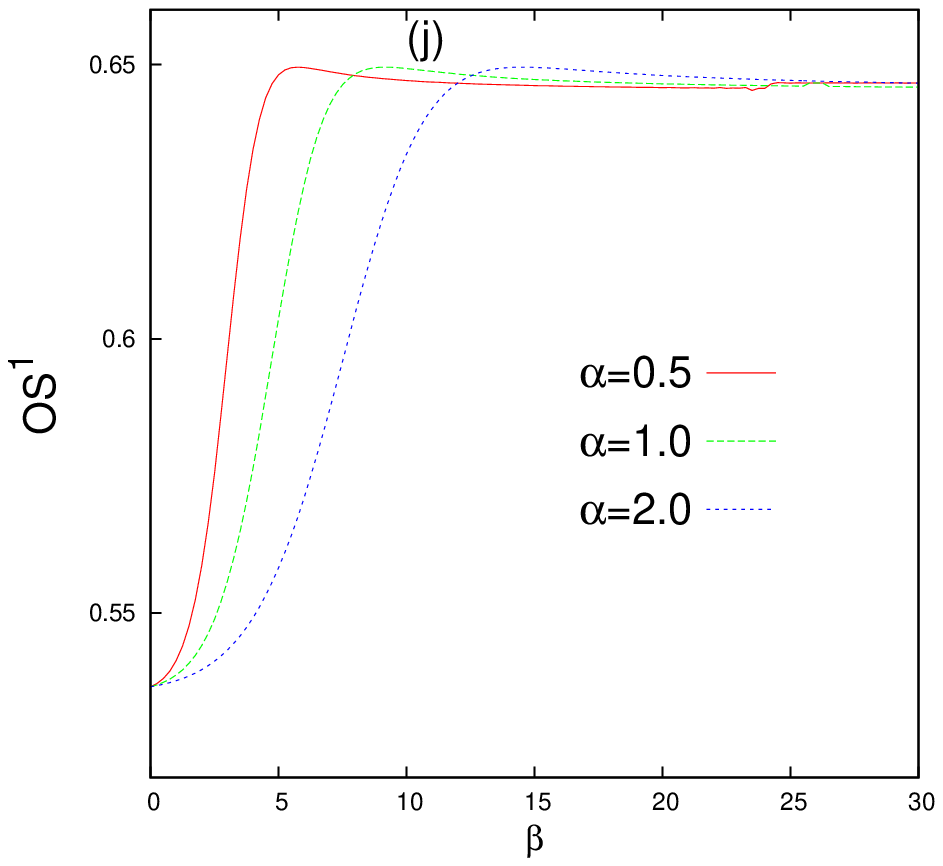}
\end{minipage}\hspace{0.15in}
\begin{minipage}[c]{0.20\textwidth}\centering
\includegraphics[scale=0.38]{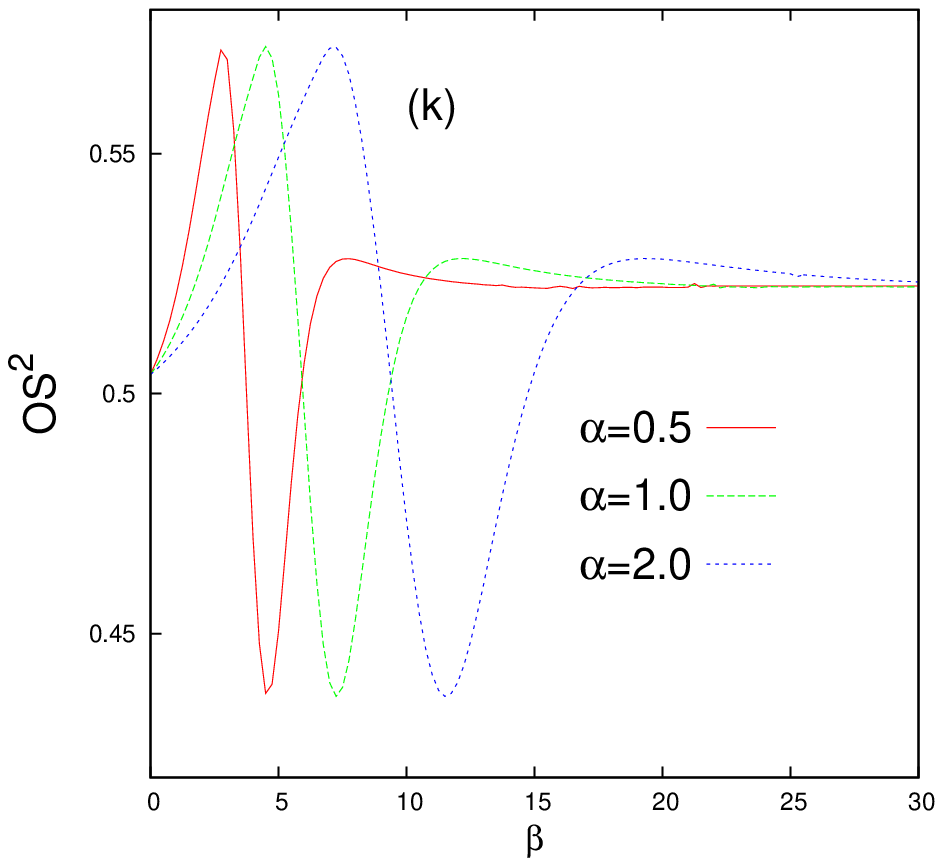}
\end{minipage}\hspace{0.15in}
\begin{minipage}[c]{0.20\textwidth}\centering
\includegraphics[scale=0.38]{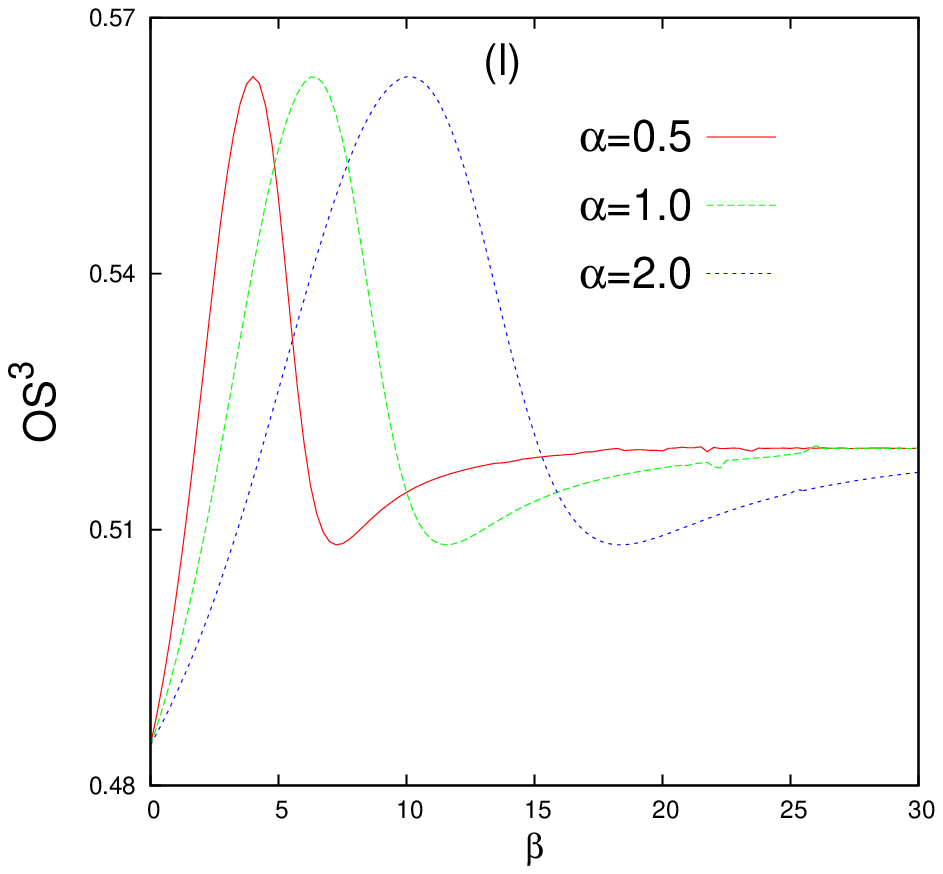}
\end{minipage}
\caption[optional]{Similar plots as in Fig.~(3), for $OS_x, OS_p, OS$  vs. $\beta$ of DW potential.
See text for details.}
\end{figure}

\subsection{Onicescu-Shannon information measures} 
We now present our results on $OS$ for four lowest states of a DW potential using Eqs.~(32), (33).
First, the top four segments of Fig.~(12) record variation of $OS_{x}$ with $\beta$ at three different 
$\alpha$, namely, 0.5, 1.0, 2.0. Except for second excited state (c), starting from a finite
non-zero value, these decrease with $\beta$, proceeds through a minimum and sharply rises 
thereafter. Positions of these minima for all these three states in panels (a), (b), (d) shift 
towards right as $\alpha$ is increased. This is again possibly due to competing effects in a DW. Like 
$S_{x}^{2}$, $E_{x}^{2}$, visibly $OS_{x}^{2}$ also behaves quite differently from remaining three 
states. Thus, initially there appears a shallow minimum followed by a small maximum 
and finally a minimum. With $\alpha$, however, positions of these extrema shift towards right 
as in other three states. Further, segments (a), (d) of Fig.~(13) establish that the pairs $OS_{x}^{0}$, 
$OS_{x}^{1}$ and $OS_{x}^{2}$, $OS_{x}^{3}$ converge at a certain $\beta$ value, indicating 
appearance of quasi-degeneracy in a DW. Next, four panels in middle row of Figure~(12) exhibit 
changes in $OS_{p}$ 
with $\beta$ at three different $\alpha$ values. In this instance also, qualitative 
behavior of $OS_{p}$ for ground (e), first (f) and third (h) excited state are similar, while
second excited state (g) stands out. Plots of former family are characterized by a prominent
maximum followed by a gradual decrease, while the lone second state offers a maximum and minimum in 
succession. Positions of the maxima shift to higher $\beta$ as $\alpha$ increases. As usual, 
the pairs $OS_{p}^{0}$, $OS_{p}^{1}$ and $OS_{p}^{2}$, $OS_{p}^{3}$ smoothly join at a 
particular $\beta$ value in middle column in (b), (e) of Fig.~(13). 

\begin{figure}             
\centering
\begin{minipage}[c]{0.30\textwidth}\centering
\includegraphics[scale=0.48]{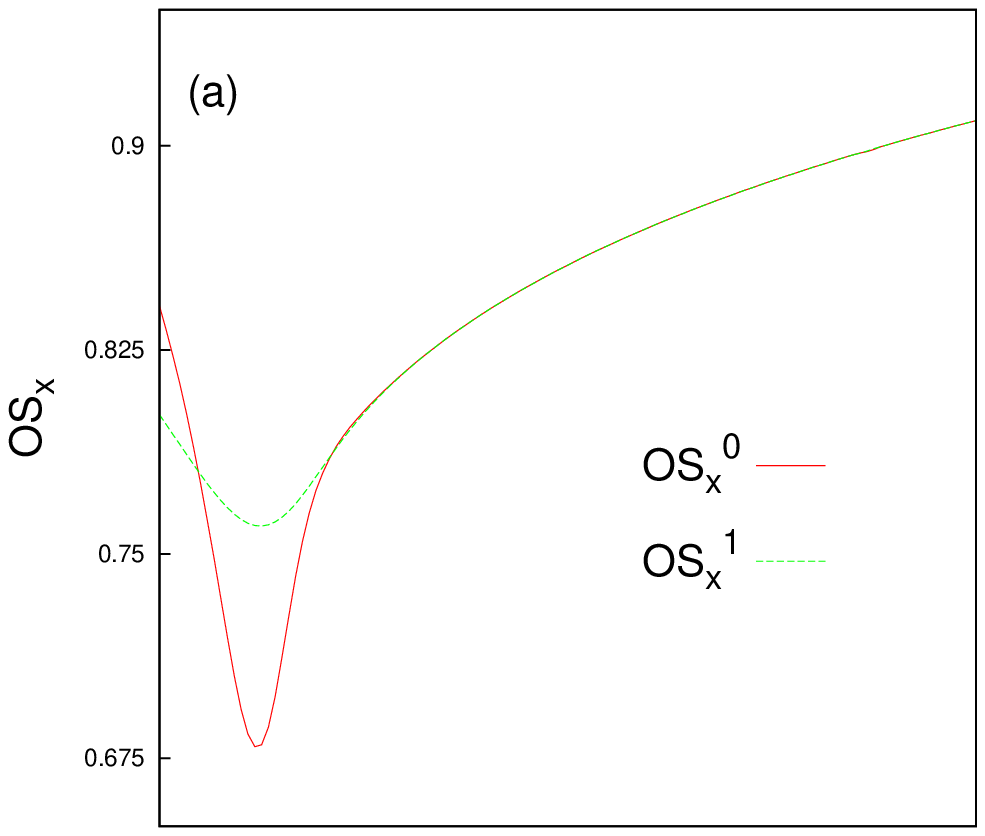}
\end{minipage}\hspace{0.05in}
\begin{minipage}[c]{0.30\textwidth}\centering
\includegraphics[scale=0.48]{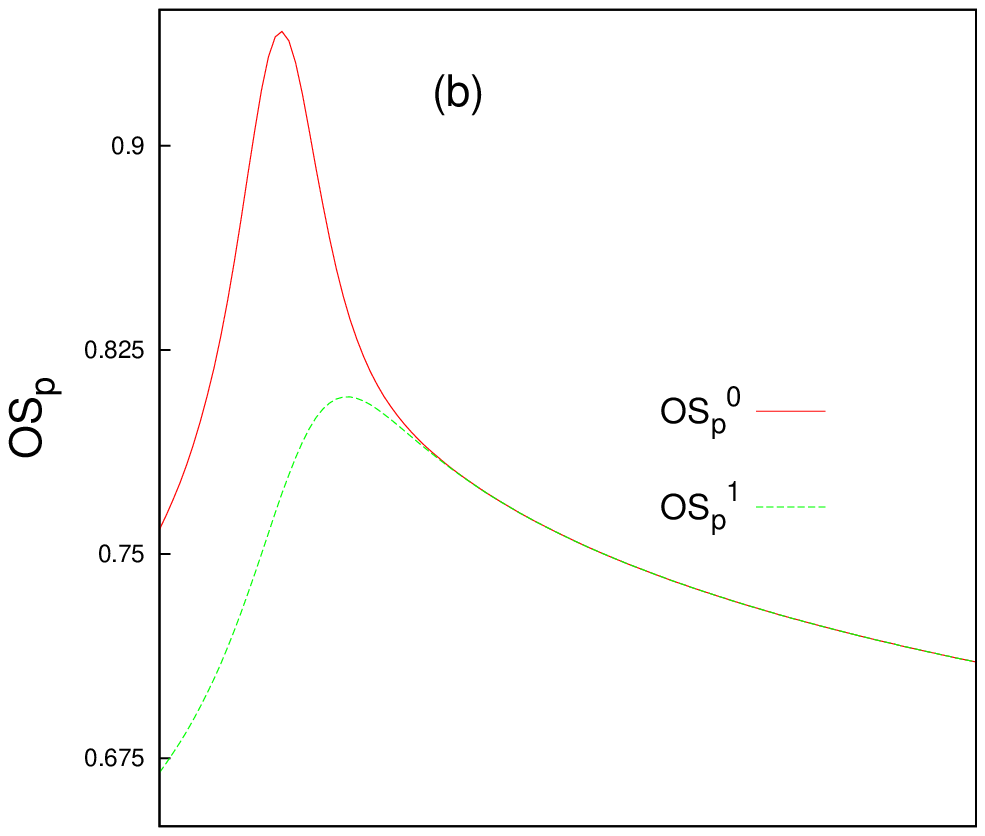}
\end{minipage}\hspace{0.05in}
\begin{minipage}[c]{0.30\textwidth}\centering
\includegraphics[scale=0.48]{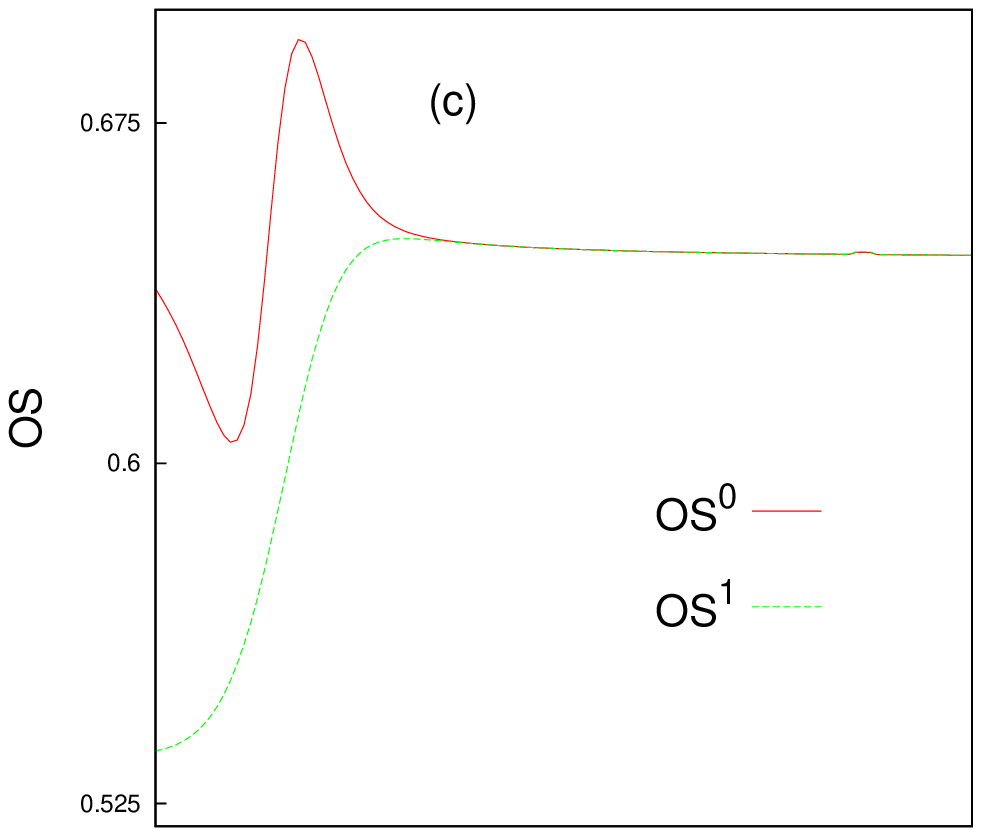}
\end{minipage}
\\[5pt]
\begin{minipage}[c]{0.30\textwidth}\centering
\includegraphics[scale=0.48]{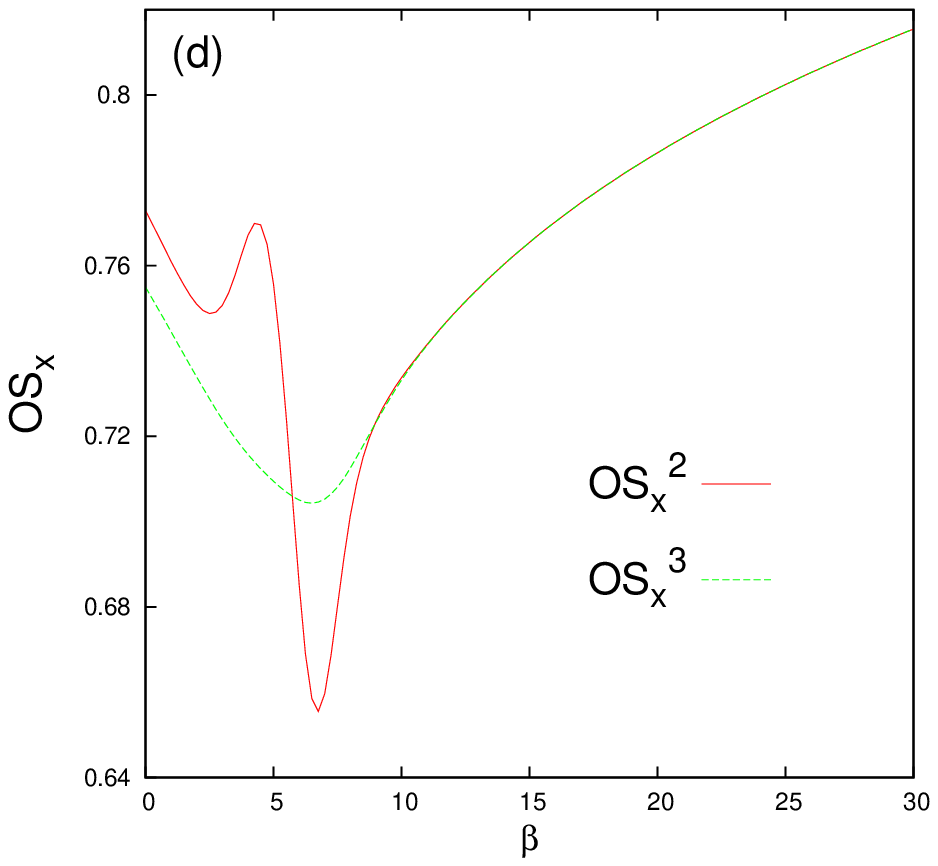}
\end{minipage}\hspace{0.05in}
\begin{minipage}[c]{0.30\textwidth}\centering
\includegraphics[scale=0.48]{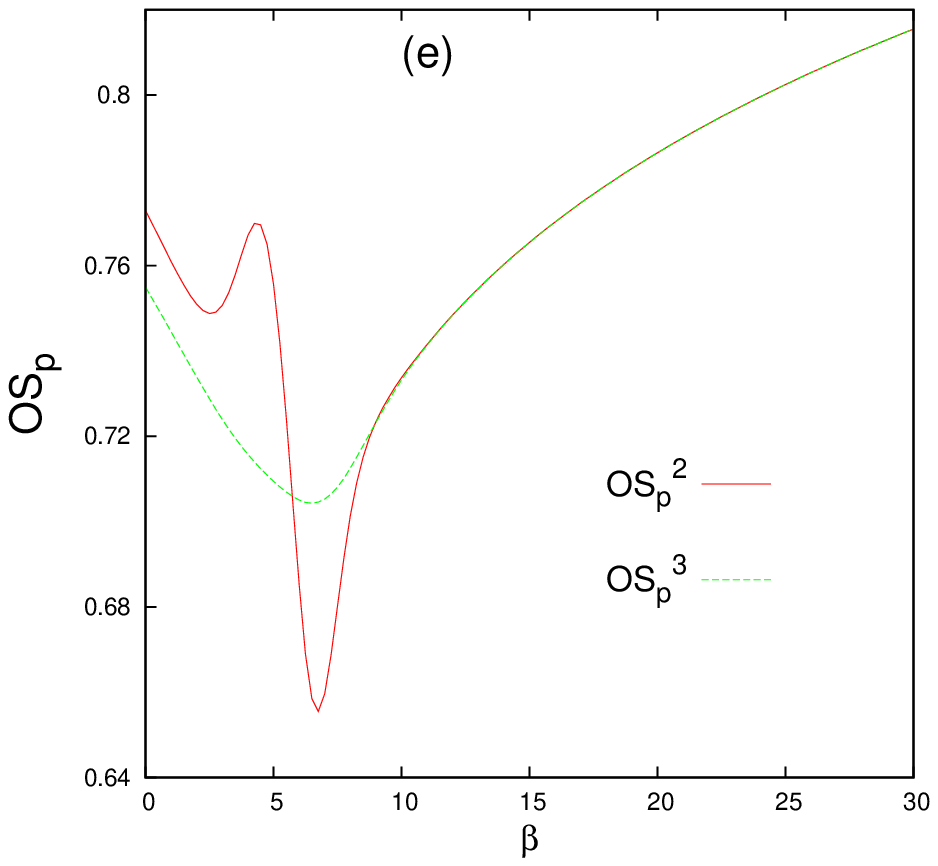}
\end{minipage}\hspace{0.05in}
\begin{minipage}[c]{0.30\textwidth}\centering
\includegraphics[scale=0.48]{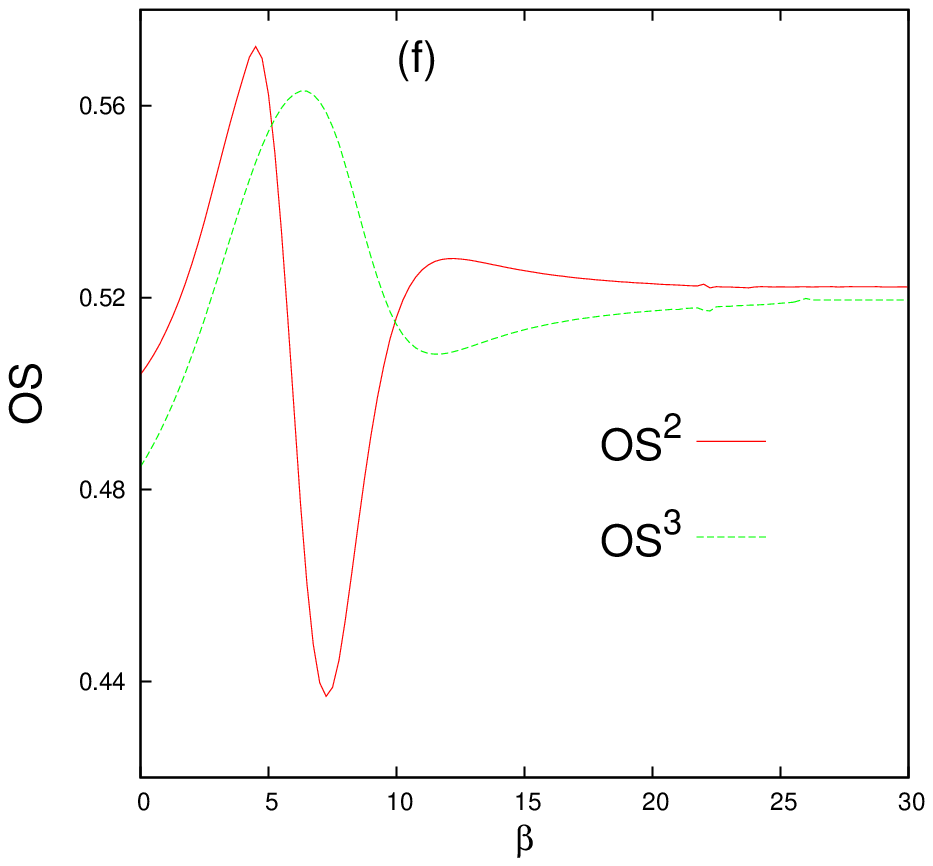}
\end{minipage}
\caption[optional]{$OS_x, OS_p, OS$  vs. $\beta$ at $\alpha=1$, of DW potential, as in Fig.~(4).
See text for details.}
\end{figure}

Figure~(12) also portrays variation of total $OS$ corresponding to first four energy states of a 
particle in a DW potential in bottom four panels, at same $\alpha$ as in previous figures. 
$OS^{0}$ in panel (i), at first, falls down to a minimum slowly, then rises to a maximum very sharply,
eventually becoming constant. These extrema move towards higher values of $\beta$ with 
increasing $\alpha$; however, these plots finally converge to same constant value of 0.6465. 
$OS^{1}$ in segment (j) greatly increases with $\beta$ initially, then converges to a stationary
value of 0.6465. Rate of increase of $OS^{1}$ with $\beta$ decreases with increase of 
$\alpha$. $OS^{2}$ in panel (k) initially rises to a maximum, then follows through a minimum and again 
converges to a constant value of 0.525. All these extrema shift toward right with increase of 
$\alpha$. $OS^{3}$ panel in (l) increases to a maximum and converges to same value (0.525) 
as in (k). In this case also, an increase of $\alpha$ leads to same result of shifting 
maxima towards right. Additionally, in two rightmost panels (c), (f) of Fig.~(13), we offer 
convergence of $OS$ for first and second 
pairs respectively. Note that while range of horizontal axis is same in all plots, 
vertical axis is \emph{not}. Also, for third excited state, convergence point of $OS$ extends
to a higher value of $\beta$ compared to the remaining states, and thus is not clearly visible 
from the figure.  

\subsection{Phase-space area}
Now we proceed for a semi-classical phase-space analysis of the particle in ground and 
first excited state of a DW potential represented by Eq.~(6). It is observed from (a), (b) of 
Fig.~(14) that, phase spaces of both these states split into two closed lobes at certain finite
but different $\beta$ values. For ground state, this happens at $\beta=2.25$. Interestingly, 
there is a clearly defined maximum for $\frac{dS_{x}^{0}}{d\beta}$ and minimum for 
$\frac{dS_{p}^{0}}{d\beta}$ at same value of $\beta$, which marks beginning of tunneling. 
This implies that, onset of tunneling is marked by an inflection point (in this case, a maximum)
in the rate of change of information with $\beta$. Worthnotingly, segment (a) of Fig.~(15) (discussed in 
following paragraph) also shows a kink at same value of $\beta$. It is hard to draw a similar 
inference for first excited state though, presumably due to presence of a node in wave function 
right at the center of barrier. Also note that, although two $\beta$ values 
corresponding to onset of tunneling and splitting of phase space into two distinct closed lobes 
match, there is no inflection in $\frac{dS_{x}^{1}}{d\beta}$. So the process of 
tunneling, at least for first excited state, still eludes an explanation in the context of 
IE in position space.

\begin{figure}    
\centering
\begin{minipage}[t]{0.43\textwidth}\centering
\includegraphics[scale=0.6]{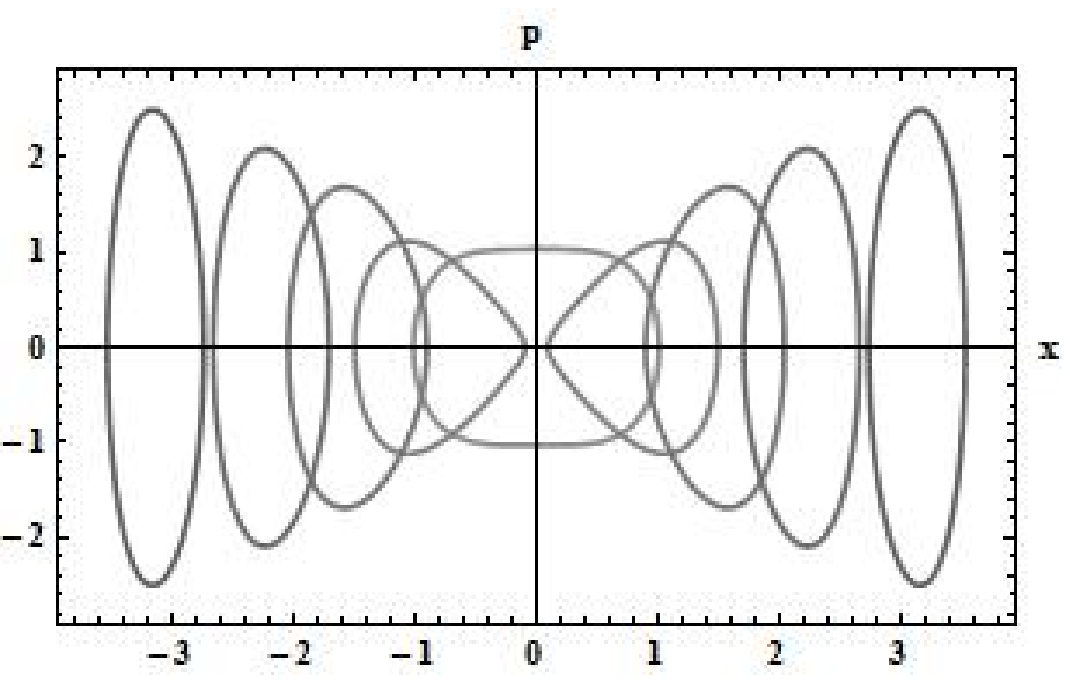}
\hspace*{10mm}{(a)}
\end{minipage}
\hspace{0.2in}
\begin{minipage}[t]{0.48\textwidth}\centering
\includegraphics[scale=0.6]{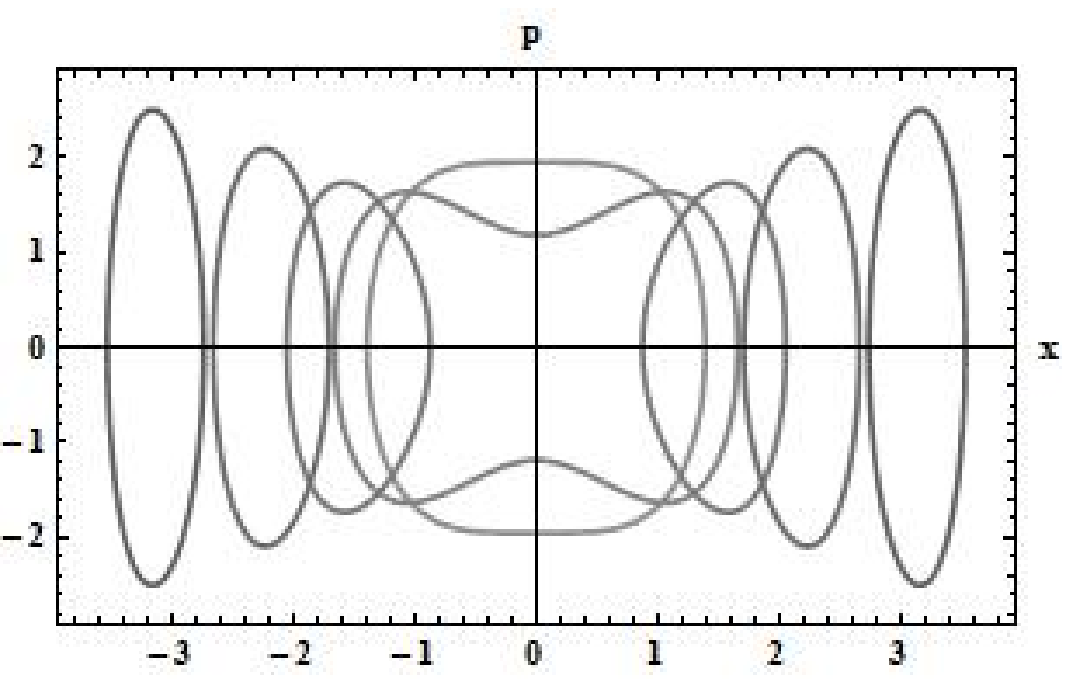}
\hspace*{10mm}{(b)}
\end{minipage}
\caption{Contour plot of phase space at five different $\beta \ (0, 2.25, 5.0, 10.0, 20.0)$ 
values, at a fixed $\alpha=1.0$ of DW potential in Eq.~(6): (a) ground state (b) first 
excited state.}
\end{figure}

Finally, a straightforward integration over phase space using Eq.~(5) leads to semi-classical 
phase-space area ($A$). These are given for first four states in segments (a), (b), (d), (e) of Fig.~(15)
respectively. $A_0$, $A_1$ qualitatively follow same pattern as their total $S$ counterparts, where 
the overall behavior remains unaffected with changes in $\alpha$; only they are shifted to right, as $\alpha$ 
progresses. For $A_0$, $A_2$, one notices a kink indicative of tunneling. Finally, the two rightmost panels (c), (f) 
compare the variation of phase-space areas for four lowest states with respect to $\beta$, for a 
constant $\alpha=1$. It is interesting to note that the kinks in $A_0$ and $A_{2}$ occur exactly 
when tunneling begins; it continues to increase thereafter with $\beta$ until no further change 
occurs. A similar correlation can also be found for first and third excited state, where the area
decreases with increasing $\beta$ and finally becoming constant; however this decrease in area 
is much faster than their corresponding decays of total $S$ in Fig.~5 (panels (i)-(l)). This again 
suggests that the relationship between semi-classical phase-space area and total IE might be particularly 
striking for ground state, but not so conspicuous for first excited state. Like total IE, 
phase-space area of ground and first excited states coalesce at a constant value of 1.565 at 
around $\beta=5.0$, while same for second and third excited state occurs at 4.705 
corresponding to $\beta=10$. This again reinforces inter-connection between phase-space area 
and total IE.

\begin{figure}         
\centering
\begin{minipage}[c]{0.28\textwidth}\centering
\includegraphics[scale=0.30]{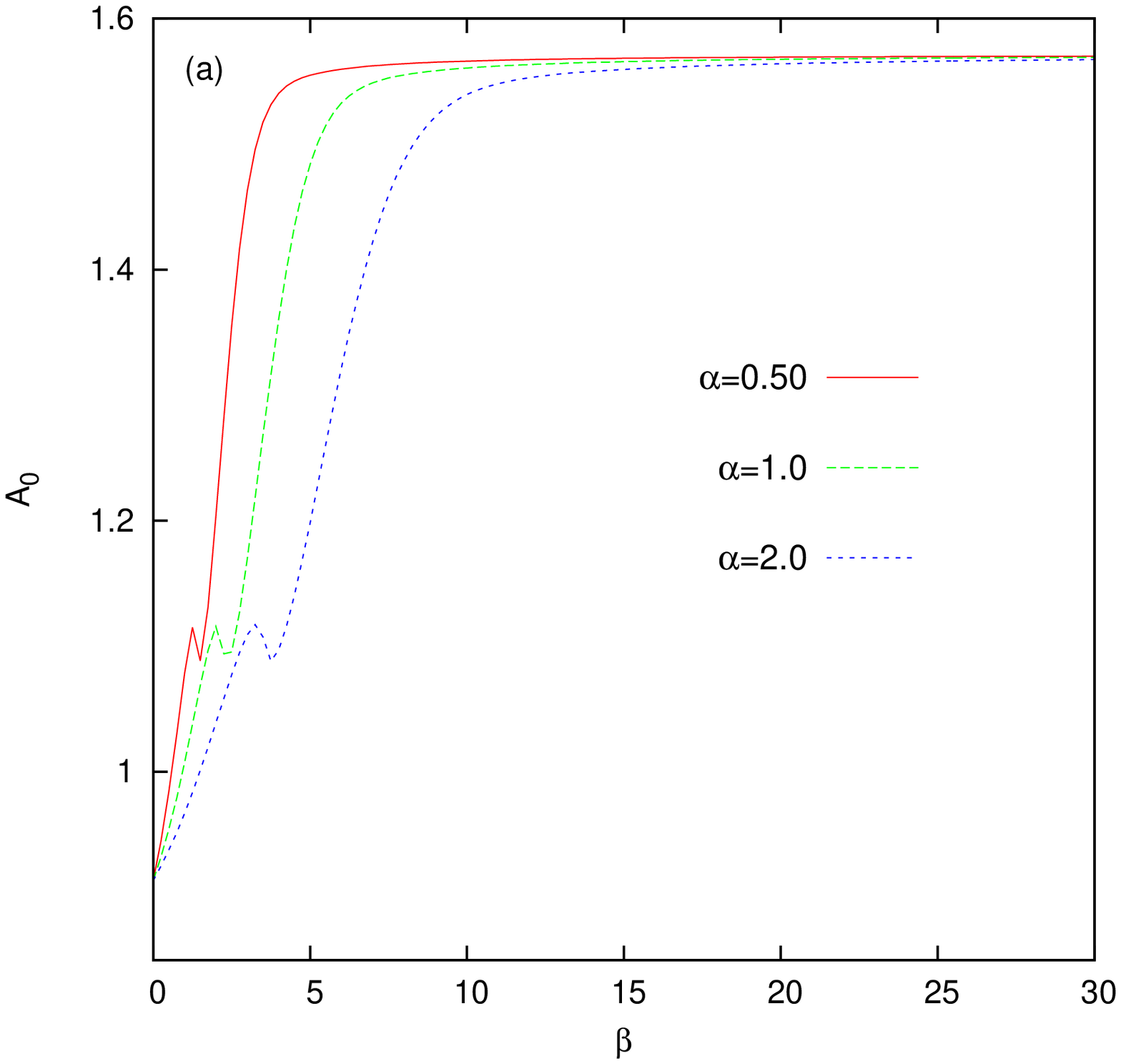}
\end{minipage}
\hspace{0.25in}
\begin{minipage}[c]{0.28\textwidth}\centering
\includegraphics[scale=0.30]{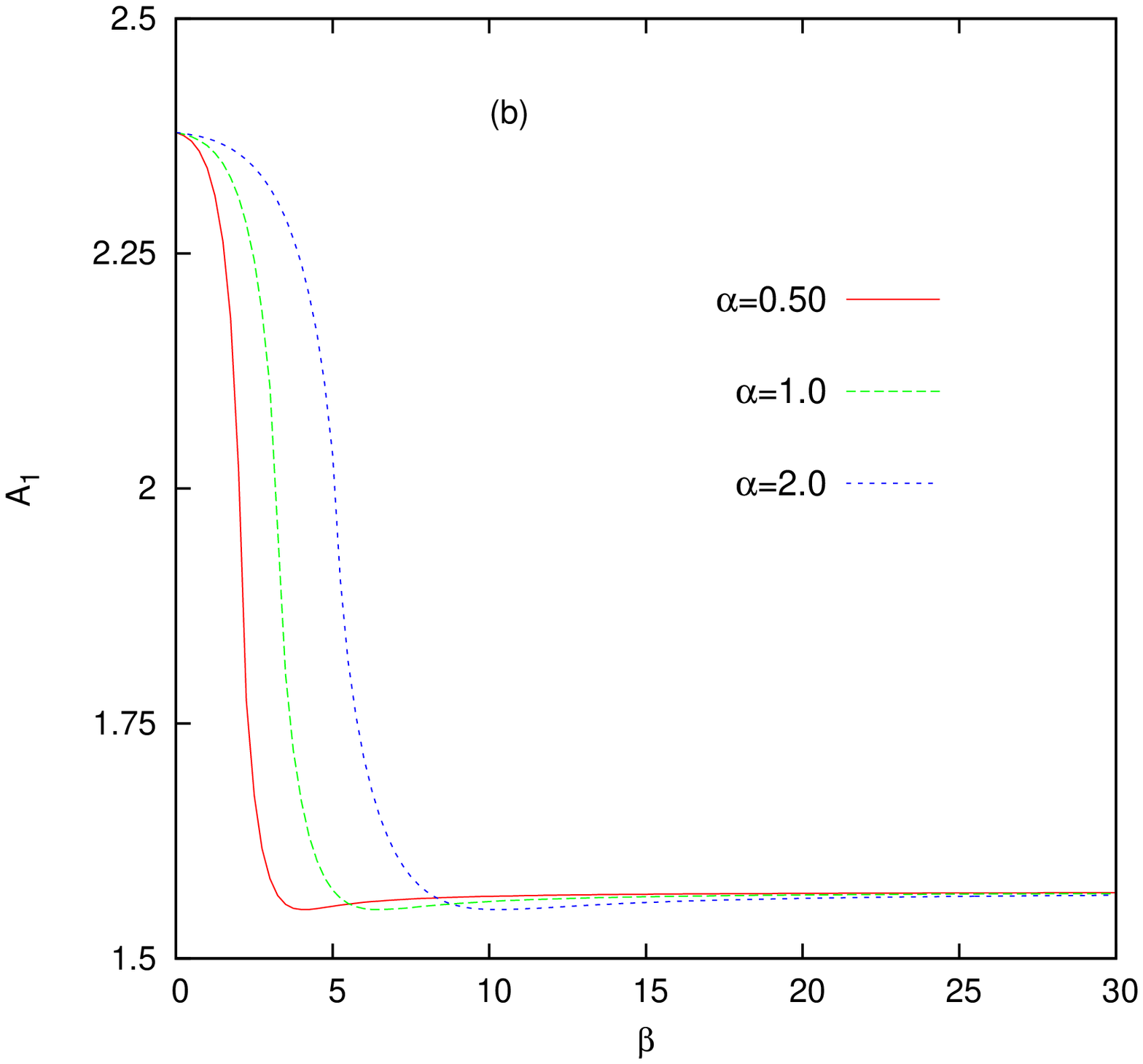}
\end{minipage}
\hspace{0.25in}
\begin{minipage}[c]{0.28\textwidth}\centering
\includegraphics[scale=0.30]{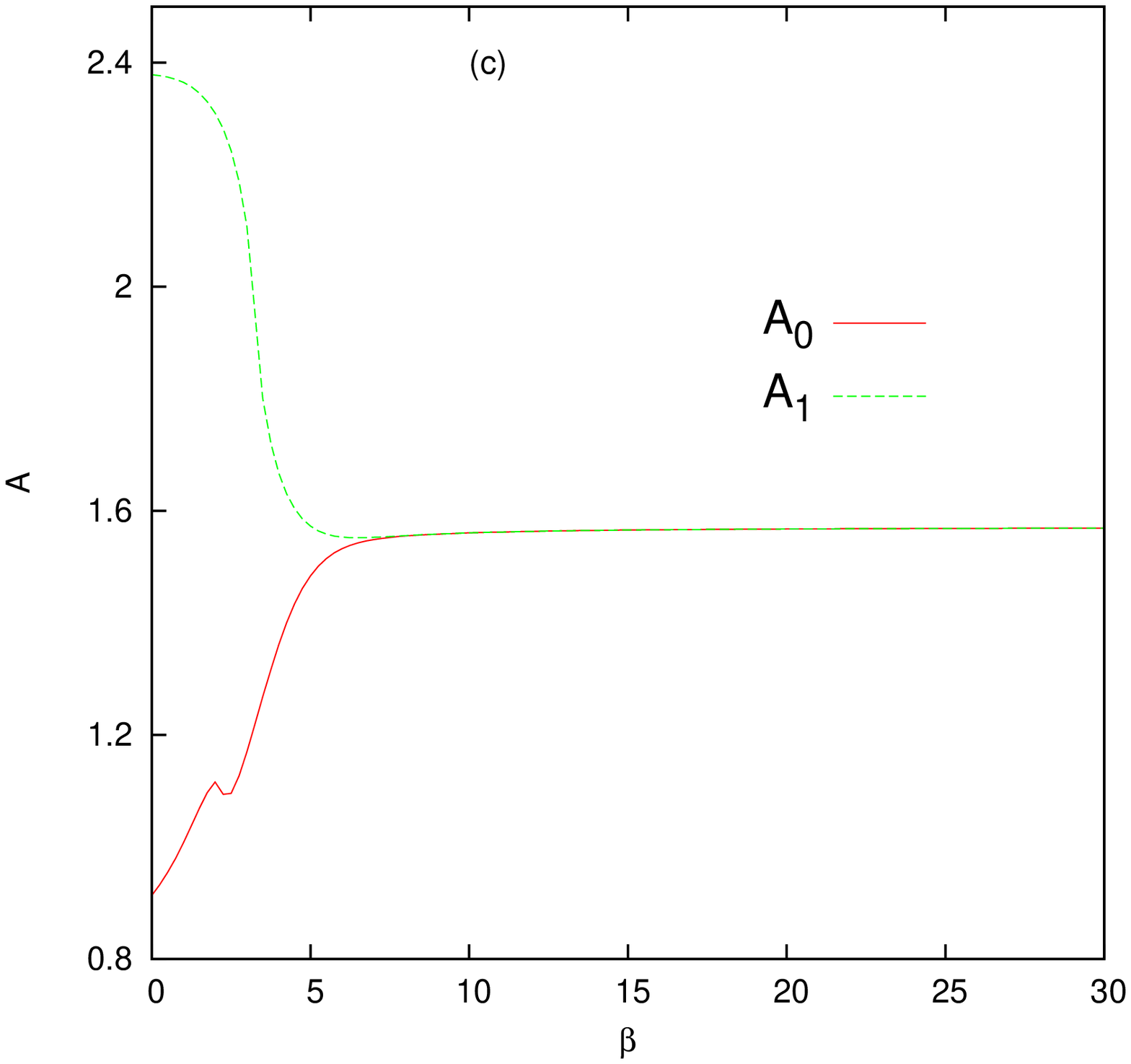}
\end{minipage}
\\[10pt]
\begin{minipage}[c]{0.28\textwidth}\centering
\includegraphics[scale=0.30]{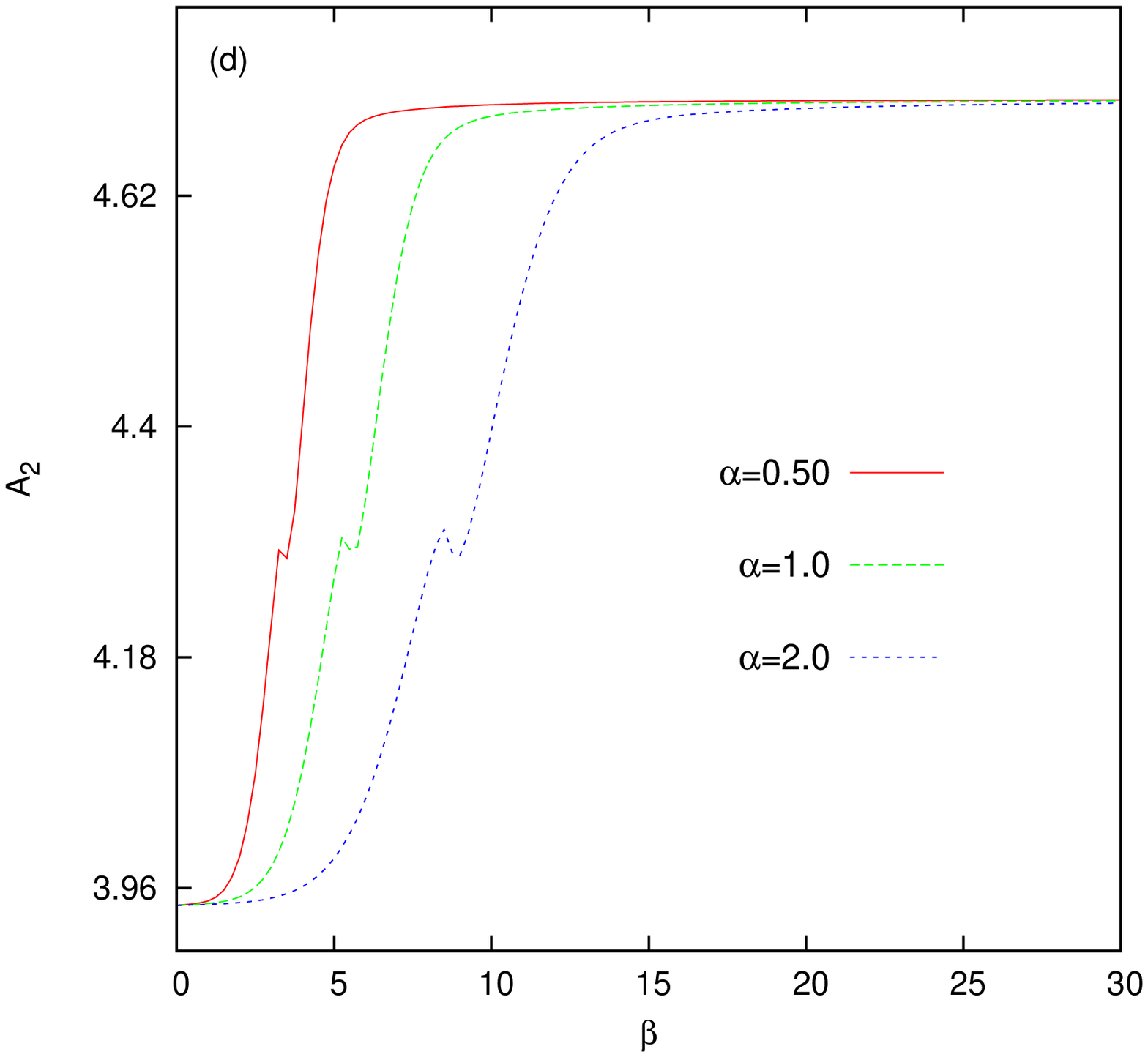}
\end{minipage}
\hspace{0.25in}
\begin{minipage}[c]{0.28\textwidth}\centering
\includegraphics[scale=0.30]{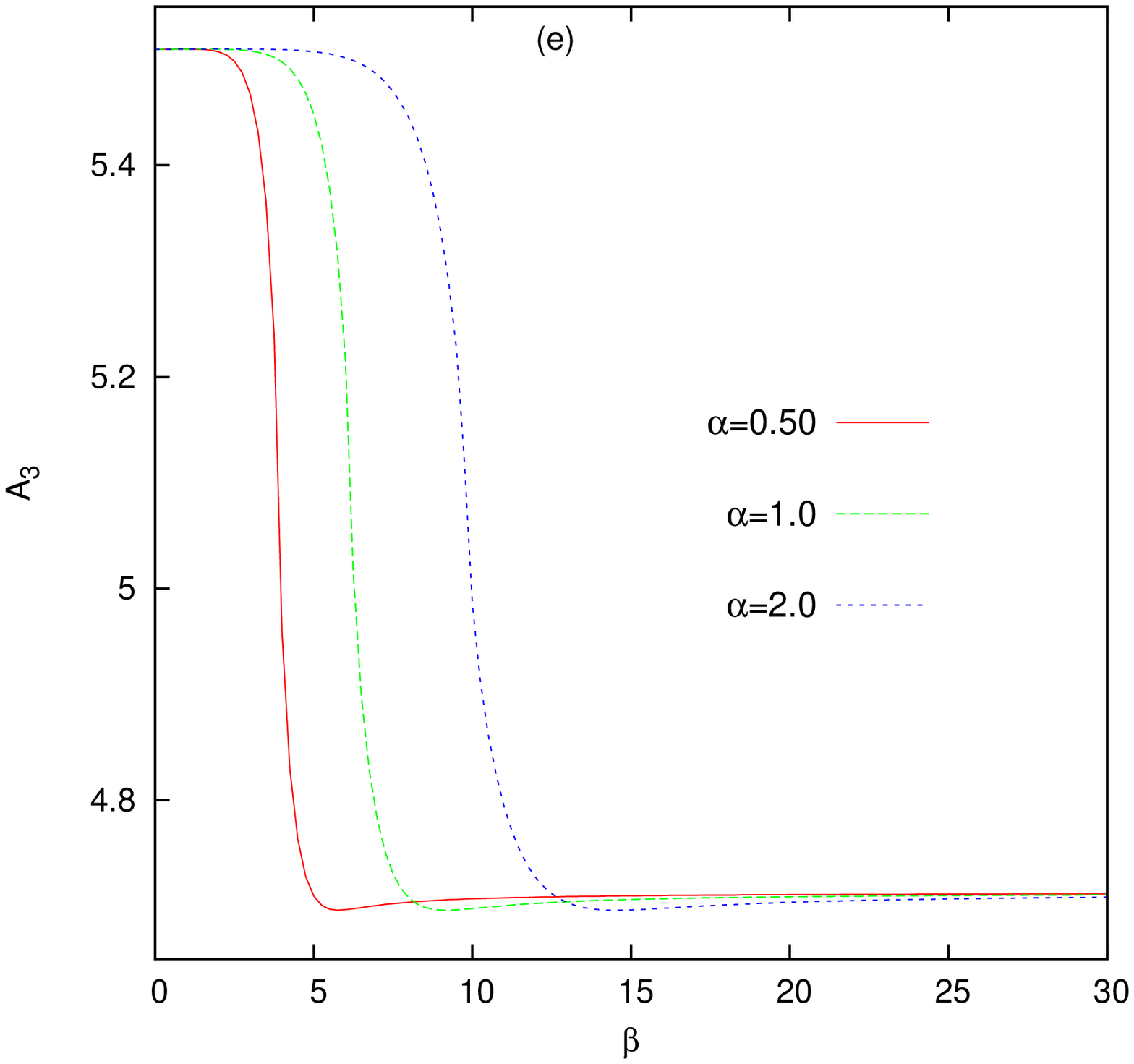}
\end{minipage}
\hspace{0.25in}
\begin{minipage}[c]{0.28\textwidth}\centering
\includegraphics[scale=0.30]{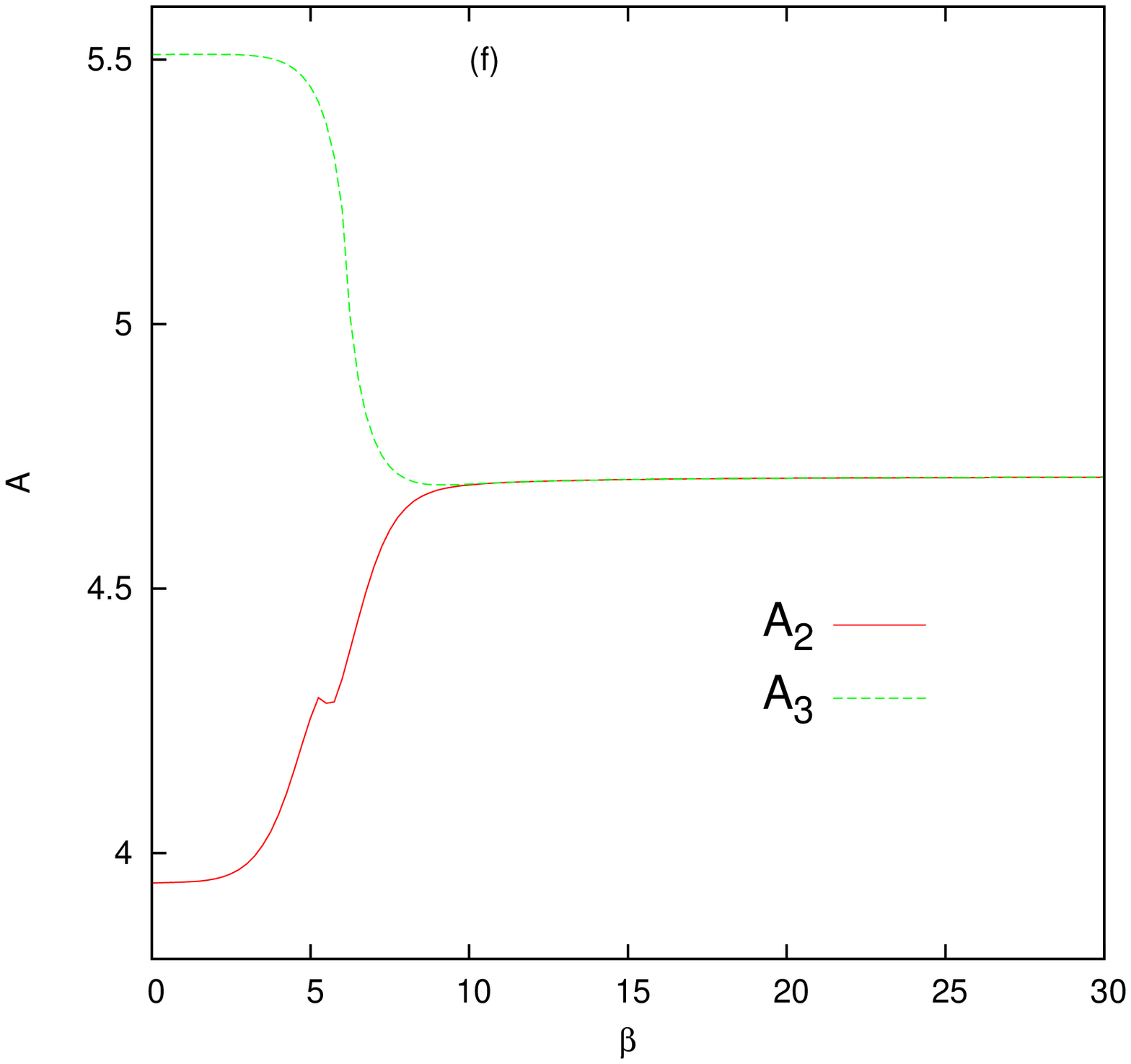}
\end{minipage}
\caption{Phase-space area with varying $\beta$ at three different $\alpha \ (0.5, 1, 2)$ for 
the DW potential in Eq.~(6): (a), (b), (d), (e) correspond to ground ($A_{0}$), first ($A_{1}$), 
second ($A_2$) and third ($A_3$) excited states respectively. Panels (c), (f) give convergence 
of phase-space area at $\alpha=1.0$: former refers to $A_0$, $A_1$, while latter corresponds 
to $A_2$, $A_3$ pair respectively. See text for details.}
\end{figure}

\section{Conclusion}
A number of entropy-based information measures such as, Fisher information, Shannon entropy, 
Onicescu energy and Onicescu-Shannon information were used to understand the quantum phenomena in 
a DW potential. Calculations in position and momentum space were carried out by means of 
variational exact diagonalization method. It is found that Fisher information measures 
are generally inadequate to explain the role of competing effects which 
causes quasi-degeneracy in such a system. Like the traditional uncertainty relation, Eq.~(1), 
these results indicate only quasi-degeneracy. Fortunately, the remaining three measures (in 
both position, momentum space as well as total) could be used to analyze quantum
phenomena like tunneling, confinement and quasi-degeneracy. The rates of changes of $S$ and 
$E$ have also been examined, which offer characteristic points of inflection at the onset of 
tunneling; one also notices exact harmonic trends at and after a certain value of $\beta$ for 
first and second quasi-degenerate pairs (threshold value of $\beta$ being unique for each pair). 

Semi-classical phase space of the particle in DW has also been investigated, for ground and 
first three excited states. One finds that, the onset of tunneling (in terms of $\beta$) 
corresponds exactly to a splitting of the closed phase-space area into two symmetric closed 
lobes, transition amongst which is classically forbidden. For ground state, another interesting 
phenomenon is revealed which marks these events in terms of the appearance of a clearly defined 
maximum in the rate of change of IE in position space. However, a similar phenomenon in 
$\frac{dS_{x}}{d\beta}$ for first excited state is missing.
 
Behavior of second excited state from the perspective of information in both position and 
momentum space is found to be considerably different from all others--sudden appearances of shoulders 
and damping of total information with increasing $\alpha$ are uncovered, which are quite unique 
to this state. Unlike other states, this leads to a yet unexplained qualitative shift in this 
state with $\alpha$--which warrants a deeper study, which may be carried out in future.  

In chemical perspective, one can say that umbrella flipping of NH$_3$ from one vibrational 
state to another happens possibly due to the fact that both these states have equivalent 
information content. Similarly, in a racemic mixture also, both compounds exist in equal 
proportion, which could be attributed to their possessing equivalent information. It will be 
interesting to investigate the behavior of IE in an asymmetric DW potential where there is also such
interplay of competing interactions.

\section{Acknowledgement}
Authors are grateful to two anonymous referees and the Editor for their kind, constructive suggestions 
and comments, which significantly improved the manuscript.  
NM thanks IISER Kolkata for a Post-doctoral fellowship. Mr.~Aakash Sarkar is thanked for useful discussion.  

\end{document}